\documentclass[fleqn,usenatbib]{mnras}
\usepackage{newtxtext,newtxmath}
\usepackage[T1]{fontenc}

\DeclareRobustCommand{\VAN}[3]{#2}
\let\VANthebibliography\thebibliography
\def\thebibliography{\DeclareRobustCommand{\VAN}[3]{##3}\VANthebibliography}

\usepackage{graphicx}	% Including figure files
\usepackage{amsmath}	% Advanced maths commands
\usepackage{wrapfig}
\usepackage[utf8]{inputenc}
\usepackage{makecell}
\usepackage[utf8]{inputenc}
\usepackage[english,french]{babel}
\AtBeginDocument{\selectlanguage{english}}
\AtBeginDocument{}
\usepackage{lscape}
\usepackage[flushleft]{threeparttable}
\usepackage{pdflscape}
\usepackage{subcaption}
\usepackage{gensymb}
\usepackage{longtable}
\usepackage[export]{adjustbox}
\usepackage{orcidlink}

\newcommand{\kms}{{\rm km\,s^{-1}}}
\newcommand{\Myr}{{\rm Myr}}
\newcommand{\kpc}{{\rm kpc}}
\newcommand{\hh}{H{\sc ii}}
\newcommand{\ho}{H{\sc i}}
\newcommand{\ha}{$\rm H\upalpha$}
\newcommand{\hb}{$\rm H\upbeta$}
\newcommand{\oo}{[O{\sc ii}]}
\newcommand{\ool}{[O{\sc ii}]\,$\uplambda$3727}
\newcommand{\ooo}{[O{\sc iii}]}
\newcommand{\oool}{[O{\sc iii}]\,$\uplambda$5007}
\newcommand{\nn}{[N{\sc ii}]}
\newcommand{\nnl}{[N{\sc ii}]\,$\uplambda$6584}

\newcommand{\nnp}{[N{\sc ii}]\,$\uplambda$6548+$\uplambda$6584}
\newcommand{\si}{[S{\sc ii}]}

\newcommand{\sip}{[S{\sc ii}]\,$\uplambda$6716+$\uplambda$6731}
\newcommand{\ergcs}{erg cm$^{-2}$\ s$^{-1}$}
\newcommand{\Arp}[1]{$\rm Arp\,{#1}$}
\newcommand{\NGC}[1]{$\rm NGC\,{#1}$}
\newcommand{\IC}[1]{$\rm IC\,{#1}$}

\title[A SITELLE view of the Arp$\,$143 interacting system]{Unveiling a cosmic tango: Integral field spectroscopy and numerical simulations of Arp$\,$143's interaction}

\author[H. Martel et al.]
{Hugo Martel$^{1}$\orcidlink{0000-0003-2917-2538}
\thanks{E-mail: Hugo.Martel@phy.ulaval.ca},
Carmelle Robert$^{1}$\orcidlink{0000-0003-2344-6593},
Laurent Drissen$^{1}$\orcidlink{0000-0003-1278-2591},
Charles-Antoine Parent$^{1}$\orcidlink{0009-0007-7505-8564},
Prime Karera,$^{1}$
\newauthor S\'ebastien Vicens-Mouret$^{1}$\orcidlink{0009-0000-0151-938X}, Salvador Duarte Puertas$^{2}$\orcidlink{0000-0002-5542-1940}, Jorge Iglesias
P\'aramo$^{3,4}$\orcidlink{0000-0003-2726-6370}
and     
\newauthor Jos\'e M. Vilchez$^{3}$\orcidlink{0000-0001-7299-8373}
\\
\\
$^{1}$D\'epartement de Physique, de g\'enie physique et d'optique, Universit\'e Laval, G1V 0A6, Qu\'ebec (QC), Canada\\
$^{2}$Departamento de F\'\i sica Te\'orica y del Cosmos, Campus de Fuentenueva, Edificio Mecenas, Universidad de Granada, E-18071 Granada, Spain\\
$^{3}$Instituto de Astrof\'\i sica de Andalucía, CSIC, Apartado de correos 3004, 18080 Granada, Spain\\
$^{4}$Centro Astron\'omico Hispano en Andaluc\'\i a, Observatorio de Calar Alto, Sierra de los Filabres, 04550 G\'ergal, Spain\\
}

\date{Accepted 2026 June 19. Received 2026 June 18; in original form 2025 November 28
}

\pubyear{2026}

\begin{document}
\raggedbottom
\label{firstpage}
\pagerange{\pageref{firstpage}--\pageref{lastpage}}
\maketitle

\begin{abstract}
We present spectral data cubes of the interacting galaxy system \Arp143, composed of the ring galaxy \NGC2445 and the lenticular galaxy \NGC2444, obtained with the imaging Fourier transform spectrometer SITELLE at the Canada-France-Hawaii Telescope. Our data allow to probe the kinematics of the interaction and the chemical properties of the ionized gas. Star-forming regions are almost exclusively found in the ring and nucleus of \NGC2445 with the exception of a few very faint ones discovered at the base of the long tidal plume extending north of \NGC2444. Analysis of the \ha\ velocity map reveals strong non-circular flows in the ring of \NGC2445, which is expanding. Its nucleus is off-centered with respect to the ring. Oxygen abundance in the ring is on average slightly sub-solar whereas it is close to solar in the nucleus. Broad-band images obtained with the Dragonfly Telephoto array allow to identify the tenuous stellar counterpart of the radio tidal plume. The interaction between the two galaxies is simulated with a chemodynamical evolution code; these simulations suggest that \Arp143 has resulted from a head-on collision between a S0 and a Sc spiral galaxy following a flyby encounter that triggered the formation of the long plume from debris of the disk galaxy. 
\end{abstract}


%%%%%%%%%%%%%%%%%%%%%%%%%%%%%%%%%%%%%%%%%%%%%%%%%%

%%%%%%%%%%%%%%%%% BODY OF PAPER %%%%%%%%%%%%%%%%%%

\section{Introduction}
Interactions play an important role in the evolution of galaxies. Numerous statistical studies based on large observational surveys suggest that galaxy-galaxy interactions are associated with morphological disturbances \citep[e.g,][]{Arp66, Arp-Madore87, Hernandez2005, Patton2016}, enhanced star formation rates (SFR) and a dilution of central metallicities \citep[e.g,][]{Ellison2008, Scudder2012}, elevated active galactic nuclei (AGN) fractions \citep[e.g,][]{Ellison2011, Satyapal2014}, and increased atomic and molecular gas fractions \citep[e.g,][]{Violino2018, Dutta2019}. These changes in global physical properties are a function of projected separations (which can be seen as a proxy for the interaction stage), mass ratios, orbital geometries, and intrinsic properties before the interaction. On the theoretical side, these signatures are in agreement with those predicted by numerical simulations taking into account broad orbital parameters \citep[e.g,][]{DiMatteo2005, DiMatteo2007, Rupke2010, Moreno2015, Moreno2019}. 

The main physical mechanism driving the interaction-induced changes in central SFR, nuclear metallicity, and AGN activity is the growth of non-axisymmetric structures that cause strong inflows of gas from the periphery to the galactic centre. Metal-poor gas funneling to the central regions enhances the central SFR and AGN activity, and dilutes the nuclear metallicity. The same gas inflow can also induce a suppression of star formation (SF) in the outer parts of the galaxies and a redistribution of metals, producing flatter abundance gradients \citep[][]{Moreno2015}. However, gas inflow alone cannot explain all properties of galaxy-galaxy interactions since, for example, SF is not always constrained to central regions. It can be intense and spatially extended \citep[][]{Martel2013,Renaud2015}. Under favourable conditions, gravitationally bound tidal dwarf galaxies (TDGs) can form up to the tip of tidal tails \citep[][]{Duc2012}. In order to get a full picture of interaction-induced effects both globally and locally, it is, therefore, necessary to spatially map the SF and metallicity distribution in a given pair of interacting galaxies. 

A complementary approach to the above statistical studies of galaxy-galaxy interaction is possible with a detailed analysis of specific targets, combining integral field spectroscopy with numerical simulations. Indeed, comparing numerical models of interacting galaxies with those of the same galaxies in isolation allows one to identify interaction-induced effects on the physical properties reproduced, as well as their temporal evolution. We followed this approach to study the interacting systems \Arp82 \citep[][hereafter Paper~I]{Karera2022} and \NGC2207/$\rm IC\,2163$ \citep{Poitras26}. We follow the same road in this paper to study the ionized gas properties in \Arp143.

\begin{figure}
    \includegraphics[width=\columnwidth]{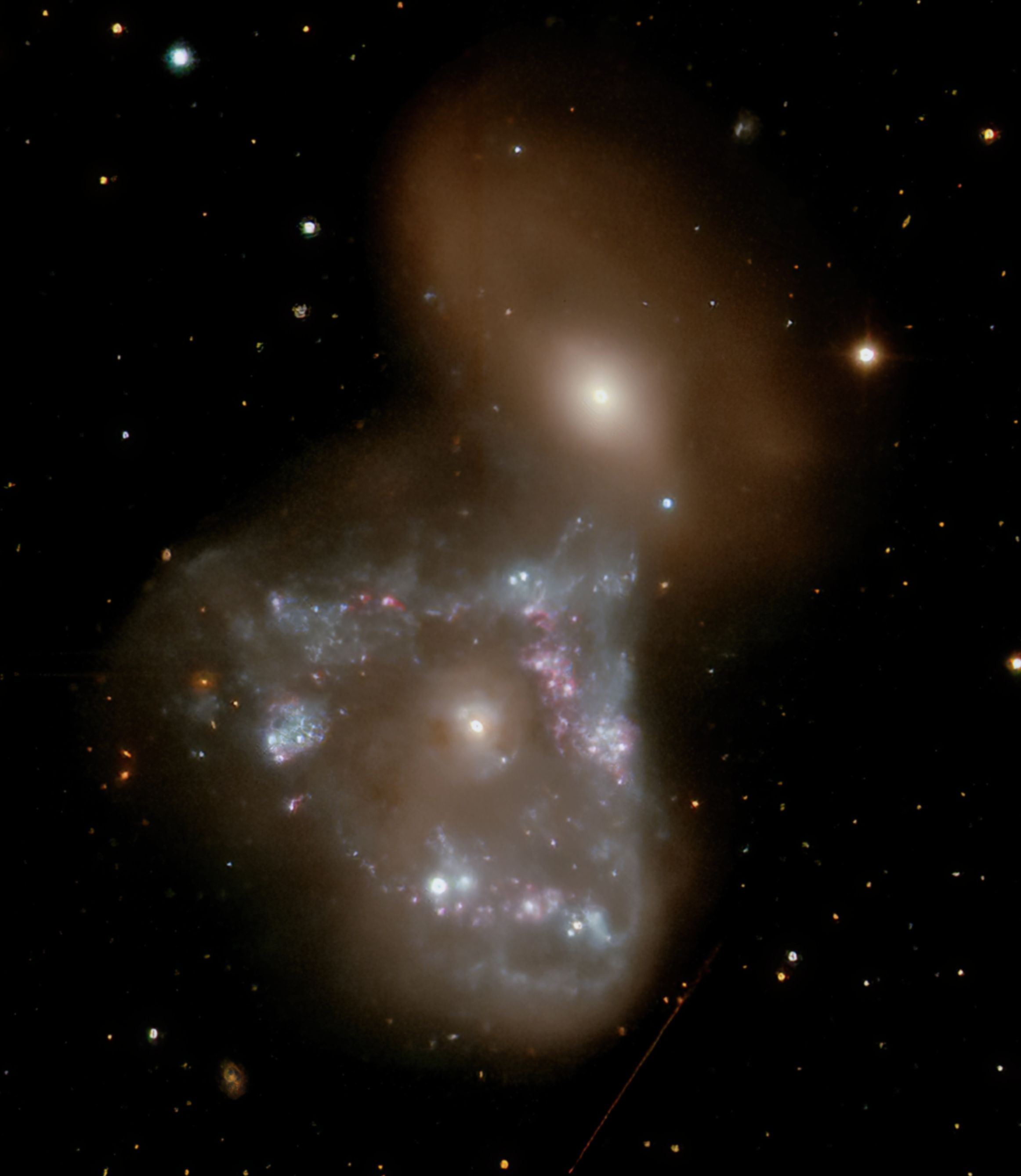}
    \caption{Composite coloured image of the central part of \Arp143 obtained with SITELLE (blue: SN2 deep image, red: SN3 deep image enhanced with the pure H$\alpha$ emission flux map, green: mean of SN2 and SN3 deep images). The image spans $3' \times 3.6'$, which is about 10\% of the total SITELLE FOV. North is up and East is left.}
    \label{fig:SITELLEdeepimages_figure}
\end{figure}

\Arp143 consists of the interacting galaxies \NGC2444 (S0) and \NGC2445 (Im pec)  one of the rare nearby collisional ring galaxy as can be seen in Fig.~\ref{fig:SITELLEdeepimages_figure}. The ring of young bright stellar clusters and \hh\ regions inside the disk of \NGC2445 is patchy and asymmetric with spokes or arms connecting it to the nucleus. The arms stand out more clearly in mid-infrared images \citep[][]{Beirao2009}. Ringlike propagating density waves are known to develop in the disk of a spiral after it collides with a more compact galaxy more or less along its axis of rotation \citep[][]{LyndsToomre1976}. \Arp143 might have resulted from a similar process as kinematics of \ho\ in the ring show signs of expansion \citep[][]{Higdon1997}. Broad-band optical/near-infrared photometry of stellar clusters in the ring indicates very young ages ($\le10^7\hbox{yrs old}$), much younger than the dynamical age of the \ho\ density wave, suggesting that SF has only recently been triggered in the ring \citep{Appleton1992}. However, the detection of a narrow and very long plume ($\rm\sim150\,kpc$) in \ho\ data \citep[][]{Appleton1987} implies that the formation of \Arp143 is more complicated. A faint optical filament was also found associated with the inner part of the plume. Numerical models of \Arp143 exist \citep[][]{Narasimhan2003,Holincheck2016} but do not reproduce the extended plume. Unpublished spectroscopy from \cite{Jeske1986}, cited by \cite{Higdon1995}, derive metallicities ($\rm 12+log[O/H]$) between 8.55 and 8.77 for the ring's brightest \hh\ regions. \cite{Romano2008} estimate a global SFR of $5.6\,\rm M_\odot\,yr^{-1}$ for \Arp143 using far-infrared luminosities. 

\hyphenation{SITELLE}
In this paper, we study the kinematics and metallicity distribution in the ionized gas of \Arp143, based on the analysis of spectral data cubes obtained using the imaging Fourier transform spectrometer SITELLE \citep[][]{Drissen2019} attached to the Canada-France-Hawaii Telescope (CFHT). The instrument has a wide field of view (FOV; $11^\prime \times 11^\prime$) sampled at 0.32$^{\prime\prime}$ per pixel. We also present broad-band images obtained with the Dragonfly Telephoto Array \citep[][]{Abraham2014}. Finally, to gain insight into the physical processes that lead the formation of this system, we present a 3D numerical model of an interaction between two galaxies, and compare the resulting morphology, kinematics, and chemical properties with the observations. Throughout this paper, we adopt a distance of $56.7\,\rm Mpc$ \citep[][]{Beirao2009}. At this distance, 1$^{\prime\prime}$ corresponds to $275\,\rm pc$.

This article is organised as follows. In Section~\ref{sec:section2} we describe our observations, data reduction, and calibration. Data analysis is presented in Section~\ref{sec:section3}. Numerical simulations are presented in Section~\ref{sec:section4}. Results are discussed in Section~\ref{sec:section5} and summarized in Section~\ref{sec:section6}.

\section{OBSERVATIONS, DATA REDUCTION, AND CALIBRATION}
\label{sec:section2} 

\subsection{SITELLE}
\label{sec:section21}
 
 \begin{figure*}
    \includegraphics[scale=0.34]{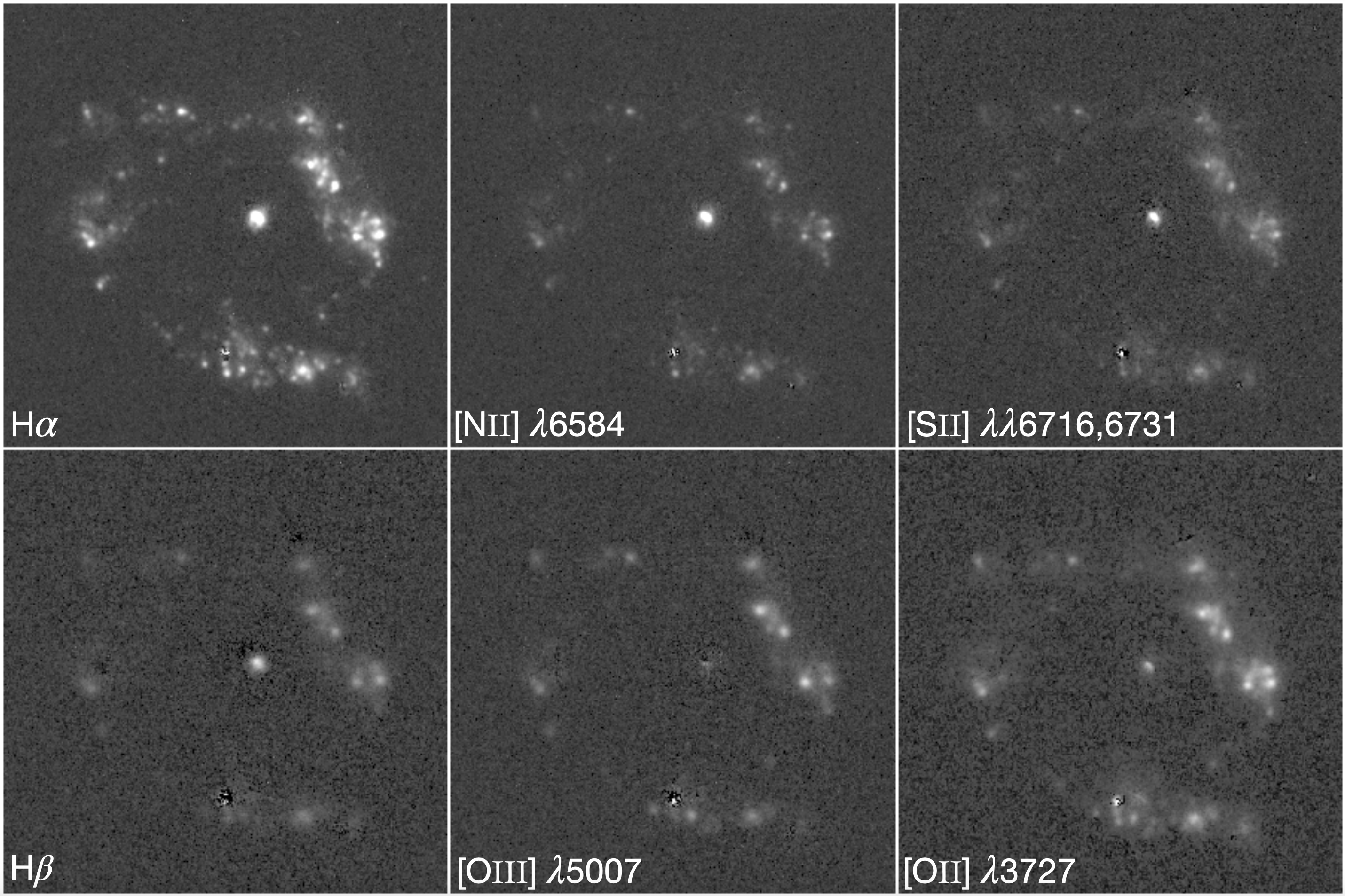}
    \caption{Emission lines flux maps of \NGC2445 extracted from the data cubes. The FOV is $1.6' \times 1.6'$, with North at the top and East to the left.}
    \label{fig:lines_figure}
\end{figure*}
 
\Arp143 was observed with SITELLE in 2016 November 28 -- 30 (SN2 and SN3 filters) and 2017 October 19 -- 20 (SN1 filter) under photometric conditions. Details and technical specifications of the observations are provided in Table~\ref{tab:table1}. Following the steps described in Paper I, data reduction, wavelength and photometric calibrations were carried out using \textsc{orbs} and \textsc{orcs} \citep[][]{Martin2015,Martin2016,Martin21}, SITELLE's dedicated data reduction and analysis pipelines. Photometric calibration is secured from images and data cubes of the spectrophotometric standard star GD71, complemented with a comparison with the Gaia spectra of two dozen stars in the FOV \citep{Vicens26}. Wavelength calibration is performed using a high spectral resolution laser data cube. In the SN3 filter, it is improved by measuring the centroid positions of the night sky OH emission lines, ensuring uncertainties on the velocity calibration across the entire field inferior to $2\,\kms$. Because of its highest spectral resolution and precise wavelength calibration, we use the SN3 data cube to derive all the kinematical properties of the system.

 \begin{table}
    \caption{SITELLE observations}\label{tab:table1}
    \centering
    \small
    \begin{tabular}{cccccc}
    \hline
    Filter & \makecell{Band \\ \small [nm]}
    & \makecell{Spectral \\ resolution} 
    & \makecell{Exp. time \\ per step \\ \small [sec]}
    & \makecell{Number \\ of steps} 
    & \makecell{Image \\ quality \\ \small [arcsec]} \\
    \hline
    SN1 & 363 -- 386 &  500 & 87 &  86 & 1.11\\
    SN2 & 482 -- 513 &  970 & 40 & 225 & 1.36\\
    SN3 & 647 -- 685 & 1900 & 25 & 337 & 0.96\\
    \hline
    \end{tabular}
\end{table}
    
After the subtraction of a median sky spectrum, spectral fits are performed for each spaxel using a sincgauss profile (a sinc function convolved with a Gaussian, to account for line broadening). Maps of parameters such as fluxes of fitted emission lines, continuum level, velocity, velocity dispersion, and their corresponding uncertainties are then produced. Flux maps of extracted \ool, \hb, \oool, \ha, \nnl, and  \sip\ are presented in Fig.~\ref{fig:lines_figure}. The total \ha+\nnp\ flux measured by SITELLE in the area shown in this figure, $8.13\times10^{-13}\rm erg\,s^{-1}cm^{-2}$, is about 12\% higher than that measured by \cite{Romano2008} using narrow-band filter imagery.

\subsection{Dragonfly}
\label{sec:section22}

The Dragonfly Telephoto array is a robotic telescope array that consists of 48 Canon $\rm400\,mm$ $f/2.8$ IS II lenses with special anti-reflective coatings on optical surfaces designed to reach surface brightness levels below $\sim28\hbox{--}29\,\rm mag\,arcsec^{-2}$ \citep[][]{Abraham2014}. It has a FOV of $2.6^\degree\times1.9^\degree$ with a pixel scale of $2.8^{\prime\prime}$. Its configuration makes it optically equivalent to a 1m $f/0.4$ refracting telescope with the same wide FOV. Half of the array is equipped with Sloan Digital Sky Survey g~filters, and the other half with r~filters. The instrument therefore always takes simultaneous data in these two filters. 

We observed \Arp143 with Dragonfly over three nights in April of 2022. A total of 55 frames ($\rm600\,s$ each) were obtained in the g-band and 77 frames (also $\rm600\,s$ each) in the r-band made it through the automated reduction pipeline \citep[][]{Danielli2020}. The frames are co-added to produce deep images in the two filters. Analysis of these data is presented in Section~\ref{sec:plume}.

\section{DATA ANALYSIS}
\label{sec:section3} 

\subsection{Properties of \hh\ regions}
\label{sec:section32}

\begin{figure*}
    \includegraphics[scale=0.55]{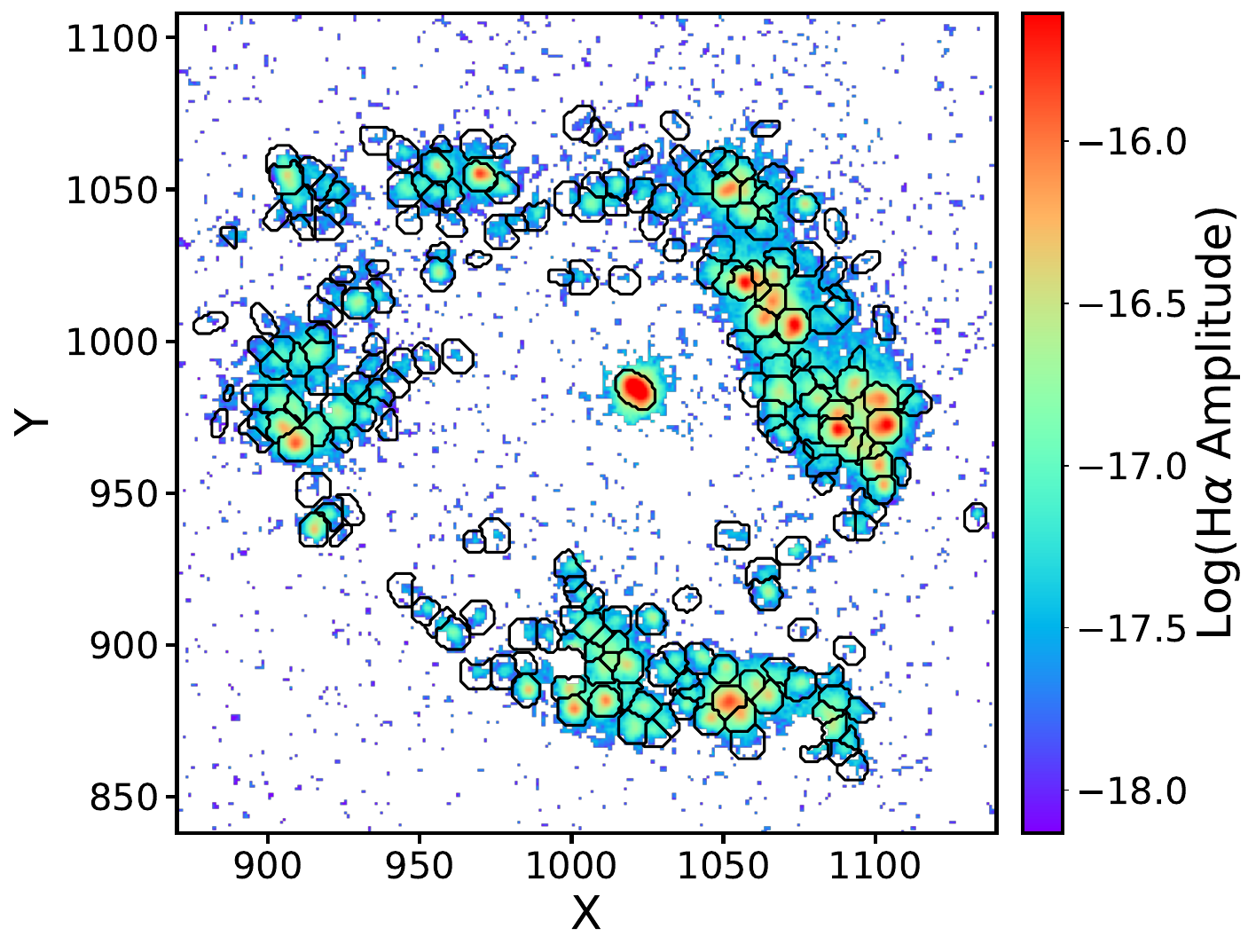}
    \caption{\hh\ region complexes domain (black contours) identified by the detection code overlaid on the \ha\ amplitude map of \NGC2445.
    }
    \label{fig:HIIregions_figure}
\end{figure*}

\begin{figure*}    
    \includegraphics[scale=0.395]
    {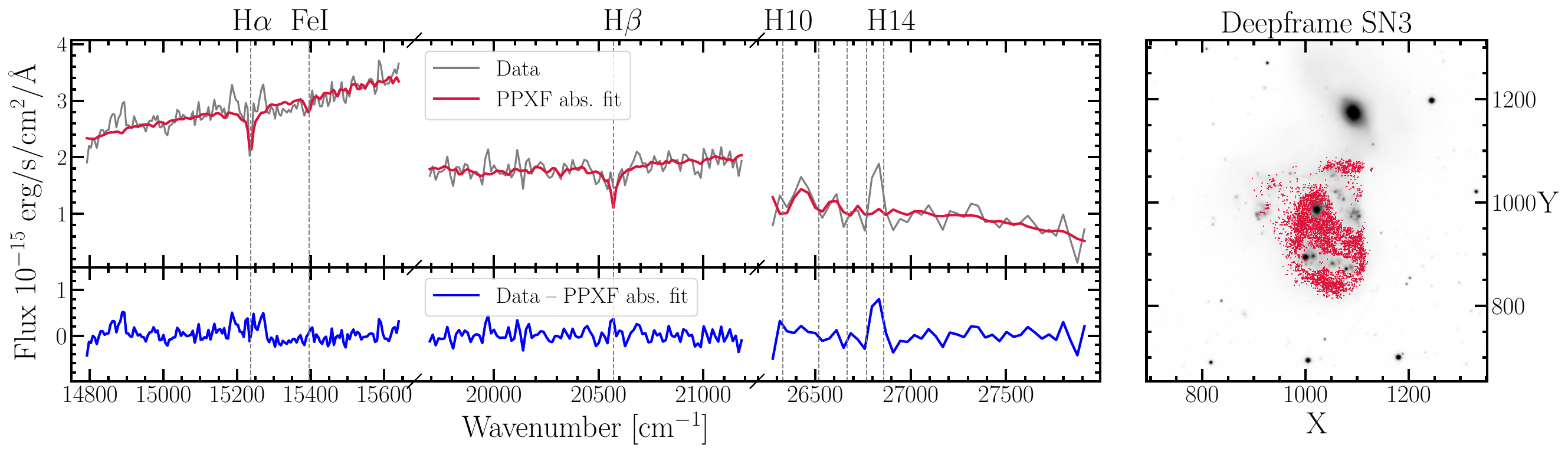}
    \caption{Left: PPXF absorption model for the global stellar population of the ring galaxy. In black, the integrated spectrum of $\sim5200$ pixels with a continuum SNR above 7 and an \ha\ emission flux below $1\times10^{-19}$ \ergcs, fter being set to the restframe velocity using our rotation model. In red and blue, the PPXF absorption fit and residual, respectively. Right: selected pixels for the stellar population model are shown in red superposed on the SN3 deep frame.} 
    \label{fig:ppxf_figure}
\end{figure*}

\subsubsection{Region detection and integrated spectra}
\label{sec:detection}

We identified the \hh\ region complexes using our latest version of the detection code adapted to the resolution of SITELLE's data \citep[see ][]{Posternak25}. It is briefly summarized here. The algorithm detects peaked regions in the \ha\ amplitude map. A Laplacian operator, with a Gaussian filter for $\sigma=1.2$, is applied on the input amplitude map to facilitate the detection of the emission peaks in a box of $3\times3$ pixels. These numbers are selected to account for the data resolution and seeing. Spatial and spectral threshold limits are then applied on the detected peaks. 
The spatial threshold is a combination of the \ha\ amplitude background image $BG$ and its spatial noise image $Noi$; $BG$ is built by convolving the \ha\ amplitude map using a $20\times20$~pixel box while $Noi$ is the standard deviation in the detection box ($3\times3$~pixels here) of the residual of the \ha\ amplitude map minus the $BG$ image. 
The spectral threshold SNR3 is the \ha\ SNR map convolved with the detection box. Peaks with an amplitude below $BG + 3\times Noi$ and an SNR3 limit below 2.3 are then rejected. This leaves us with 265 peaks to study in an area covering the galaxies and the northern plume (i.e. from $x=800$ to 1250 and $y=780$ to 1850).   

In order to define the domain of emission region, we first create a zone of influence (ZoI) around each peak. This is done by assigning a peak to each pixel within a maximum of 500 pc radius, by maximizing the $A/r^2$ parameter (where $A$ is the peak amplitude and $r$ its distance). Next, a 2D Gaussian + plane profile is fitted (using an Markov Chain Monte Carlo routine)  centered on each peak using the \ha\ amplitude of the pixels within the ZoI. Finally, each emission region domain is spatially demarcated by the intersection of the $4\sigma$ Gaussian fit with the ZoI. A $4\sigma$ limit, instead of a classical $3\sigma$ limit, is selected to make sure that this domain will encompass most of the emission line flux in the SN2 filter with a poorer seeing (see Table~\ref{tab:table1}). A study of the profile fit, using the profile $\chi^2$ value within $3\sigma$ and the average standard deviation in a $3\times3$ pixel box for pixels in the ZoI, allows to keep a sample of 227 emission regions as good \hh\ region complex candidates. These regions are shown in Fig.~\ref{fig:HIIregions_figure} along with their respective domain. Most of the regions are located in and near the ring. Not shown in this figure are two other emission regions detected in the northern plume, labeled H1 and H2 as discussed in Section~\ref{sec:plume} and shown in Fig.~\ref{fig:PlumeHII} below; since only \ha\ (and a weak \ool\ line for H1) was detected in their spectrum, they are not included in the following analysis. 

The emission region domains defined using the \ha\ amplitude map, was projected over the SN1 and SN2 datacubes using the WCS coordinates of the pixels within each domain. Emission line fluxes for each region can then be measured with \textsc{orcs} using the integrated spectrum for all its domain pixel. Prior to the line flux measurement, we subtracted a stellar population model, created as described in the next subsection. This subtraction, which mainly affects the H$\beta$ line, allows a more accurate measurement of the extinction, SFR, and metallicity. A background emission (which includes the diffuse ionized gas, or DIG) is not subtracted here. This operation was tested and shown to leave fewer regions with a proper signal-to-noise ratio (SNR) for the analysis, therefore we decided not to do it. Nevertheless, the presence of DIG will not be ignored in our analysis.   

\subsubsection{Stellar population subtraction}
\label{sec:stelpop}

A subtraction of the local stellar population underneath each  emission regions is not possible with our data due to the low SNR in the continuum of individual pixels. Nevertheless, we were able to create a spectrum for the global stellar population by combining the spectrum of individual pixels with a high continuum SNR and very weak emission lines. Since the emission regions detected are located in the ring galaxy, we also masked the companion galaxy. 
Imposing a minimum value of 7 for the continuum SNR and a maximum value of  $1\times10^{-19}$\ergcs for the  \ha\ flux, 
about 5200 pixels have been selected allowing a global SNR in the continuum above 20. Their integrated spectrum and location over the ring galaxy can be seen in Fig.~\ref{fig:ppxf_figure}.
Prior to the summation of the individual pixel spectrum, we took into consideration their relative velocity, based on our rotation model centered on the nucleus of \NGC2445 (see Sec.~\ref{sec:velocity}). As seen in Fig.~\ref{fig:ppxf_figure}, the stellar population spectrum obtained still contains emission lines, but we use PPXF \citep{Cappellari17} to create a pure stellar population model from this spectrum (in red in Fig.~\ref{fig:ppxf_figure}). The version of PPXF used has been modified to take into account SITELLE's sinc instrumental profile for the emission lines. As expected, this model reproduced well the Balmer absorption lines, and the FeI line at 6495\AA\ ($\rm15\,396\,cm^{-1}$). 

Splitting the pixels used to create the stellar population spectrum in two groups, one group surrounding the galaxy nucleus in a few kpc circular aperture and one with the remaining more distant pixels, gave spectra with lower SNR, not ideal for a good fit with PPXF, but also with no real evidence for a distinct stellar population in the inner and outer galaxy. 

\begin{figure}   
\includegraphics[width=\columnwidth]{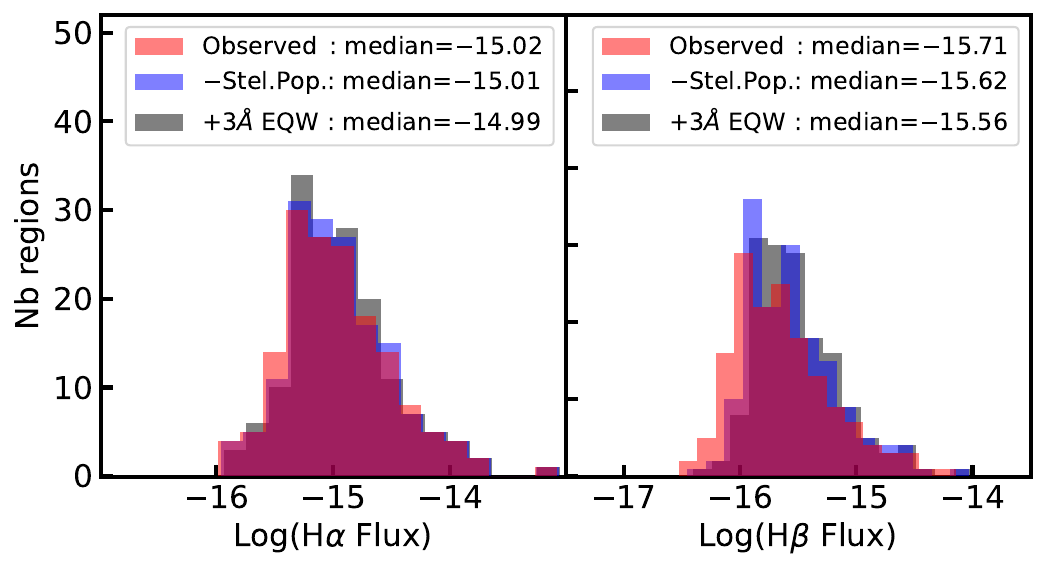}
    \caption{Comparison of the \ha\ and \hb\ flux of the emission regions for the observed values (red), with the stellar population correction (blue) and when a 3\AA\ EQW is simply added. Only regions with an $\rm SNR>3$ for both lines are considered. Median values for the 3 distributions are given on the plots.}
    \label{fig:HaHbcompflux_figure}
\end{figure}

For each emission region detected, the stellar population model was matched to the region's continuum level and line-of-sight velocity, prior to its subtraction. We are aware that this is still an approximation since the observed continuum could include as well a contribution from the cluster underneath an \hh\ region and from the nebular gas. Fig.~\ref{fig:HaHbcompflux_figure} shows the effect of the stellar population model subtraction on the flux of the \ha\ and \hb\ line. As a validity test for our stellar population model, this figure also shows the line flux values if a 3\AA\ equivalent width (EQW) is simply added to the observed lines. Median values of the distribution are indicated on the plots. The effect of the stellar subtraction is more important for \hb, as expected, and not as strong as if a 3\AA\ EQW is considered, as often used to represent the effect of a young or an old population \citep{Gonz05}.  

\subsubsection{Extinction}
\label{sec:extinc}

\begin{figure*}    
\includegraphics[scale=0.38]
{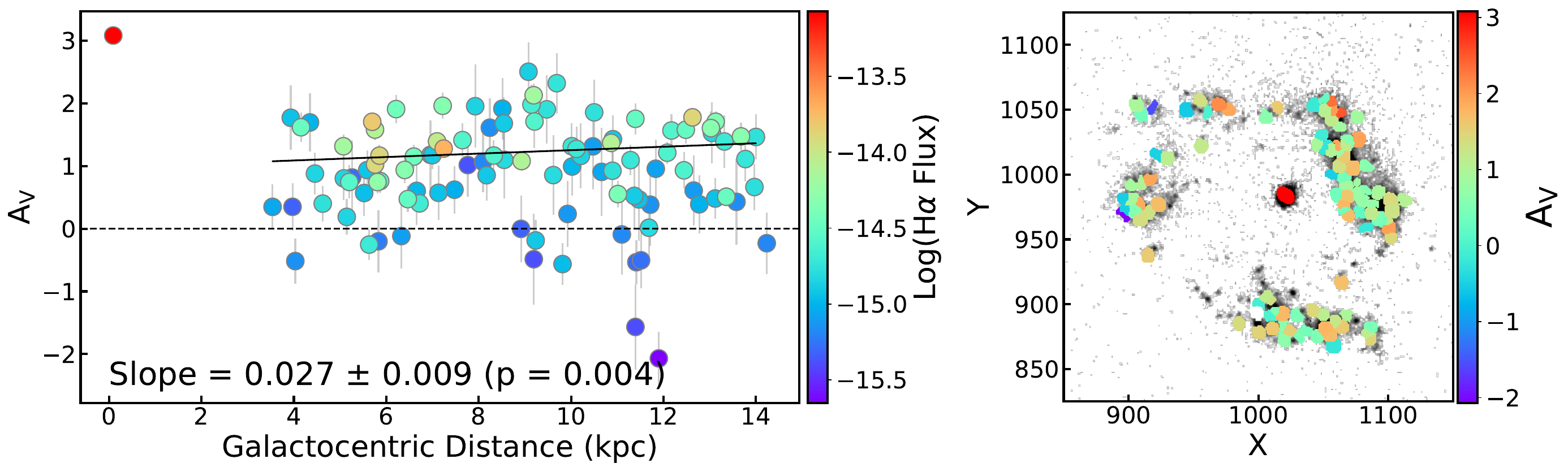}
    \caption{Dust extinction as a function of the galactocentric distance. Only regions with a $\rm SNR>5$ in \ha\ and \hb\ are shown. Left: each region is colour-coded by its \ha\ flux.  The dashed line indicate a 0 mag extinction. The black line represents a linear regression done while excluding the nucleus and regions with a negative extinction. The slope of the regression is given on the plot (along with its p-reliability test value). Right: map of the extinction plotted within the domain of the regions, and superposed to the \ha\ flux map.  
    }
    \label{fig:ext_figure}
\end{figure*}

\begin{figure*}    
\includegraphics[scale=0.38]
{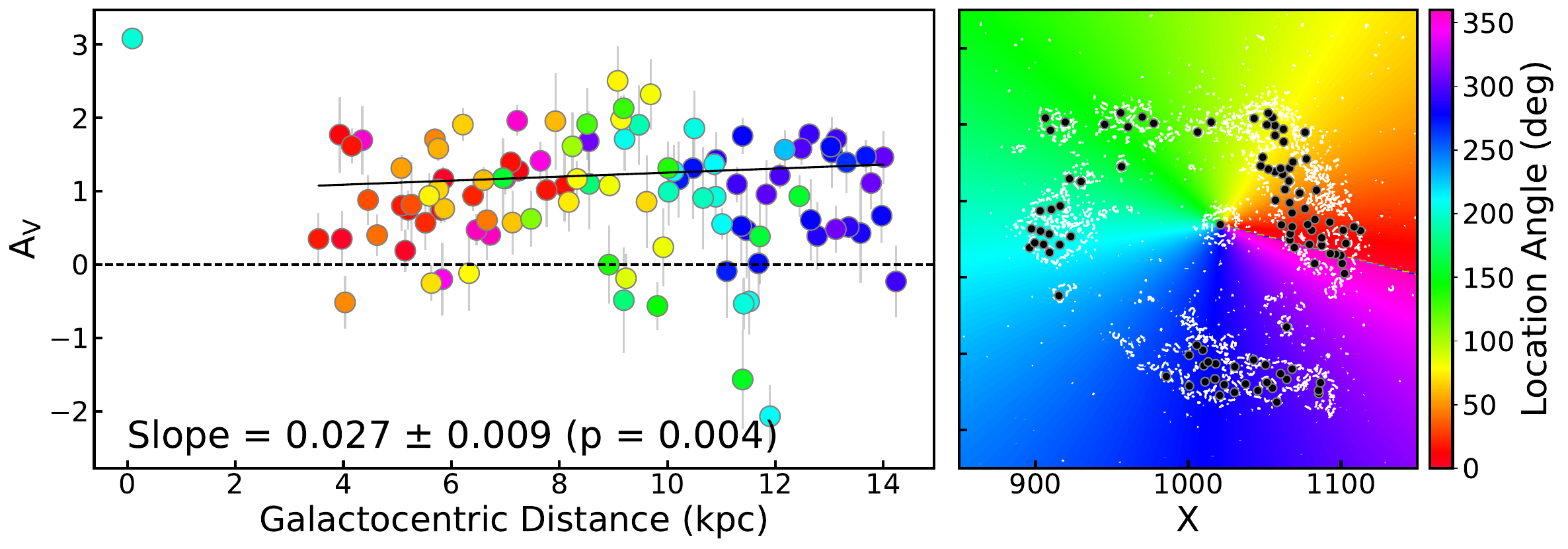}
    \caption{Dust extinction colour-coded by the location angle. Left: as in the previous figure, regions extinction is shown as a function of their GCD, with the linear regression, but here colour-coded with their location angle. As shown on the map to the right, the location angle is set to $0^\circ$ on the dynamical axis of \NGC2445 and increases toward the northeast direction (from red-green-blue) considering the inclination and position angle given by the disk rotation analysis (see text). Contours of the \ha\ flux (in white) are superposed to the location angle map, along with the position of the regions (in black) considered here.}
    \label{fig:ext2_figure}
\end{figure*}

The impact of the foreground Galactic extinction is considered negligible here ($\rm A_V=0.118\,mag$\footnote{As provided by the NASA/IPAC Extragalactic Database (NED)}). The internal extinction for each region is calculated using their Balmer decrement, considering a theoretical \ha/\hb\ value of 2.87, as expected for the case B recombination with an electronic temperature and density $T_e=10^4\,\rm K$ and $n_e=100\,\rm cm^{-3}$ \citep[][]{Osterbrock&Ferland2006}, the extinction law of  \citet{Fitz90} for the LMC within PyNeb \citep{Luridiana15}, and with $\rm R_V=3.1$.

Fig.~\ref{fig:ext_figure} presents the values of $\rm A_V$ for the 103 detected regions with a $\rm SNR>5$ for both lines \ha\ and \hb\ as a function of their galactocentric distante (GCD; calculated centered on \NGC2445 nucleus, using the inclination and position angle as given in Table~\ref{tab:param2445}). The nucleus presents the largest extinction with $\rm A_V=3.08\pm0.11\,mag$, which is not surprising given the ubiquitous dusty lanes visible in the HST images of this region (see Fig.\ref{fig:specNGC2445nucvel}). The median extinction value, excluding the nucleus and regions with negative values, is $\rm1.13\,mag$ with a standard deviation of $\rm0.54\,mag$.

Fig.~\ref{fig:ext_figure} is color-coded  with the integrated \ha\ flux of the region, confirming that negative extinction values are mainly found in low luminosity regions where the \hb\ flux is more inclined to be affected by the DIG; as we did not subtract a local background emission, it is expected that some DIG is present and will affect more the properties of a faint \hh\ region. Along with the nucleus, three emission regions present an extinction above 2 mag. These regions present strong emission lines in both \ha\ and \hb. As seen in the $\rm A_V$ map of Fig.~\ref{fig:ext_figure}, two of them are located on the far northwest side of the ring, closer to the companion galaxy. This location also corresponds to the area where pixels show a higher velocity dispersion (see Fig.~\ref{fig:velocities_figure} below). Overall, as shown by the $\rm A_V$ map, the dust attenuation is rather inhomogeneous in the ring. A very small gradient with a positive slope of $0.03\pm0.01$ is given by a linear regression (without considering again the nucleus and negative extinction region; see  Fig.~\ref{fig:ext_figure}). Although a p-reliability test is indicating that this slope value is significant, the weak gradient could simply be due to a dustier group of regions southwest in the ring as illustrated by Fig.~\ref{fig:ext2_figure}. This figure is colour-coded using the location angle of the regions. This angle is set to $0^\circ$ on the dynamical axis of \NGC2445, increasing toward the northeast. Its calculation is based on the disk rotation analysis (see Sec.~\ref{sec:velocity}) using the inclination $i$ and position angle PA found for the nucleus of \NGC2445; as given in Table~\ref{tab:param2445}. Dark blue dots in Fig.~\ref{fig:ext2_figure} are located near $275^\circ$, i.e. in the southwest part of the ring, extending further away from the nucleus than the other regions.  

Emission line fluxes of the detected regions are presented in Table~\ref{tab:tableA1} along with their extinction value. The line fluxes in the table have been corrected for the stellar population but not the extinction.

\subsubsection{Excitation properties}

\begin{figure*}
    \centering
    \includegraphics[scale=0.36]{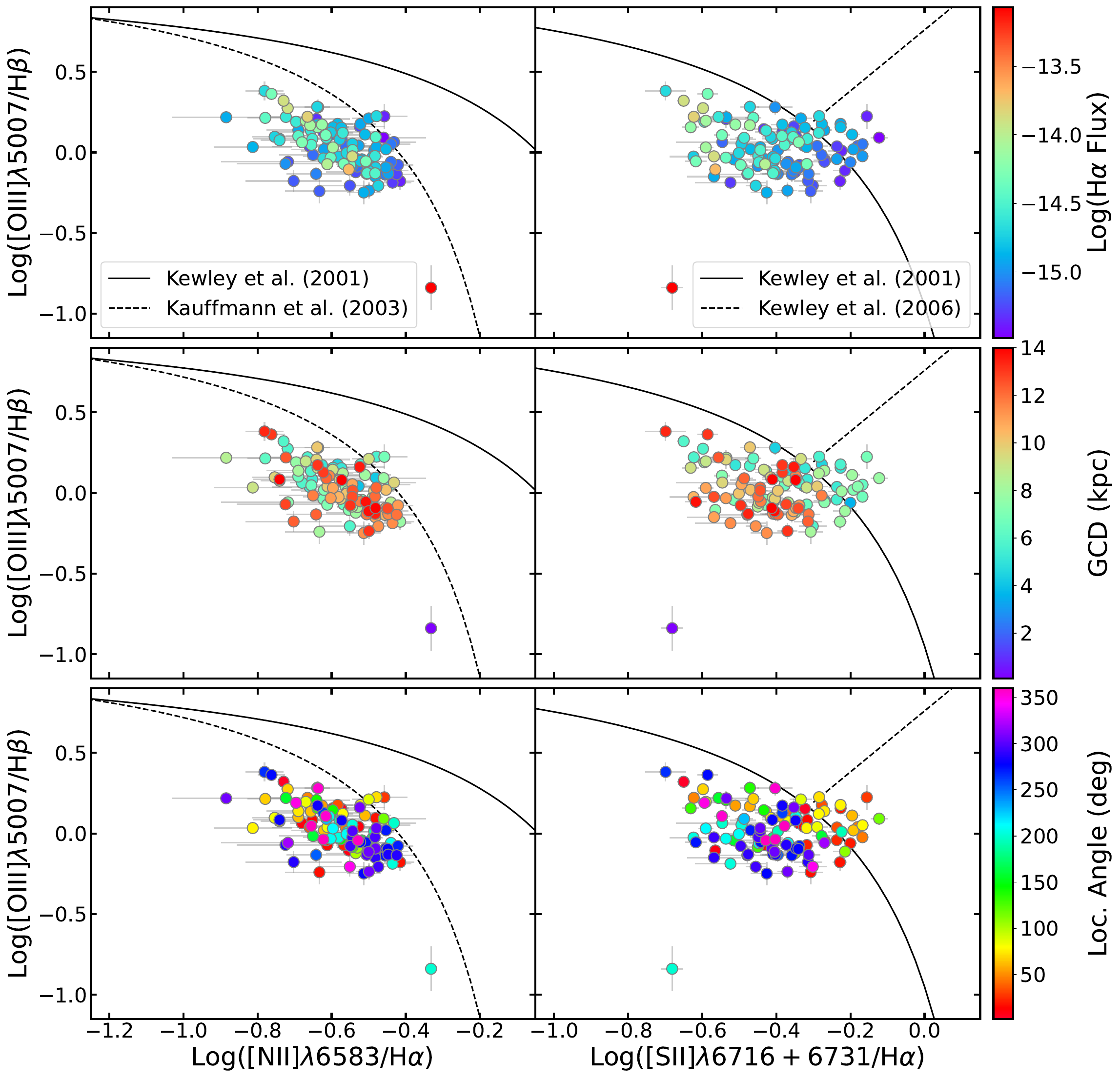}
    \caption{\nn\ and \si-BTP diagrams colour-coded with the integrated \ha\ flux of each regions (top), GCD (middle), and location angle (bottom). Only regions (114) with an $\rm SNR>3$ for all the lines used here are shown. The emission line fluxes have been corrected for the underlying stellar population but not the extinction (since lines used for a ratio are very close in wavelength). Overlaid are the diagnostic curves from \citet{Kewley2001, Kewley06} and \citet{Kauffmann2003}. 
    The outlier region, with a low ratio \ooo/\hb, is the nuclear region. High- and low-ionization emission regions are mostly found bellow the diagnostic curves.}
    \label{fig:BPT_figure}
\end{figure*}

\begin{figure*}
    \centering
    \includegraphics[scale=0.45]{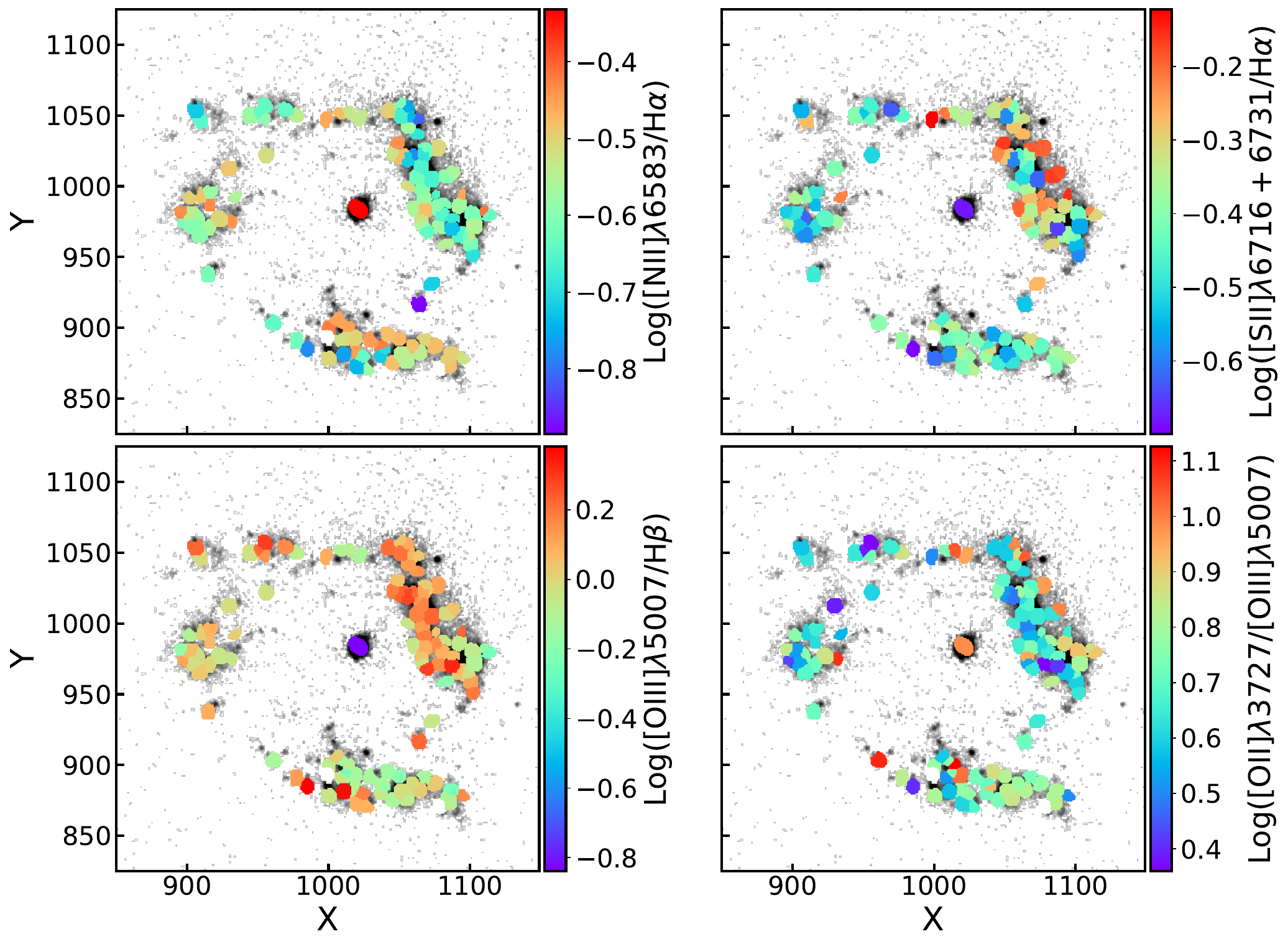}
    \caption{Maps of emission lines ratios for the emission regions present in the BPT diagrams (Fig.~\ref{fig:BPT_figure}). Regions are colour-coded with the logarithmic value of the ratios and superposed on the \ha\ flux map.}
    \label{fig:mapratio_figure}
\end{figure*}

The classical Baldwin–Phillips–Terlevich (BPT) diagrams \citep[][]{BPT1981} of the detected regions with corrected fluxes are shown in Fig.~\ref{fig:BPT_figure}; only the emission regions with an $\rm SNR>3$ for all the lines are shown in these plots (in total, 114 regions). Three versions of these plots are presented using different colour coding (\ha\ flux, GCD, and location angle). The nucleus clearly stands out, with a weak \oool\ emission line, indicative of a high metallicity (see Sec.~\ref{sec:abun} and Sec.~\ref{sec:nuclei}). 
 
In the \nn-BPT diagram, most regions fall below the limit of \citet{Kauffmann2003}, confirming that they are \hh\ region complexes. In this diagram, the regions seem rather well mixed in terms of their \ha\ flux, GCD, and location angle. But as shown in Fig.~\ref{fig:mapratio_figure}, the \nn/\ha\ map suggests a higher ratio on average if from the southwest part of the ring. 

The distribution in the \si-BPT diagram (Fig.~\ref{fig:BPT_figure}, right-hand panels) shows more emission regions in the shock zone, i.e. above the \citet{Kewley2001} limit, but below the \citet{Kewley06} AGN limit. Since there is a tendency for low-luminosity regions to sit closer to the shock zone,
%??? {\color{red}This might just be a S/N bias. As the flux are getting smaller, lines tends to sit close to the noise level (and not reaching 0), giving lines ratio close to 1 (log close to 0)}
it may again be an indication of \hh\ region complexes contaminated by the DIG. On the other hand, as shown by the GCD and location angle colour-coded version of the \si-BPT, regions in the shock zone are preferably found in the northwest part of the ring, closer to the companion galaxy (i.e. these are colour-coded in light blue-green for a small GCD, and colour-coded in orange-yellow for a location angle between 0 and $100^\circ$). The \si/\ha\ map of Fig.~\ref{fig:mapratio_figure} confirms this behavior, along with a higher ratio on average for \ooo/\hb. These northwest regions also displays a high velocity dispersion on average (see Fig.~\ref{fig:velocities_figure} below). 

Fig.~\ref{fig:mapratio_figure} also presents the map for the \ool/\oool\ ratio (extinction corrected). Along with some scattered regions (many located on the outer edge of the northwest part of the ring), the nucleus sticks out in this map with a high value, as expected for a high metallicity region. The southwest part of the ring also displays a more uniform mid-to-high value of this ratio, possibly in relation with the softness of the ionization environment in this case.

\subsubsection{Oxygen abundance determination}
\label{sec:abun}

The chemical properties of the ionized gas in the detected emission regions are analyzed using various indicators from the literature. We used the N2 indicator, based on the \nnl/\ha\ emission line ratio, and the 03N2 indicator, adding the \oool/\hb\ ratio, calibrated by \citet[][hereafter PP04]{pp04} and \citet[][hereafter PM09]{pm09}. We also used two indicators calibrated by \citet[][hereafter P16]{p16}: their R indicator that combines the emission line ratios \nnl/\ha, \oool/\hb, and extinction corrected \ool/\hb, and their S indicator, based on the line ratios \nnl/\ha, \oool/\hb, and extinction corrected \sip/\hb.

\begin{figure*}
\includegraphics[scale=0.4]
{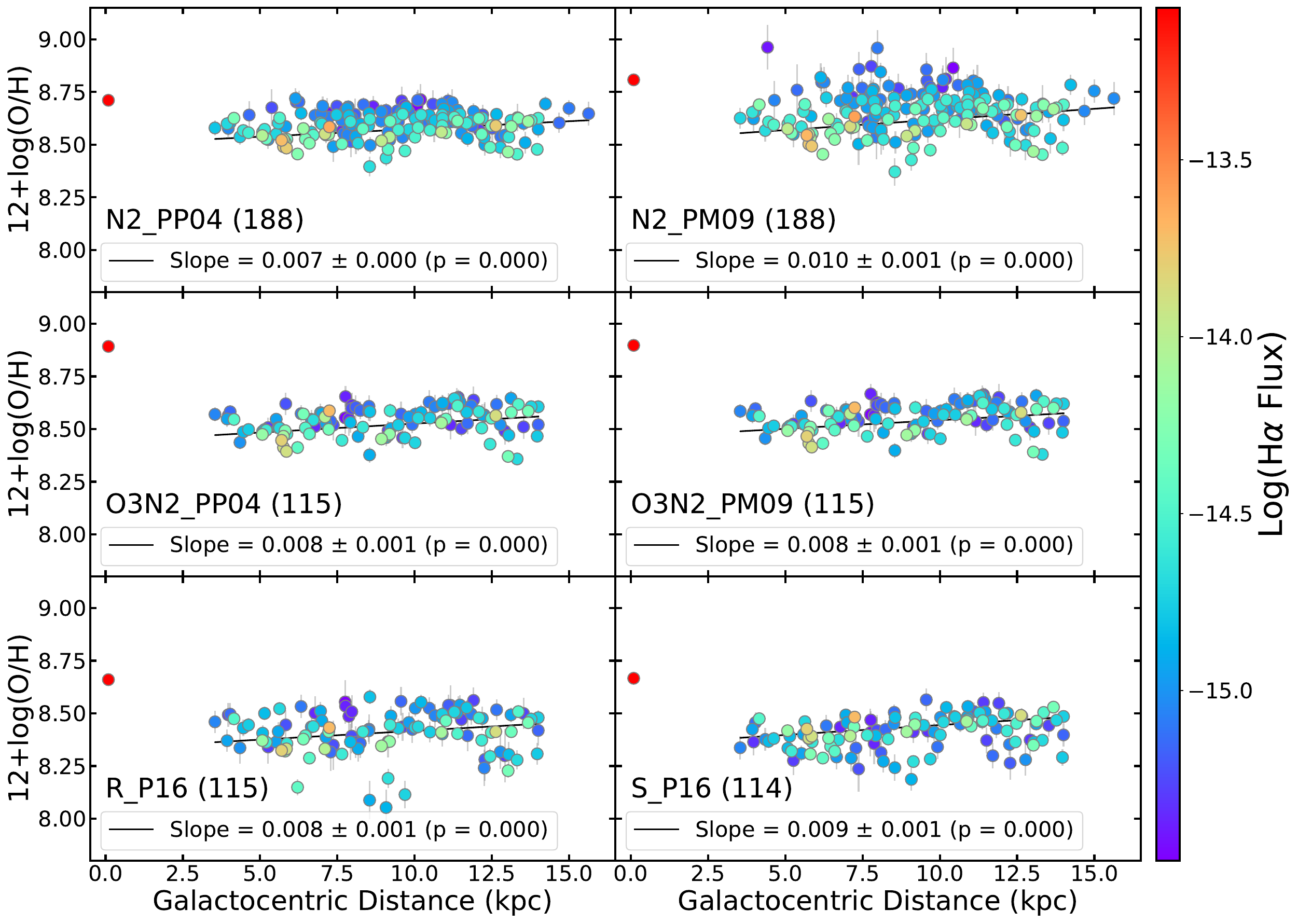}
    \caption{Oxygen abundances  colour-coded with the integrated \ha\ flux of the regions.
    The indicators used (see text) are identified on each plot. Only regions with an SNR > 3 for the lines used for an indicator have been considered (the number of regions are given in parenthesis on each plot). The black lines represent the linear regression done while excluding the nucleus; slope are given on each plot (along with the p-reliability test value).}
    \label{fig:metGCD_Ha}
\end{figure*}

Fig.~\ref{fig:metGCD_Ha} presents the O/H abundances,
calculated with the 6 selected indicators, as a function of the GCD and colour-coded with the \ha\ flux of the regions. Only regions with SNR > 3 for all the lines considered for an indicators are shown. Compared to the \hh\ region complexes in the ring, the nucleus presents a higher metallicity in general for all the indicators, near $\rm12+log(O/H)=8.75$ on average, but with an offset according to the different indicators (see values in Table~\ref{tab:met_table}). Regions in the ring have a median metallicity near $\rm12+log(O/H)=8.5$ with a standard deviation below 0.10 according to the indicator (again see Table~\ref{tab:met_table} for the offset among indicators). These values are compatible with those of \citet{Jeske1986}. Large deviations from the median are seen in the case of N2 from PPO4 and R from P16, but many of these points are low luminosity regions. Nevertheless, three bright regions near a GCD of 9~kpc have an R indicator below 8.25, again these are located close by in the northwest part of the ring, i.e. close to the companion galaxy. 

\begin{figure}
\includegraphics[width=\columnwidth]
{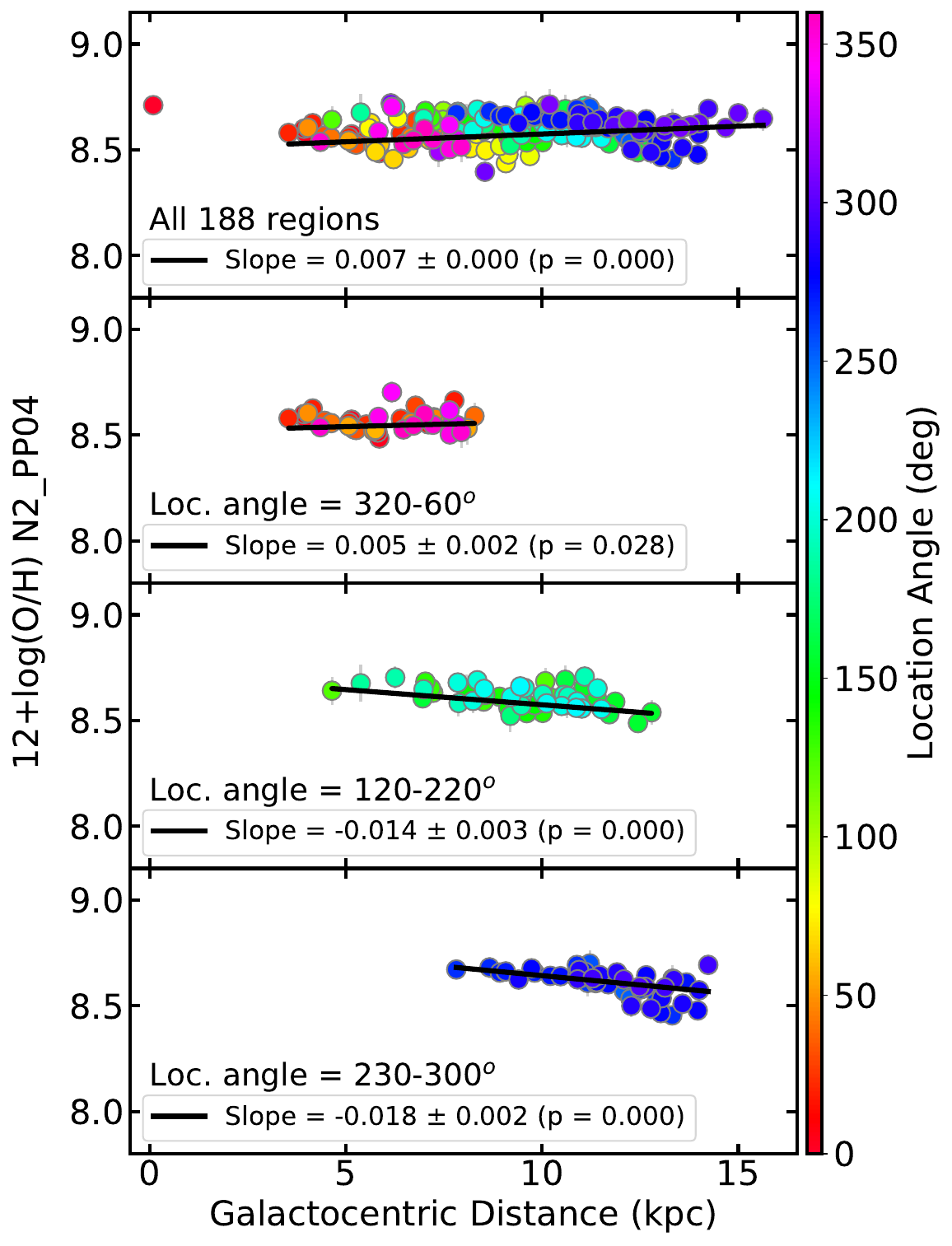}
    \caption{Top panel: N2 PP04 oxygen abundances as in Fig.~\ref{fig:metGCD_Ha} but colour-coded with the location angle. Bottom panels: same data but for sub-samples in a specific location angle range, as indicated on the plots. The black lines represent the linear regression done while excluding the nucleus (top panel) of for regions in the location angle range considered; slopes are given on each plot (along with its p-reliability test value).}
    \label{fig:metgrad}
\end{figure}

\begin{figure}
\includegraphics[width=\columnwidth]
{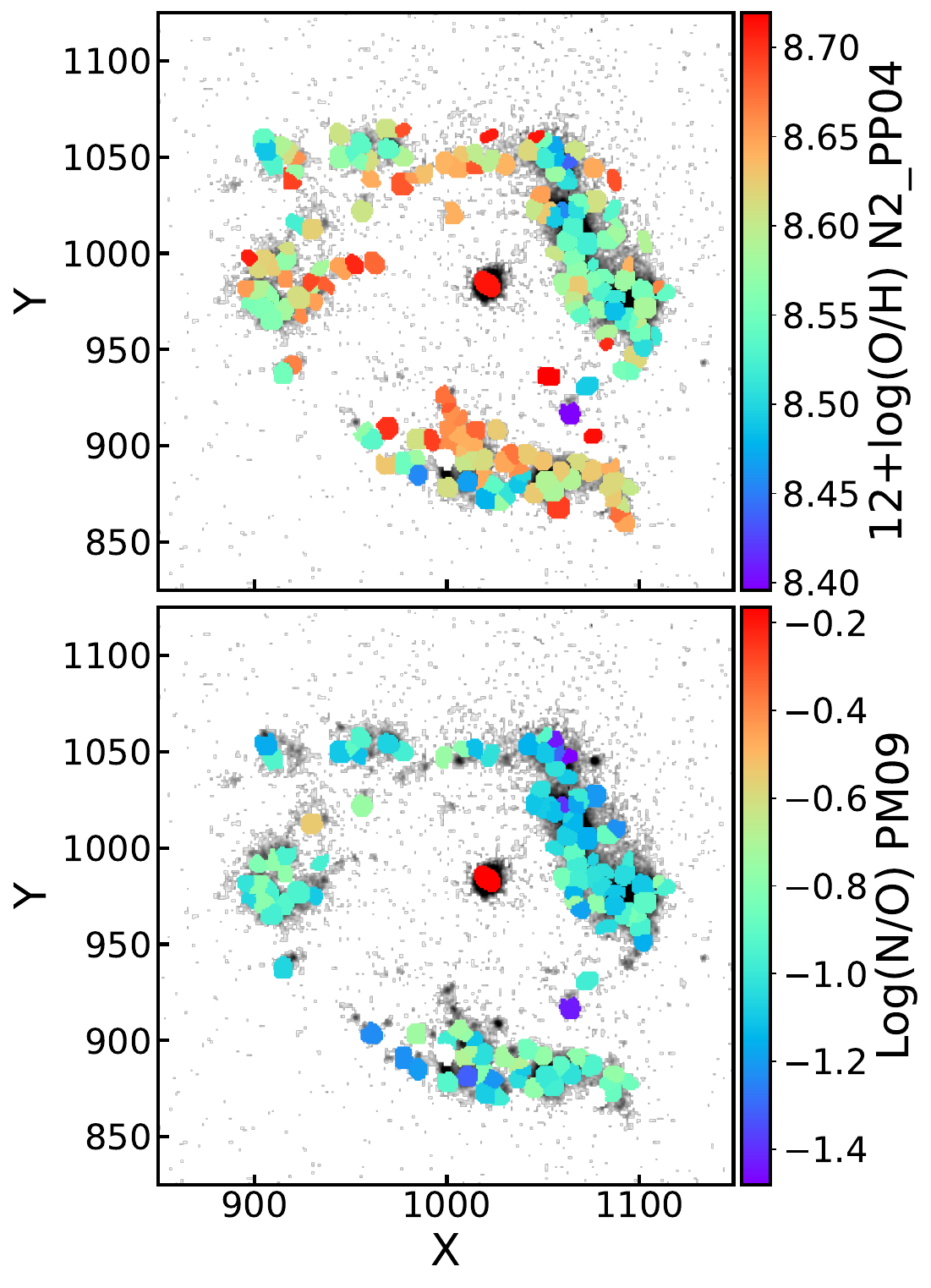}
    \caption{Maps of the oxygen abundance, using the N2 indicator from PP04, and of the nitrogen-to-oxygen ratio, using PM09 indicators. Regions considered all an SNR > 3 for the lines 
    used as in Fig.~\ref{fig:metGCD_Ha}.
    }
    \label{fig:metmap}
\end{figure}

Linear regressions done on the different metallicity plots (Fig.~\ref{fig:metGCD_Ha}; without the nucleus) indicate a small, but reliable, positive gradient between 0.007 and 0.010, according to the indicators. This gradient is rather weak and based on the standard deviation of the data we prefer to say that globally in the ring the metallicity is flat with the GCD. Nevertheless, looking at the behavior of the gradient when using the location angle to colour code the region, some parts of the ring seems to present a negative gradient instead. Fig.~\ref{fig:metgrad} shows this effect for the N2 of PP04 (the other indicators  behave in a very similar way). In this figure, we zoomed on specific complexes located in 3 parts of the ring: in the northwest with a location angle between $320^\circ$ and $60^\circ$, in the northeast between $120^\circ$ and 220$^\circ$, and in the southwest between $230^\circ$ and $300^\circ$. This figure indicates a negative gradient, although faint, in the northeast and southwest part of the ring, while the northwest part, i.e. close to the companion, has a very faint positive gradient similar to the ring overall gradient. The metallicity map (using N2 from PP04, as an example) in Fig.~\ref{fig:metmap} particularly shows a more uniform lower value (light blue-green) near the companion galaxy compared to the southwest part of the ring that display a clear change from a higher value (red) at small distance from the nucleus to a smaller value (orange-green) further away. 

\begin{table}
\caption{\hh\ region complexes $\rm12+1og(O/H)$ and Log(N/O)}
\label{tab:met_table}
    \centering
    \small
    \begin{tabular}{lccc}
    \hline
    Indicator & Nucleus & \multicolumn{2}{c}{Other complexes} \\
    & & Median & STD \\
    \hline
    O/H N2 PP04   &           $8.71\pm0.01$ &           8.60 & 0.06 \\
    O/H N2 PM09   &           $8.81\pm0.01$ &           8.66 & 0.10 \\
    O/H O3N2 PP04 &           $8.89\pm0.03$ &           8.54 & 0.07 \\
    O/H O3N2 PP04 &           $8.90\pm0.03$ &           8.56 & 0.06 \\
    O/H R P16     &           $8.55\pm0.10$ &           8.42 & 0.10 \\
    O/H S P16     &           $8.67\pm0.01$ &           8.42 & 0.07 \\
    \noalign{\medskip}
    N/O PP09      & \llap{$-$}$0.17\pm0.06$ & \llap{$-$}0.99 & 0.06 \\
    N/O P16       & \llap{$-$}$0.59\pm0.04$ & \llap{$-$}1.17 & 0.04 \\
    \hline
\end{tabular}
\end{table}

\begin{figure*}
\includegraphics[scale=0.4]
    {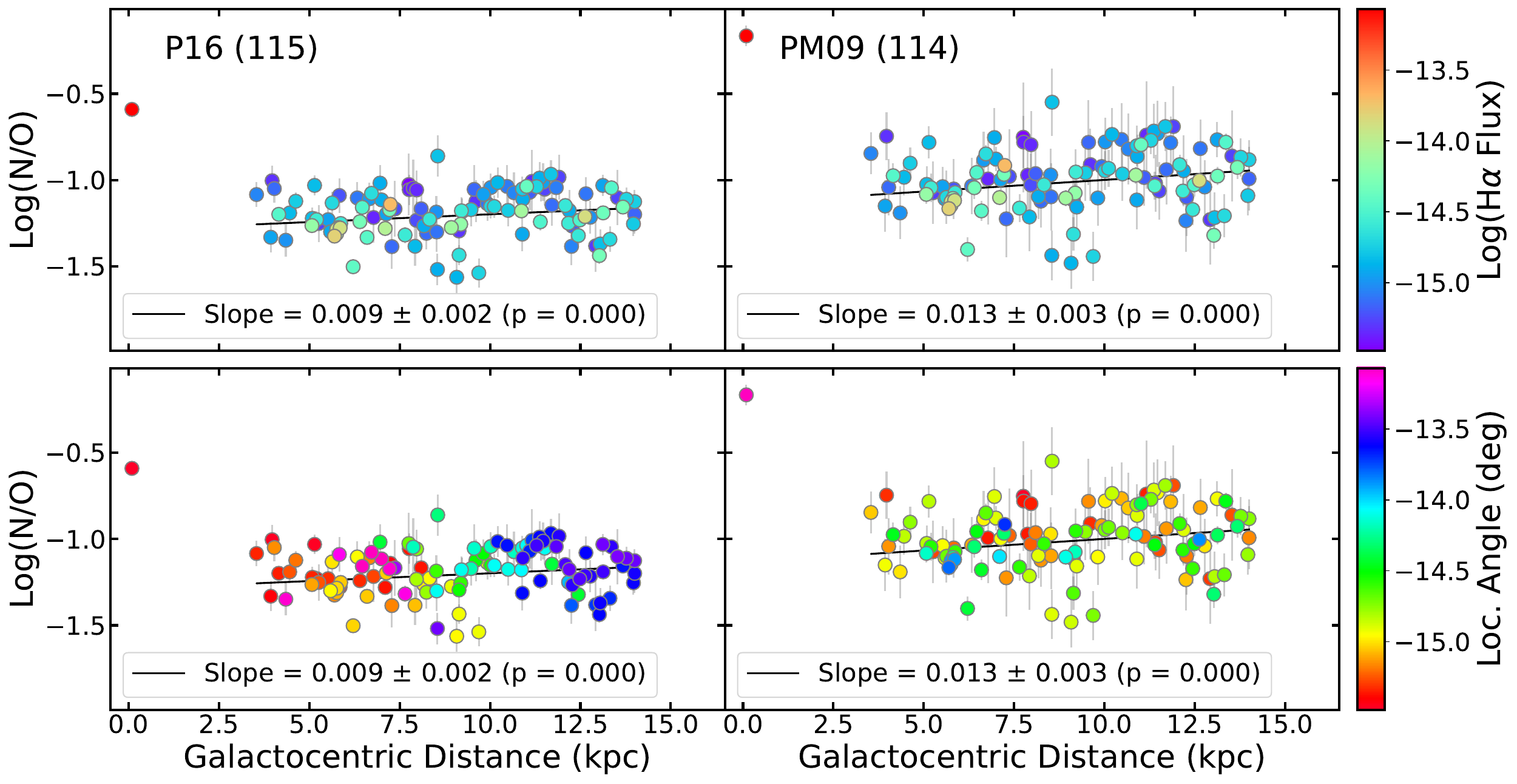}
    \caption{The N/O ratios using equations of P16 (left) and PP04 (right) as a function of the GCD, colour-coded the integrated \ha\ flux of the regions (top) and the location angle (bottom).}
    \label{fig:noGDC}
\end{figure*}

We also looked at the N/O ratio considering the equations of P16 and PM09. This ratio behaves a lot like the metallicity, i.e. a higher value in the nucleus compared to the ring (see values in Table~\ref{tab:met_table}), with a hint of an overall positive log(N/O) gradient for the other regions (see Fig.~\ref{fig:noGDC}). Again this gradient is reverse in the southwest part of the ring (plot not shown here), but the search for a N/O gradient in the other ring parts is not conclusive. The log(N/O) map shown in Fig.~\ref{fig:metmap} is confirming a mixture of low and mid log(N/O) values in the northwest ring part between the two galaxies, a rather homogeneous mid log(N/O) value in the northeast part, and a small gradient in the southwest part.

\begin{figure}
\includegraphics[width=\columnwidth]
    {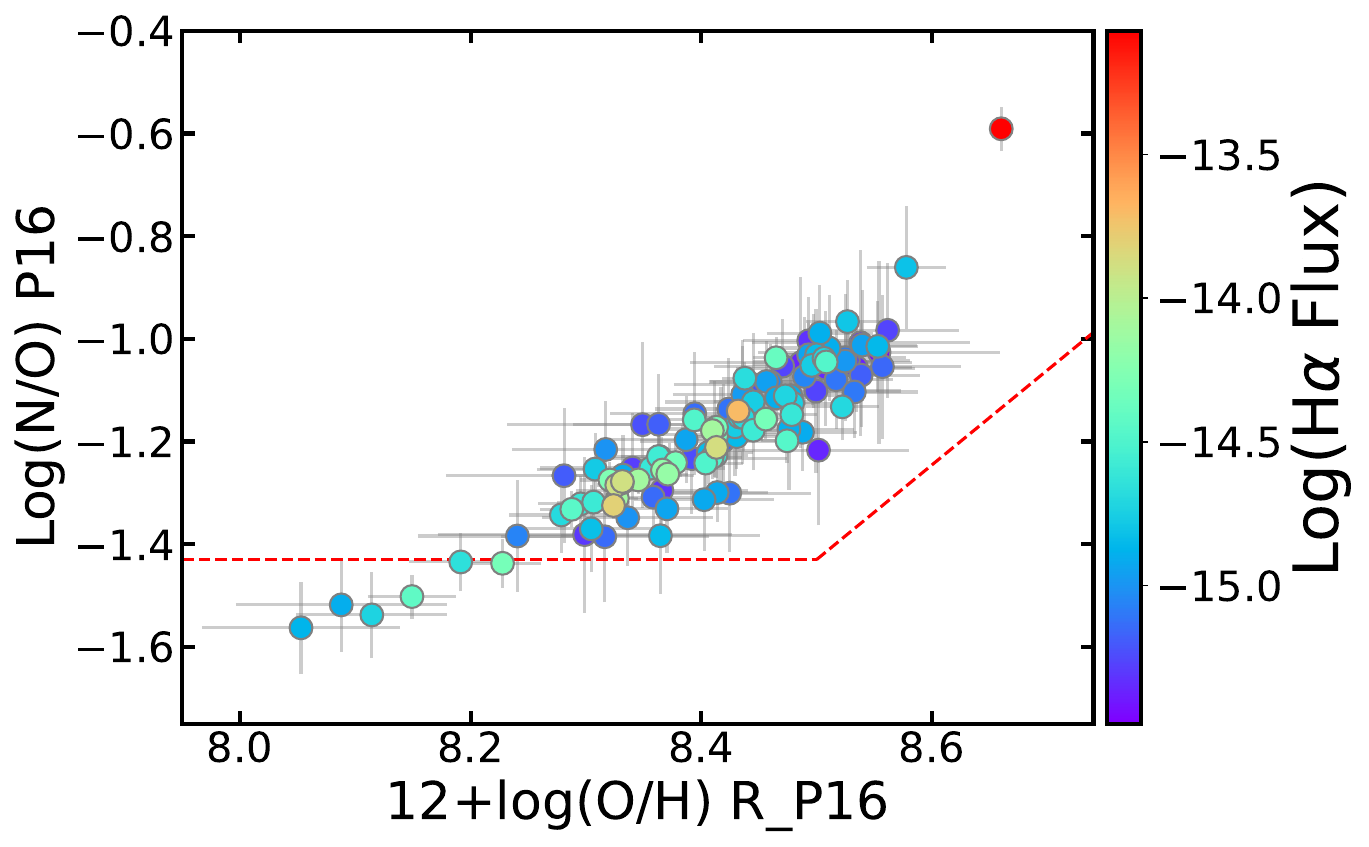}
    \caption{N/O ratio versus metallicity using equations of P16, colour-coded with the integrated \ha\ flux of the regions. The red dashed line shows the relation of
    \citet{AM13}.}
    \label{fig:NOvsMet}
\end{figure}

Fig.~\ref{fig:NOvsMet} presents the N/O vs metallicity plot, using the P16 calibration. Overall, it shows an increasing N/O ratio with metallicity. This general behavior is explained by the fact that in the higher-metallicity complexes, where the production of secondary nitrogen becomes more important, the nitrogen abundance increases faster than the oxygen abundance \citep{Henry00}. In Fig.~\ref{fig:NOvsMet} we also show lines (red dotted) of the N/O median behavior at low and high metallicity measured by \citet{AM13} for a large sample of SLOAN galaxies; these lines may be interpreted as limits for pure primary (horizontal line) and secondary (incline line) N production. Even though the \hh\ region complexes in \NGC2445 follow the general behavior of these demarcations, they are shifted to the left while some regions are located below the pure primary limit. As explained by \citet{AM13}, shifts in the distribution may be the consequence of a lower metallicity and mass, and/or an important star formation activity that induce gas flows, affecting the relative abundances; situations that include \NGC2445. N/O values bellow the pure primary limit are seen mostly in low mass and low metallicity galaxies \citep[see Fig. 8 of][and references therein]{Zaragoza-Cardiel2022}. According to \citet{Garnett90}, it may be the consequence of a non-constant SFR over the galaxy evolution, where a limited number of sporadic bursts result in a low and fluctuating N/O ratio at low metallicity. A comparison with the Cartweel galaxy is also interesting: \citet{Zaragoza-Cardiel2022} describe \hh\ regions in the galaxy ring with very low N/O (from $-$1.4 to $-$1.7) that display a decreasing trend with metallicity (from 7.7 < log(O/H)+12 < 8.4). They explain this behavior considering post-collisional infall of metal-poor gas at the location of the ring. In the case of \NGC2445, 5 out of the 6 complexes below the pure primary limit are located in the northwest ring, where we also see a higher velocity dispersion. As for the Cartwell galaxy, these \hh\ regions could be formed using falling-in material with a relatively lower metallicity (between 8 and 8.2) between the interacting galaxies, but it could also be that following the interaction, these \hh\ regions are formed using low metallicity material initially at a large galactocentric distance, or/and because local metal rich material has escaped following a more violent star formation event.

The average mildly subsolar oxygen and nitrogen abundances estimated in the ring of \NGC2445 agree well with those found in other star-forming regions of ring galaxies \citep[][]{Bransford1998}. Some studies \citep[][]{Bransford1998,Kouroumpatzakis2021, Elagali2018} have suggested that the metallicity in the ring is likely due to gaining gas from a low metallicity companion whereas the observed central metallicity is related to the initial metallicity of the more evolved ring galaxy progenitor. This could be the case here as we see lower values in the northwest part of the ring close to the companion galaxy. But at the same time, it seems less likely since \NGC2444 was already more evolved than \NGC2445 prior to the interaction (see Section~\ref{sec:section4}). Alternatively, and as seen at less in the southwest part of the ring that extend to a large GCD, the lower metallicity of complexes in the ring with respect to the nuclear star-forming knot may originate from the negative metallicity gradient in the progenitor as expected in normal spiral galaxies with gas flows and star formation activity influencing the end result. Variations in the chemical properties of the knots located in the ring can be explained by local variations in their intrinsic properties if they are created almost simultaneously by the propagating wave. In addition, some complexes may have been displaced from their initial positions in the progenitor disk by the head-on collision.

\subsubsection{Star formation}

The sum of the dust-corrected \ha\ luminosity of the \hh\ region complexes is $5.8\times10^{41}\rm erg\,s^{-1}$ which translates to a global SFR of $3.2\,\rm M_\odot\,yr^{-1}$ using the \citet{Murphy2011} calibration. The nuclear star-forming region accounts for 47\% of this luminosity (extinction corrected). The remaining pixels (when masking all the detected emission regions) counts for 25\% of the \ha\ flux in the whole disk of the ring galaxy. If we correct this flux using the average extinction calculated for the emission regions, and add it to the flux of the detected complexes, the SFR increases to $3.8\,\rm M_\odot\,yr^{-1}$. 

The SFR calculated here for \Arp143 using the \ha\ emission is lower than the SFR estimated by \citet{Romano2008} using FIR data (i.e. $\rm SFR_{FIR}=5.60,\rm M_\odot\,yr^{-1}$). These authors also give an \ha+\nn\ flux (uncorrected for extinction) of $7.24\times10^{-13}$\ergcs. Considering a contribution of 20\% for \nn\ lines to this flux, they obtain an $\rm A_V$ of $\rm1.91\,mag$ in allowing their \ha\ SFR to match their SFR$_{\rm FIR}$. This extinction value is rather close to our average value (without the nucleus) of $\rm1.13\,mag$.  

An interesting feature of ring galaxies is the correlation between the number (N) of ultra-luminous X-ray sources (ULXs) they host and their SFR \citep[][]{Wolter2018,Mapelli2010}. Using the SFR-N$\rm _{ULX}$ relation of \citep[][i.e. their fig.~3 for a low metallicity galaxy]{Mapelli2010} the above estimated SFR suggests that \NGC2445 hosts $\sim8$ ULXs as detected by \citet{Wolter2018}.

\begin{figure}\includegraphics[width=0.9\columnwidth]{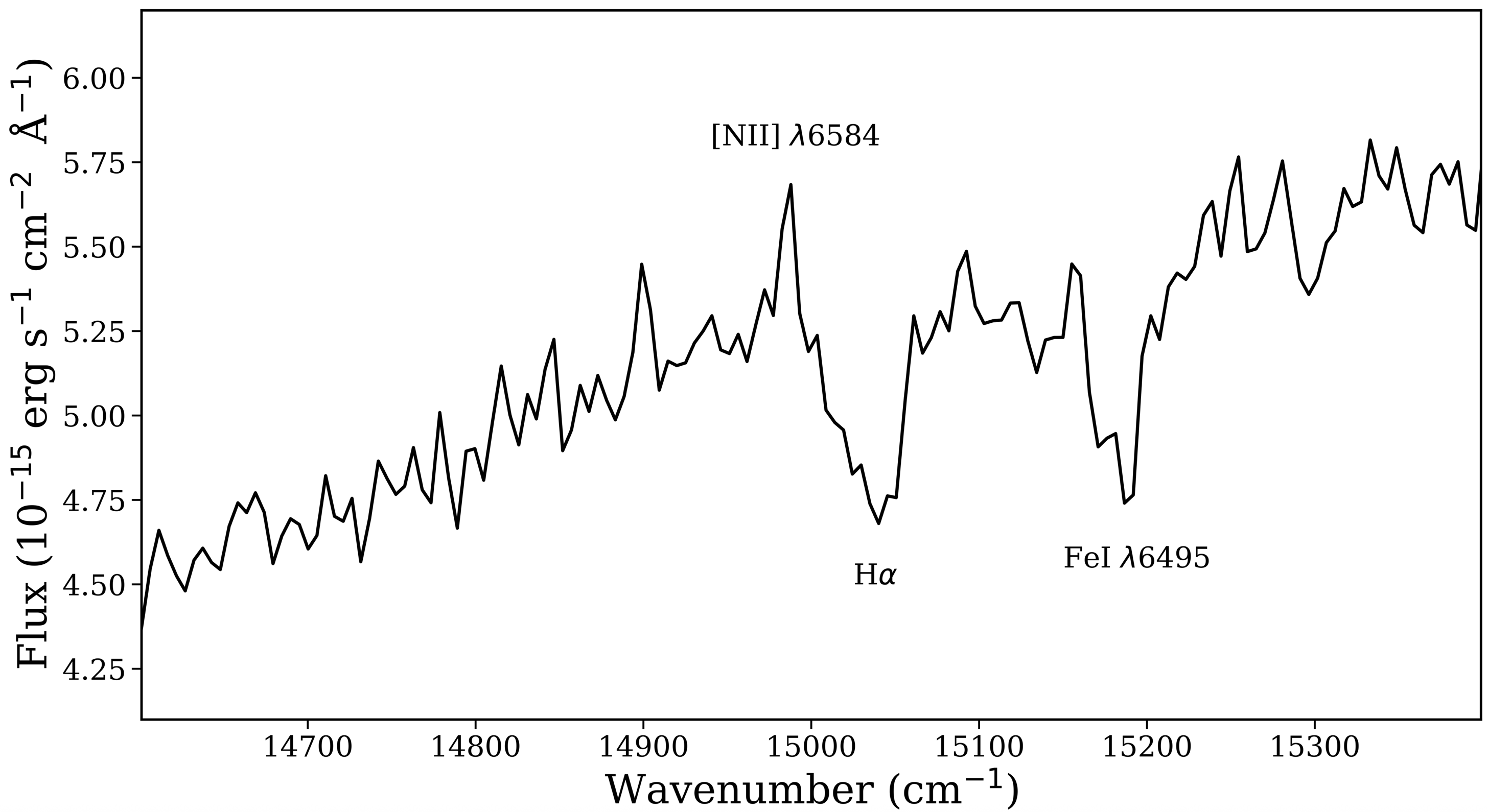}
    \caption{SN3 spectrum of the nucleus of \NGC2444 (integrated within a radius of $4.5''$).}
    \label{fig:specNGC2444}
\end{figure}

\begin{figure}
    \includegraphics[width=\columnwidth]{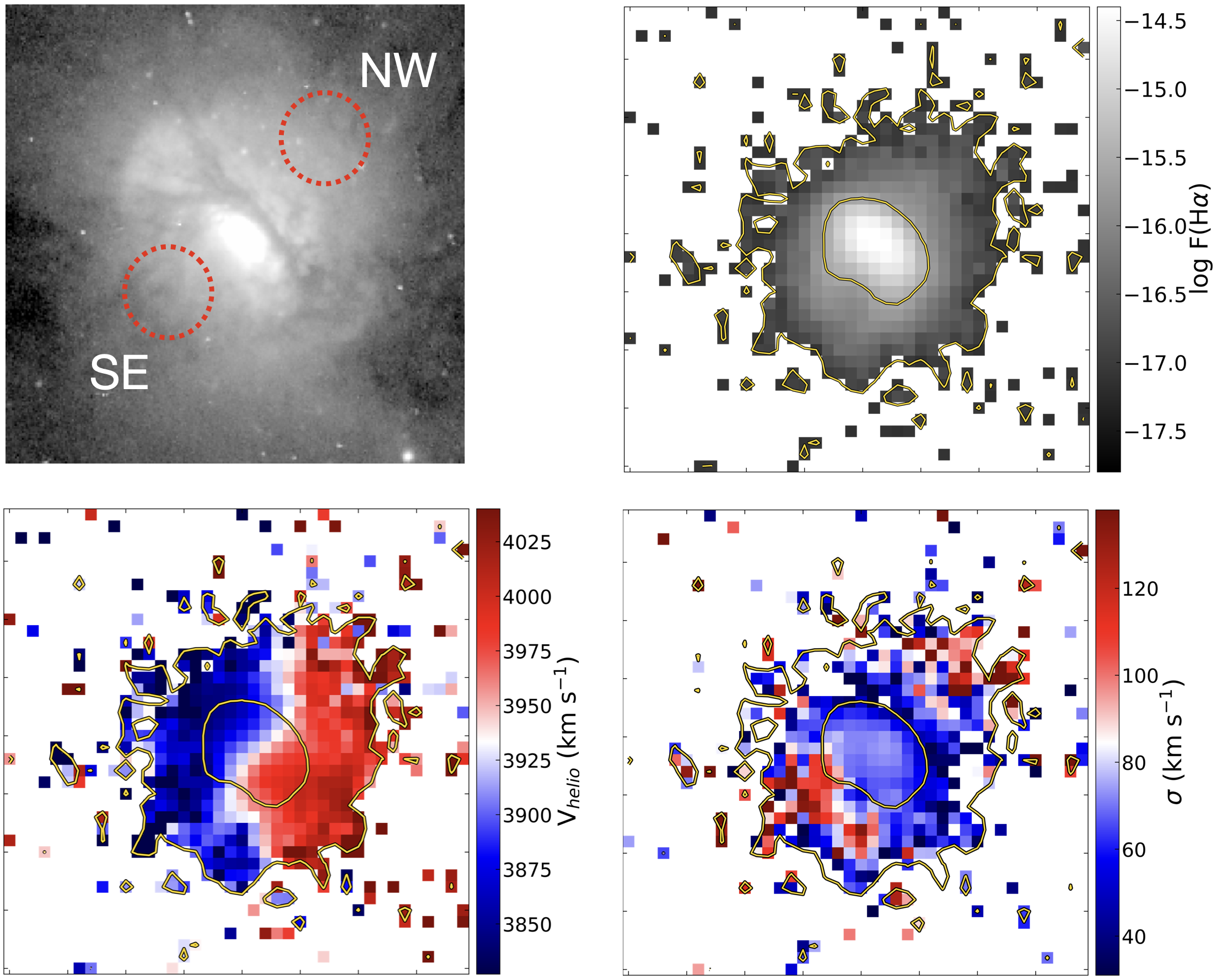}
    \caption{Upper panels: HST/ACS image of the nucleus of \NGC2445 (filter F606W) and the corresponding SITELLE H$\alpha$ flux. Lower panels: heliocentric velocity and velocity dispersion maps. Isocontours with $\rm F(H\alpha)=0.8$ and $27\times10^{-17}\rm erg\,s^{-1}cm^{-2}\,pixel^{-1}$ are drawn on all maps. FOV is $13''\times13''$. The location of the SE and NW lobes discussed in the text is indicated in the HST/ACS image.}
    \label{fig:specNGC2445nucvel}
\end{figure}

\begin{figure}  \includegraphics[width=\columnwidth]{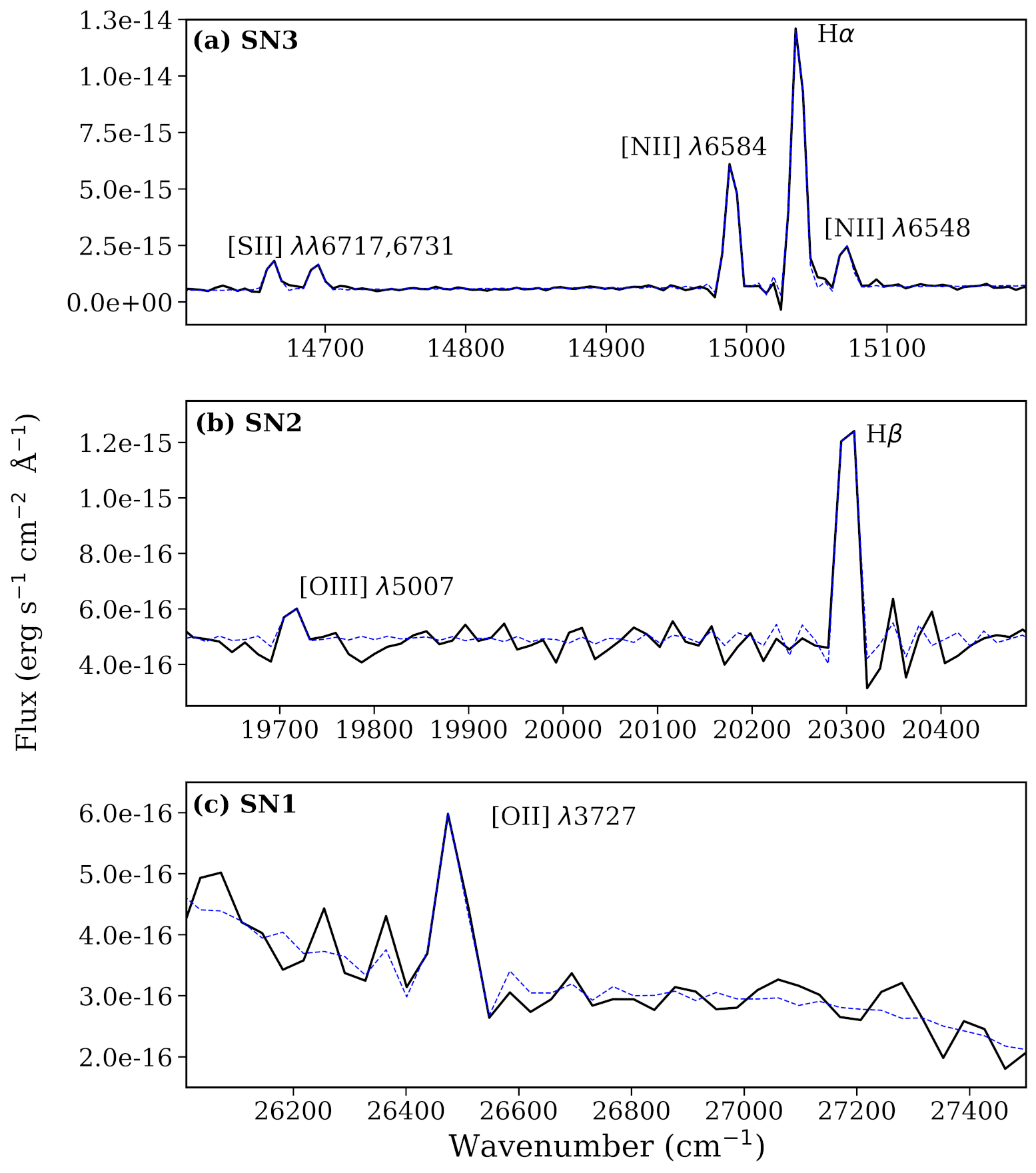}
    \caption{Spectrum of the nucleus of \NGC2445,  integrated within a radius of 1.2$''$, along with the fit (dashed blue line).}
    \label{fig:specNGC2445nucspec}
\end{figure}

\begin{figure}  \includegraphics[width=\columnwidth]{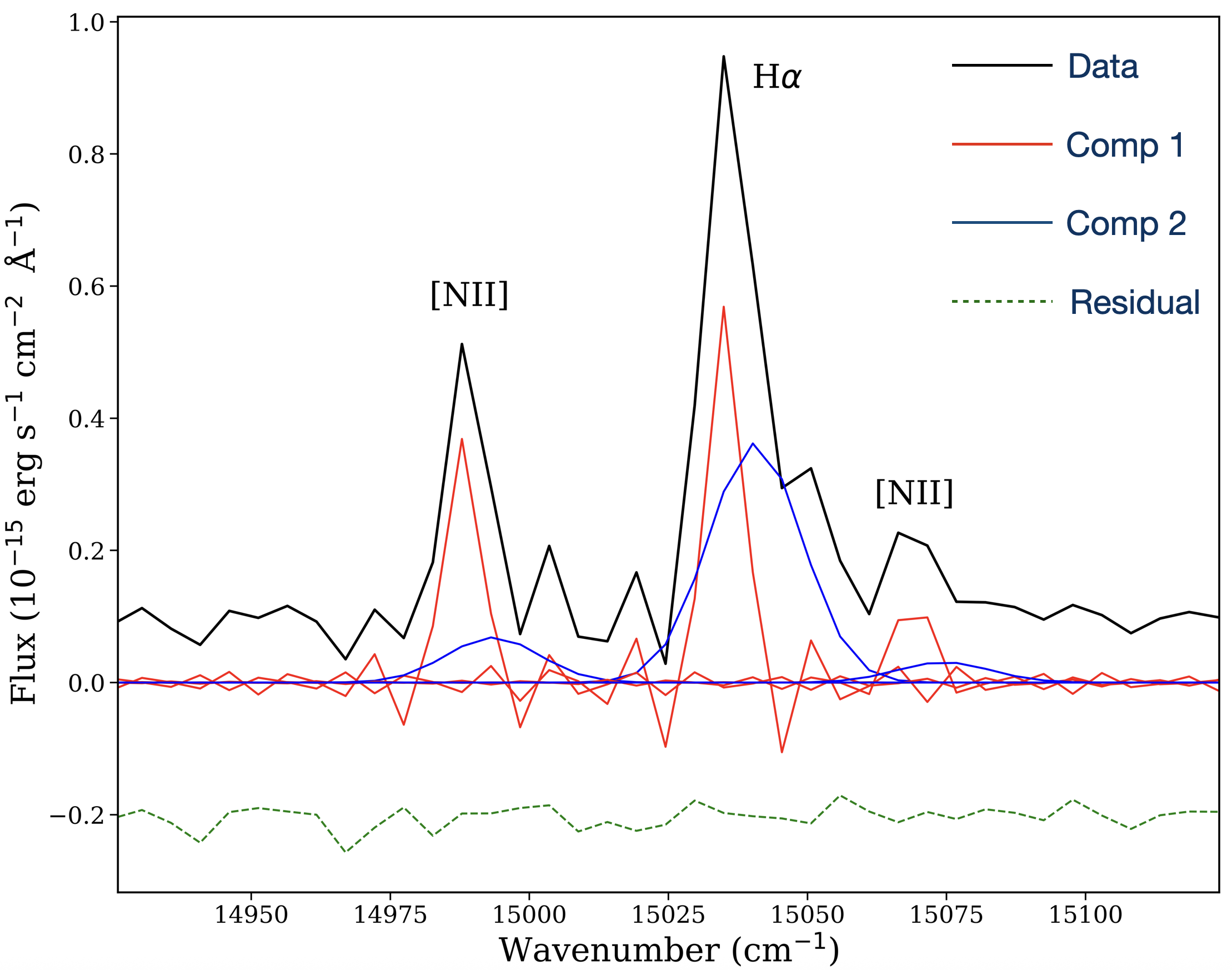}
    \caption{Spectrum of the SE lobe of \NGC2445 (integrated within a radius of 1.2$''$), centered on the \ha\ and \nn\ lines, along with the two-component fit.}
    \label{fig:specNGC2445SEspec}
\end{figure}

\begin{figure}   \includegraphics[width=\columnwidth]{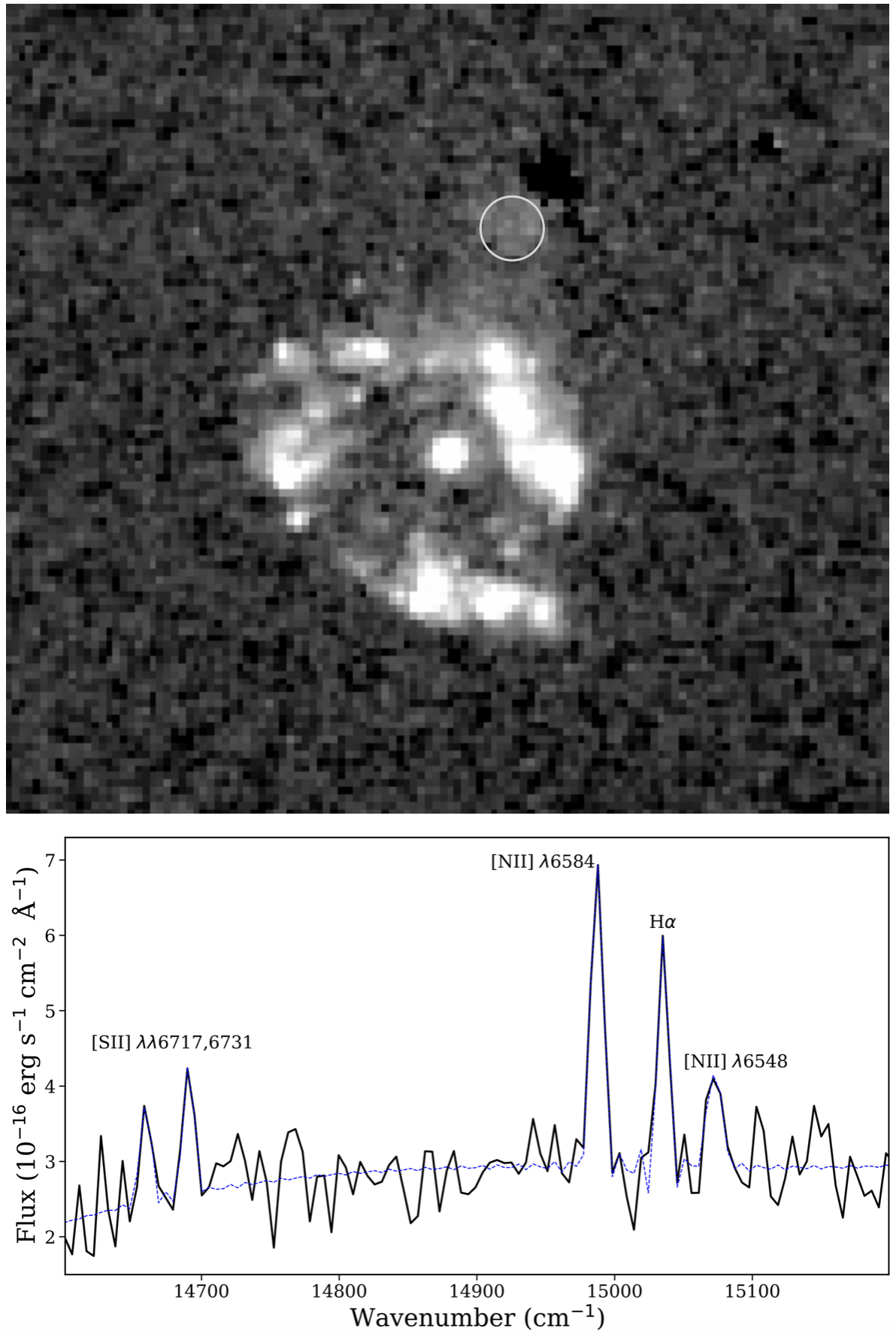}
    \caption{Top: Combined \ha\ and \nnl\ emission-line map of \Arp143, extracted using a 5-pixel binning. A structure of ionized gas is visible between the two galaxies. The white circle indicates the region where the spectrum below is extracted. Notice that the nucleus of \NGC2444 appears as a negative flux because its spectrum is dominated by stellar absorption. Bottom: Integrated spectrum, in the SN3 band, of the circular region highlighted in the top image, along with the fit (dashed blue line).}
    \label{fig:specbridge}
\end{figure}

 \begin{figure*}
    \includegraphics[scale=0.34]{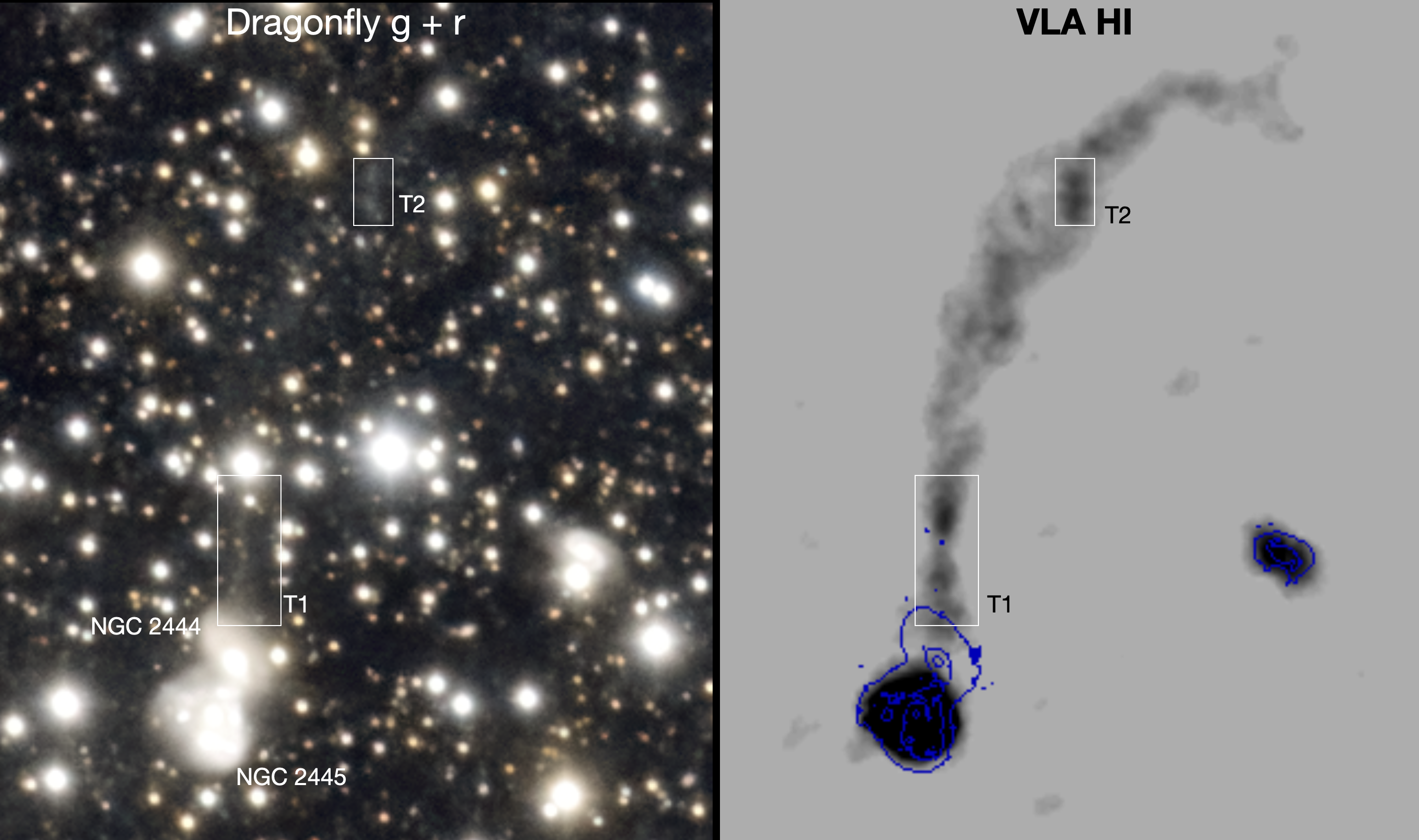}
% {DragonflyHIdeepimages.png}
    \caption{Left: Composite coloured image of \Arp143 obtained with Dragonfly Telephoto Array (blue: g band deep image, red: r band deep image, green: mean of g and r deep images). The blue tidal plume is visible. Right: VLA $\rm C+D$ \ho\ map of the same field \citep[][]{Appleton1987}. The images span $14' \times 17'$. North is up, East is left. The rectangles labeled T1 and T2 denote the location of detected blue star-forming knots in the stellar tidal plume that have an obvious counterpart in the \ho\ map. The SITELLE image of region T1 is reproduced in Fig.~\ref{fig:PlumeHII}; T2 is outside the SITELLE FOV.}    \label{fig:DragonflyHIdeepimages_figure}
\end{figure*}

 \begin{figure*}
    \includegraphics[scale=0.36]{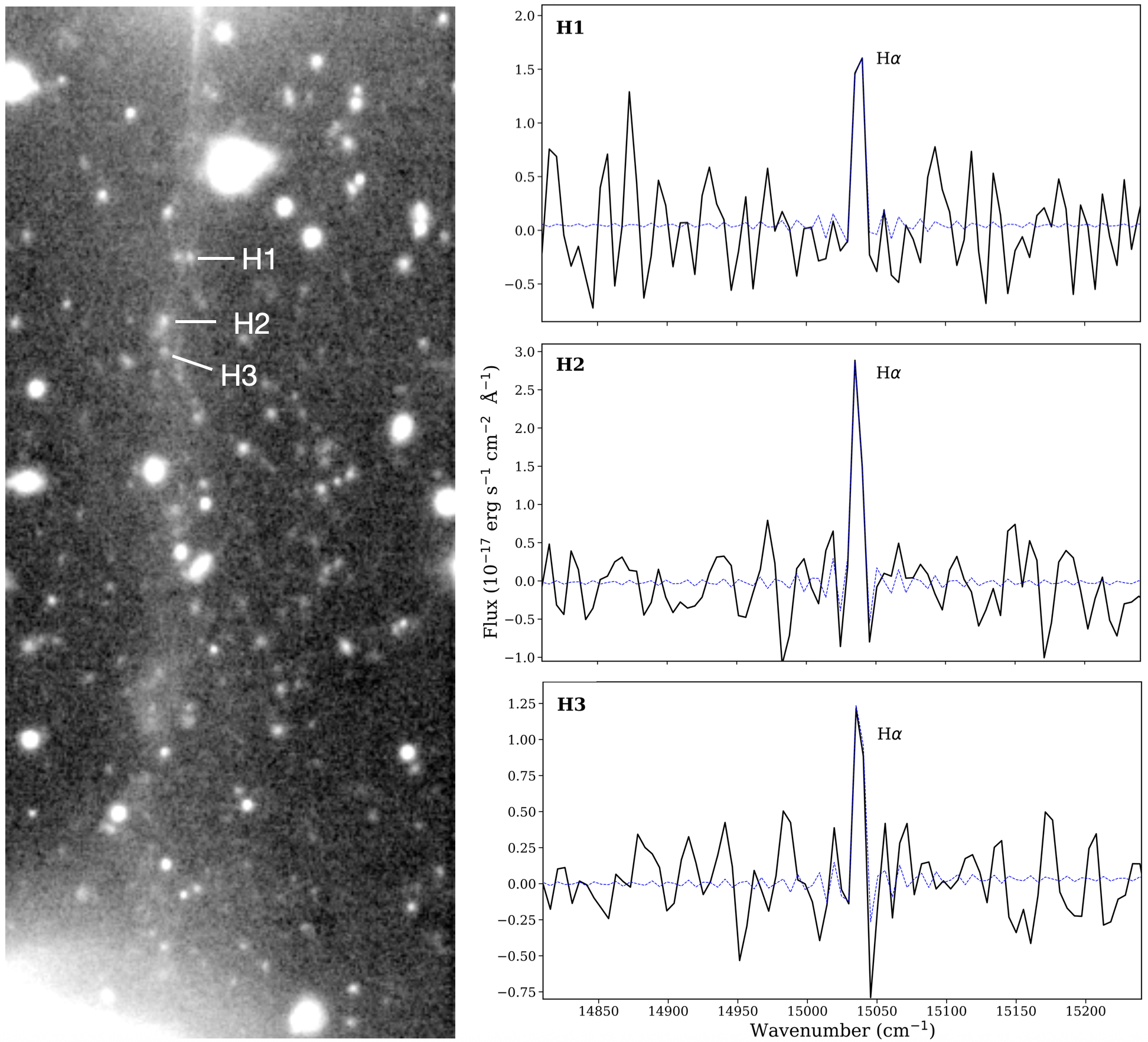}
% {DragonflyHIdeepimages.png}
    \caption{Left: SITELLE zoomed-in view ($1.2' \times 2.7'$) of the base of the plume (labeled T1 in Fig.~\ref{fig:DragonflyHIdeepimages_figure}) where star-forming knots H1, H2 and H3, detected in H$\alpha$, are identified. Right: Spectrum of the clumps along with an \textsc{orcs} fit to the \ha\  line (dotted blue).}
    \label{fig:PlumeHII}
\end{figure*}

\subsection{The nuclei}
\label{sec:nuclei}
The two nuclei present very different spectral characteristics. That of \NGC2444 (Fig.~\ref{fig:specNGC2444}) is dominated by broad stellar absorption lines of \ha\ and FeI, in agreement with the S0 classification of the host galaxy. The source of the \nnl\ emission feature is either the upper end of the ionized ridge, discussed in the next section, or some residual nuclear star formation, as predicted by the simulations (see below); the accompanying \ha\  line in emission is thus likely drowned by the strong absorption component.

On the other hand, the nucleus of \NGC2445 presents convincing characteristics of a nuclear starburst. It is a strong X-ray \citep{Wolter2018} and radio \citep[][]{Nikiel14} source, but line ratios in the far infrared using the Spitzer Space Telescope argue against the presence of an AGN \citep[][]{Beirao2009}; this is also the case of the visible line ratios presented below.

Fig.~\ref{fig:specNGC2445nucvel} shows an HST/ACS image of the central 3.6 square kpc surrounding it, along with the corresponding SITELLE \ha\ flux, heliocentric velocity, and velocity dispersion maps. The presence of strong dust lanes is conspicuous in the HST image, in agreement with the high value of $\rm A_V$ discussed in the previous section. The ionized gas presents a bright elongated structure along with two fainter sidelobes to the SE and NW directions. The spectrum of the bright structure (the inner $\rm330\,pc$; Fig.~\ref{fig:specNGC2445nucspec}) is typical of high-metallicity star-forming regions, and the width of the lines in the SN3 filter ($\sigma=80\,\kms$) can completely be accounted for by rotation. However, the sidelobes present higher velocity dispersions, and the line profiles cannot be well reproduced with a single component, even when a significant velocity dispersion is taken into account. The best fit to the spectrum of the SE lobe, shown in Fig.~\ref{fig:specNGC2445SEspec}, requires two distinct velocity components: a red, narrow (${\rm V}=3950\,\kms$, $\sigma=20\,\kms$) component and another, bluer and much wider (${\rm V}=3840\,\kms$, $\sigma=170\,\kms$). The situation is similar in the NW lobe, with two velocity components separated by $\sim150\,\kms$. It is tempting to attribute these characteristics to a nuclear outflow similar to that of the much closer starburst galaxy M82, best illustrated by fig.~25 of \citet[][]{Yoshida19}. On the other hand, the numerical simulation (Sec.~\ref{sec:modelkin}; Fig.~\ref{fig:model_velocity_fields}) accurately predicts enhanced velocity dispersions in the vicinity of the nuclear region due in part to a significant {\it infall} of material on the nucleus.

\subsection{The ionized ridge}
\label{sec:ridge}
Fig.~\ref{fig:specbridge} presents a combined \ha\ and \nnl\ emission-line map of the pair, extracted using a 5-pixel binning scheme to increase the sensitivity to low surface brightness structures. It shows the presence of faint ionized gas between the nucleus of \NGC2445 and its star-forming ring, but, more interestingly, also a ridge of emission between the two galaxies (with an average H$\alpha$ surface brightness of $1.2\times10^{-17}\rm erg\,s^{-1}cm^{-2}arcsec^{-2}$) leading to the nucleus of \NGC2444. It is clearly offset to the East of the blue continuum streak of young stars visible in Fig.~\ref{fig:SITELLEdeepimages_figure}. It does not either coincide with the maximum of the radio emission ridge detected by \cite{Nikiel2014}, but rather with its fainter extension leading to the nucleus of \NGC2444 (see their fig.~1). The integrated spectrum of the upper part of this ridge, shown in the bottom panel of Fig.~\ref{fig:specbridge}, displays broad lines ($\sigma\simeq 85\,\kms$ for \nn\ and \si, $60\,\kms$ for \ha) with \nnl/\ha\ $\simeq 1.7$ and \sip/\ha\ $\simeq1.2$, higher than anywhere else in \Arp143. Both ratios decrease to $\sim0.9$ in the southern part of the ridge and dramatically drop to $\sim0.2$ in the \hh\ region complexes of \NGC2445 closest to the ridge. The heliocentric velocity of this region, 3950 km~s$^{-1}$, is very similar to that of the northwestern edge of \NGC2445 (see Fig.~\ref{fig:velocities_figure} below). In the SN2 cube, the \oool\  line is very faint and \hb\ is lost in the noise, while the \ool\ clearly stands out in the SN1 cube; this leads to an \ooo/\oo\ ratio of $\sim 0.15$. The ridge is thus a region of low ionization, in the shocks region of diagnostic diagrams, and highly turbulent (as suggested by the line widths in the SN3 filter). Although its nature is unclear, the ionized ridge likely emanates from \NGC2445 from which it has been stripped away during the interaction.

\subsection{The extended plume}
\label{sec:plume}

The Dragonfly deep $\rm g + r$ images, shown in Fig.~\ref{fig:DragonflyHIdeepimages_figure}, reveal the optical counterpart of the puzzling, very extended plume, located to the north of the interacting galaxies and first revealed in the \ho\ data of \citet{Appleton1987}. The plume, although of very low surface brightness, is as extended as its \ho\ counterpart and blue in colour in the Dragonfly images; its base is also detected in the SITELLE deep frames (Fig.~\ref{fig:PlumeHII}). The colour and gaseous content of the plume suggest that it is very likely associated with the debris of the ring galaxy; this is strongly supported by the numerical simulations presented in Section \ref{sec:section4}. This kind of tidal feature is not uncommon in collisional ring galaxies : it is, for example, detected in the Cartwheel galaxy \citep[][]{Higdon1996}, \Arp141 \citep[][]{Hibbard2001} and in \NGC922 \citep[][]{Elagali2018,Delgado2023}. The plume needs not to be aligned with the intruder but rather develops in the wake of its host.

\cite{Appleton1987} also pointed out the existence of a bubble-like shell in the \ho\ plume, about two-thirds of its length ($\sim 100\rm\,kpc$) northward of \NGC2444. The brightest part of the shell, identified as T2 in Fig.~\ref{fig:DragonflyHIdeepimages_figure}, 
is also clearly visible in the Dragonfly image as a series of blue, elongated knots. These could be star-forming regions, but a spectroscopic confirmation is needed as the knots are located outside the FOV covered by the SITELLE data. Within this field however, a careful inspection of the SN3 data cube reveals the presence of three barely resolved regions of active star formation in the lower section of the plume (labeled T1 in  Fig.~\ref{fig:DragonflyHIdeepimages_figure}). The SITELLE deep image (filter SN3) of this section identifying these three regions, along with their spectrum, is shown in Fig.~\ref{fig:PlumeHII}, while their properties are listed in Table~\ref{tab:table2}. Their heliocentric velocity is very close to the derived systemic velocity of \NGC2445, which is consistent with the global movement of the plume in the simulations, perpendicular to the line of sight. Their \ha\ luminosities, assuming no extinction, amount to 3 to 5 times that of the Orion nebula.

\begin{table}
    \caption{Properties of the \hh\ regions in the plume}\label{tab:table2}
    \centering
    \small
    \begin{tabular}{ccccc}
    \hline
    Region & RA & Dec & F(\ha)$^1$ & V$_{\rm helio}$ [$\kms$]\\
    \hline
    H1 & 07:46:52.17 & +39:04:45 &  $8.5\pm1.8$ & $3905\pm14$ \\
    H2 & 07:46:52.55 & +39:04:35 & $12.9\pm1.9$ & $3930\pm10$ \\
    H3 & 07:46:52.54 & +39:04:30 &  $4.9\pm1.1$ & $3919\pm16$ \\
    \hline
    \end{tabular}
\begin{tablenotes}
      \item $^1$ In units of $10^{-17}\rm erg\,s^{-1}cm^{-2}$
\end{tablenotes}
    \end{table}

\subsection{Kinematics}
\label{sec:section31}
\subsubsection{Velocity distributions and rotation model}
 \label{sec:velocity}

 \begin{figure*}
    \includegraphics[scale=0.25]{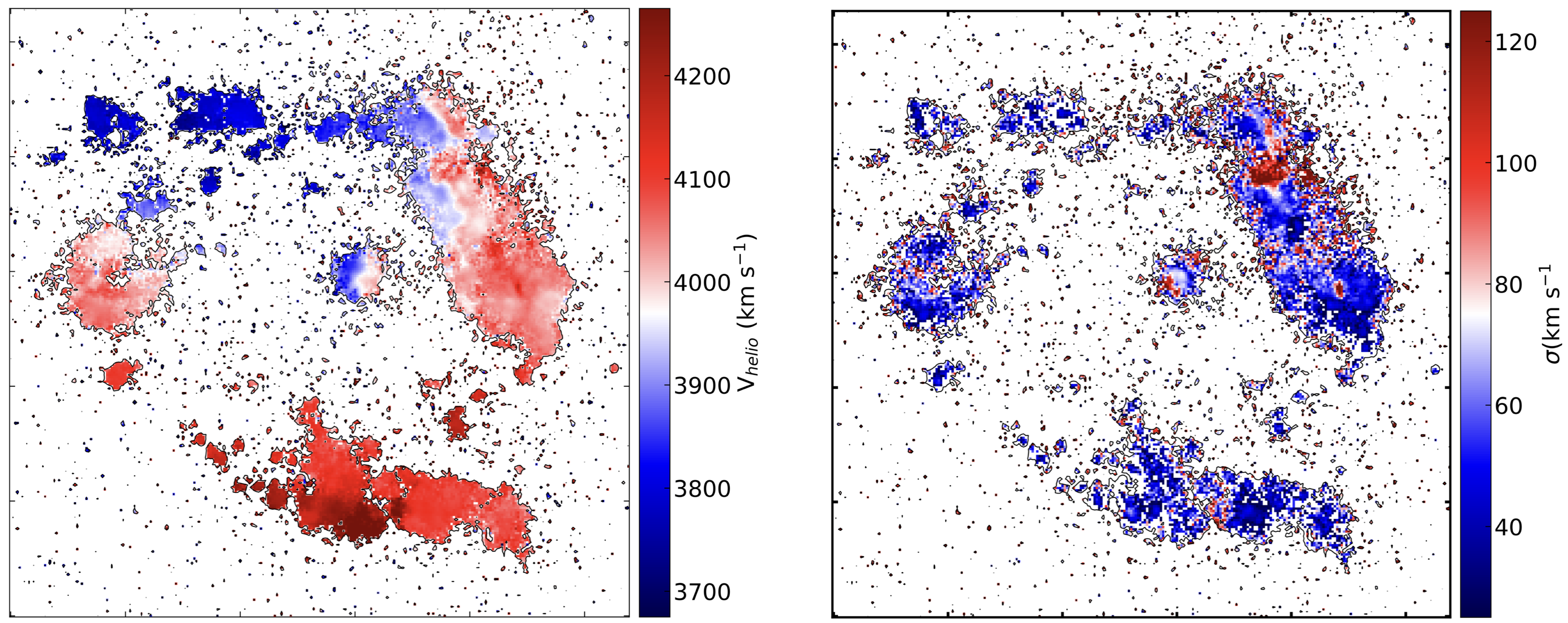}
    \caption{Heliocentric velocity (left) and velocity dispersion (right) of \NGC2445. Contours correspond to a flux of $6\times10^{-18}\rm erg\,s^{-1}\,cm^{-2}\,pixel^{-1}$. FOV is $86''$ on a side, centered at RA 07h46m55.51s, Dec $+39^\degree 00\arcmin49\arcsec$.}
    \label{fig:velocities_figure}
\end{figure*}

\begin{figure*}
    \includegraphics[scale=0.41]{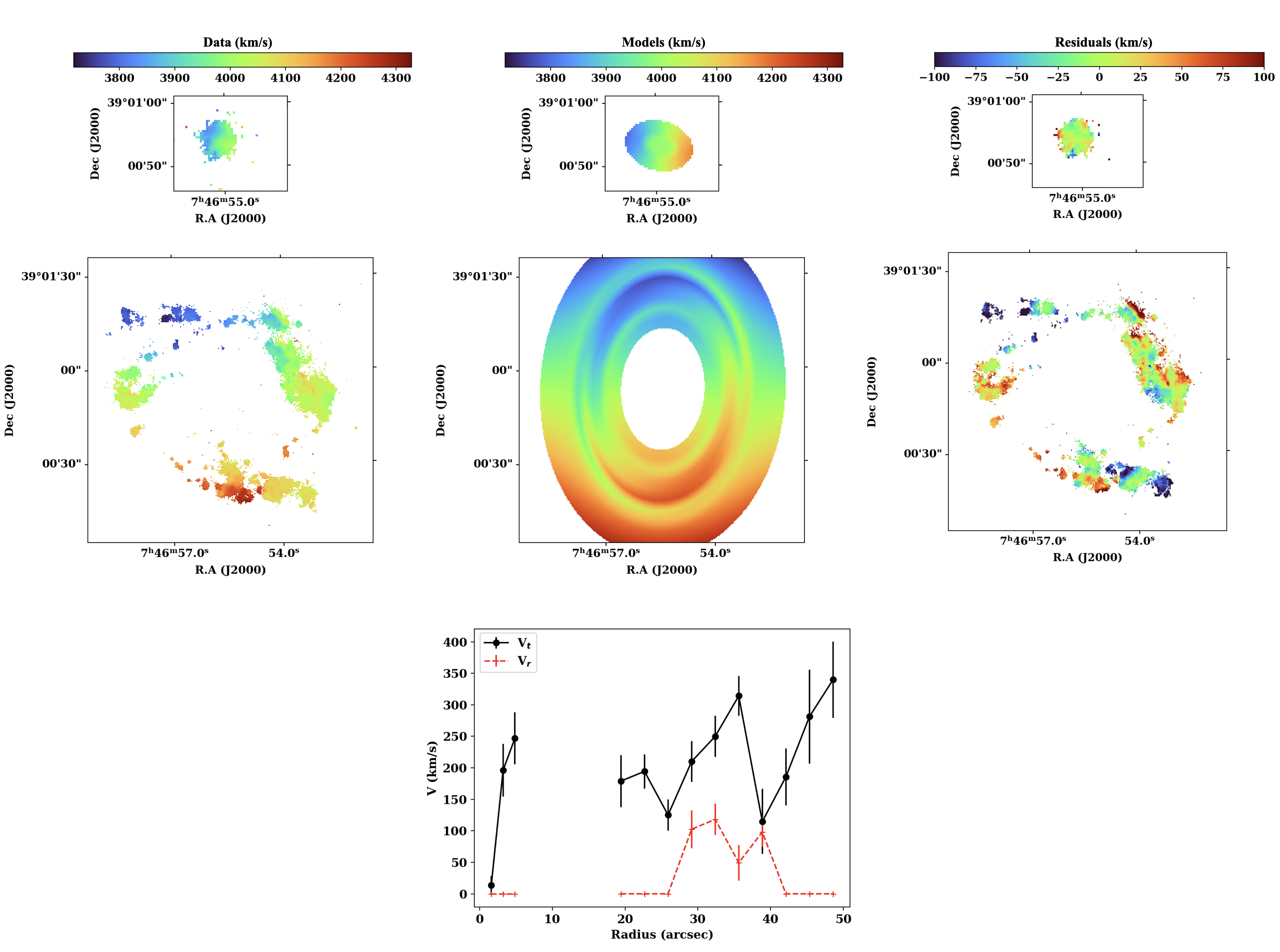}
    \caption{\ha\  velocity field (left), best fitted velocity models (middle), residuals (right) to the nucleus (top row) and ring structure (middle row) of \NGC2445. Rotation-only model is applied to the nucleus whereas the ring is fitted with a rotation+expansion model (see text). Radial profiles (bottom row) of the circular and expansion velocities are represented by the continuous (black) and dashed (red) lines respectively.}
    \label{fig:velocitymodel_figure}
\end{figure*}

Fig.~\ref{fig:velocities_figure} presents the heliocentric velocity and velocity dispersion maps of \NGC2445. Typical uncertainties on the velocity range from $0.9\,\kms$ per pixel in the core of the bright \hh\ regions to $30\,\kms$ for the faintest pixels included within the boundaries shown in this figure.   

 A very perturbed rotation pattern is evident in Fig.~\ref{fig:velocities_figure}, with a difference of $\sim600\,\kms$ between the extreme approaching and receding sides of the ring, which can be joined by a line passing $\sim4\,\rm kpc$ to the southwest of the nucleus and misaligned by $25^\circ$ with respect to the rotation pattern of the nucleus (shown in the 3rd panel of Fig.~\ref{fig:specNGC2445nucvel}).

\begin{table}
 \caption{Best-fitting kinematic parameters for \NGC2445.}
  \label{tab:param2445}
    \centering
      \small
      \begin{tabular}{lcc}
      \hline
      & Nucleus & Ring \\
      \hline
      $x_c$ & 7h46m54.9s & 7h46m54.4s \\
      $y_c$ & $+39\degree00'53.2''$ & $+39\degree00'53.6''$ \\
      $V_{\rm sys} [{\kms}]$ & $3992.7\pm5.7$ & $3998.2\pm6.9$ \\
      $i$ & $45.9\degree\pm9.5\degree$ & $45.9\degree\pm1.0\degree$ \\
      PA$_{{\rm kin},d}$ & $255.8\degree\pm3.7\degree$ & $176.7\degree\pm3.2\degree$ \\
      \hline
      \end{tabular}
\begin{tablenotes}
      \item Note: $x_c$ and $y_c$ are positions of the kinematic centre, and PA$_{{\rm kin},d}$ is the position angle of the kinematic major axis. 
\end{tablenotes}
\end{table}
 
 A simple rotation-only model is fitted to the nucleus ionized gas (within $\sim 4$~kpc) using \textsc{diskfit} \citep[][]{Spekkens2007,Sellwood2010,Sellwood2015} to determine its kinematic parameters. The model can be represented by the following expression : 
 \begin{equation}
    V_{\rm model}=V_{\rm sys}+V_t\sin i\cos \theta\,,  
	\label{eq:rotonlymodel}
 \end{equation}
 where $V_{\rm sys}$ is the systemic velocity, $V_t$ is the circular velocity, $\theta$ is the azimuthal angle relative to the major axis of the disk and $i$ is the disk inclination. Note that \textsc{diskfit} assumes a flat disk, and hence, returns a single value for the position angle, inclination, systemic velocity, and disk centre for the model. The model and residuals can be seen in Fig.~\ref{fig:velocitymodel_figure}. The kinematic parameters are listed in Table~\ref{tab:param2445}. Uncertainties on the models parameters were estimated by generating 100 bootstrap realizations of each velocity model \citep[][]{Sellwood2010}. The ionized gas nucleus is offset by $2.1^{\prime\prime}$ ($\sim577\,\rm pc$) southwest of the optical nucleus. 

Since collisional rings are predicted to be expanding caustic waves, a rotation + expansion model is applied to the 2D velocity field of the ring using \textsc{diskfit}. The rotation + expansion model can be expressed as :
\begin{equation}
    V_{\rm model}=V_{\rm sys}+\sin i\big[V_t\cos\theta+V_r\sin\theta\big]\,,  
	\label{eq:rotexpmodel}
\end{equation}
where $V_r$ is the radial velocity. Results of the ring modelling are also shown in Fig.~\ref{fig:velocitymodel_figure} and the associated kinematic parameters are listed in Table~\ref{tab:param2445}. The derived position angle of the ring is different from the direction of its elongation. The same result was obtained for \NGC2535 in Paper~I, and points to the fact that \NGC2445 is intrinsically elliptical. As collisional rings can be circular, the ellipticity was induced most likely following a previous encounter with its companion as shall be discussed in Section~\ref{sec:section4}. Another interesting result of the modelling is that the kinematic centre of the ring is offset from the kinematic centre determined for the ionized gas in the nucleus ($\sim5^{\prime\prime}$ or $\rm1.375\,kpc$ eastward). Besides, the determined position angles are different, hinting that those structures evolved separately. The kinematic centre of the ring is also different from the optical nucleus ($\sim3.4^{\prime\prime}$ or $\rm935\,pc$ shift southeast). The optical nucleus is closest to the northwestern part of the ring structure. This asymmetry is a direct consequence of ongoing interaction with \NGC2444. Fig.~\ref{fig:velocitymodel_figure} also shows the radial profiles of circular and expansion velocities of the models. $V_r$ reaches a maximum of $118\,\kms$ with a corresponding $V_t$ of $250\,\kms$ at a radius of $32^{\prime\prime}$, or $\rm9\,kpc$ considering the adopted distance. Assuming that the caustic wave have been travelling through the disk at a constant speed, we estimate an age $\rm\sim75\,Myrs$ for the ring structure.

Therefore, the kinematics in this system is certainly complex, with different parts of the ring galaxy moving with different velocities. This, together with off-plane and local non-circular motions induced by the interaction might explain the large residuals observed in the ring and the complex morphology of \NGC2445 as seen in Fig.~\ref{fig:SITELLEdeepimages_figure}. Some material may be falling on/from the companion, in the direction of the bridge, whereas the tips of the ring, located northeast and southwest, seem to bend away from the companion.

\subsubsection{Velocity dispersion}
 \label{sec:vsigma}

Apart from the complex radial motion in the ring discussed above, our observations also reveal high velocity dispersions (above $50\,\kms$; Fig.~\ref{fig:velocities_figure}) in a fair fraction of the bright star-forming complexes in the ring (in particular in the northwest part of the ring, closest to \NGC2444), as expected for interacting galaxies \citep[][]{Bournaud2011}. These values are significantly larger than those measured for \Arp82 \citep{Karera2022} where the pair of galaxies has only experienced a close passage.

Bulk motions with strong off-plane components are expected, as well as outflows from active star-forming regions. The most noticeable, high-$\sigma$ feature in \Arp143 is located in the northwest section of the ring\footnote{RA = 07:46:54.1, Dec = +39:01:09}, in an approximately circular zone (diameter $\sim1.5\,\rm kpc$) but with a diffuse opening to the west. It presents an average $\sigma$ of $135\,\kms$, a velocity $\sim100\,\kms$ higher than its surroundings, and a high \si/\ha\ ratio of 0.55. This structure is diffuse, outside the bright \hh\ regions and is not associated with a bright cluster in HST/ACS images.

\begin{table*}
    \caption{Initial properties of the simulated galaxies. $M$, $\ell$, and $h$ are the mass, disc scale length, and scale height (component indicated as index) respectively. Masses are in units of $\rm 10^9M_\odot$. Scales are in units of $\rm kpc$.} \label{tab:initial1}
    \centering
    \small
    \begin{tabular}{llcccccccc}
    \hline
    \smallskip
    Galaxy & Representing 
           & $M_{\rm DM}$ & $M_{\rm gas}$ & $M^{\rm bulge}_{\rm star}$ 
           & $M^{\rm disc}_{\rm star}$ & $\ell_{\rm gas}$& $h_{\rm gas}$  & $\ell^{\rm disc}_{\rm star}$  & $h^{\rm disc}_{\rm stars}$ \\
    \hline
    Gal1 & \NGC2444 & 3154 & 0.7375 & 36.800 & 36.800 & 1.8 & 0.18 & 1.8 & 0.18 \\
    Gal2 & \NGC2445 & 1577 & 16.800 & 10.000 & 47.200 & 3.6 & 0.36 & 1.8 & 0.18 \\
    \hline
    \end{tabular}
\end{table*}

\section{NUMERICAL MODEL}
\label{sec:section4}
\subsection{Simulation code}
\label{sec:section41}

We simulated the interaction between \NGC2444 and \NGC2445 using the tree/SPH galactic code GCD+ \citep[][]{Kawata2003,Kawata2013}. The code simulates the evolution of galaxies, accounting for self-gravity, hydrodynamics, radiative cooling, SF, SN feedback, metal enrichment, and metal diffusion. The overall method and parameters that govern SF and feedback are described in Paper I. We use a SF efficiency $C_{*}=0.01$, a SF density threshold $n\rm_{th}=10\,{\rm cm}^{-3}$, a stellar wind power $E\rm_{SW}=10^{36}\rm erg\,s^{-1}$, and a SN energy output $E\rm_{SN}=10^{51}\rm erg$ of which only 10 per cent contribute to the feedback, the rest being radiated away. The value of $E\rm_{SW}$ is typical of a massive star like an O star (a combination of $\dot M=3\times10^{-6}\rm M_\odot\,yr^{-1}$
and $V=1000\,\kms$).

Dark matter (DM), stellar, and gaseous components of the simulated galaxies are represented by particles. All baryonic particles (gas and stars) have equal masses. Each star particle represents a population of stars born at the same time, with masses in a narrow mass range, and 61 different masses ranges are used, such that the star particles together reproduce a Salpeter IMF \citep{Salpeter1955}. A $\Lambda$CDM standard cosmology with $h=0.73$, $\mathrm{\Omega_0}=0.266$, $\mathrm{\Omega_b}=0.044$, $\mathrm{\lambda_0}=0.734$ is assumed hereafter.

\subsection{Initial conditions}
\label{sec:section42}
    
DM halo particles are distributed following a Navarro-Frenk-White (NFW) profile \citep[][]{Navarro1996} with a concentration parameter $c = 20$. Stellar/gaseous disks of the progenitors are created with an exponential density profile: 
\begin{equation}
    \rho = \frac{M}{4 \pi h\ell^2} \mathrm{sech}^2 \left( \frac{z}h \right) \mathrm{exp}\left(-\frac{R}{\ell} \right),
	\label{eq:expprofile}
\end{equation}
where $R$ and $z$ are the radial and vertical coordinates, $\ell$ is the scale length, $h$ is the scale height (set to $\ell$/10), and $M$ is the mass of the stellar/gaseous component. A classical bulge component is added to the numerical models of the galaxies. The initial structural parameters of the progenitors are given in Table~\ref{tab:initial1}. We adopt the nomenclature used in Paper~I for the progenitors, i.e,  they are labelled Gal1 and Gal2 followed by the suffix ``int'' or ``isol'' depending on whether they are evolved in interaction or isolated. We set up initial central iron abundances $\rm[Fe/H]$ of stellar/gaseous disks for both Gal1 and Gal2 to $\rm-0.2\,dex$. $\rm[Fe/H]$ is distributed radially with an initial gradient of $\rm-0.04\,dex\,kpc^{-1}$. The $\upalpha$-elements abundances are given by:
\begin{equation}
    \rm[\upalpha/Fe] = 
\begin{cases}
    -0.16\rm [Fe/H],& \text{stars;} \\
    0, & \text{gas.}
\end{cases}
\end{equation}

\begin{table}
    \caption{Number of particles and particle masses for each component. Subscript b (baryons) refers to both star ans gas particles. Masses are in units of ${\rm 10^3M_\odot}$}\label{tab:resolution}
    \centering
    \small
    \begin{tabular}{lcccccc}
    \hline
    \smallskip
    Galaxy & $N_{\rm DM}$ & $N_{\rm gas}$ & $N^{\rm bulge}_{\rm star}$ 
    & $N^{\rm disc}_{\rm star}$ & $m_{\rm DM}^{\phantom0}$ & $m_{\rm b}^{\phantom0}$ \\
    \hline
    Gal1 & 1515168 &  3540 & 176640 & 176640 & 2082 & 208.3 \\
    Gal2 &  757584 & 80640 &  48000 & 226560 & 2082 & 208.3 \\
\hline
    \end{tabular}
\end{table}

Table~\ref{tab:resolution} gives the number of particles for each component, and the particle masses. Gas and star particles have identical masses $m_{\rm b}$.

The progenitors are initially separated by a distance sufficiently large to ensure that they start in a quasi-isolation stage. However, the separation between the interacting galaxies is not set to extremely large values since spiral arms may eventually evolve into central stellar bars within $\sim1\,\rm Gyr$ \citep[][]{Fanali2015}, before the interaction starts. Initial positions, velocities, and inclination angles are given in Table~\ref{tab:initial2}. The initial parameters were chosen to optimize the match with the observational data in a try-and-see approach where the quality of the numerical model is estimated by eye. In particular, the initial masses of stellar/gaseous components of Gal2 were chosen to reproduce those of \NGC2445, derived from observations. Our choice of the stellar mass of \NGC2444 is based on the $N$-body simulation of \citet{Narasimhan2003} who propose a mass ratio of two for the progenitors (the spiral galaxy being less massive) to account for the tidal disruption observed for the ring galaxy and the relative velocities of the galaxies in \Arp143. We assign a gas fraction of one per cent for the early-type progenitor Gal1. The ratio between stellar and DM halo masses is the same as the one used in Paper I. 

\begin{table}
    \caption{Initial positions $\bf R$ and velocities $\bf V$ of the galaxies, and orientation of their disks. Indices 1 and 2 refer to Gal1 and Gal2, respectively. $\bf R$ and $\bf V$ are expressed in cartesian coordinates $(x,y,z)$, while angles $\theta$ and $\phi$ indicate, in spherical coordinates, the direction of the angular momentum vector of the disks.}\label{tab:initial2}
    \centering
    \small
    \begin{tabular}{lr}
    \hline
    ${\bf R}_1\, [{\rm kpc}]$ & (0,0,0)\\
    ${\bf R}_2\, [{\rm kpc}]$ & ($-69,-17,0$)\\
    ${\bf V}_1\, [{\kms}]$ & (0,0,0)\\
    ${\bf V}_2\, [{\kms}]$ & ($460,-50,100$)\\
    $(\theta,\phi)_1$ & $(277^\circ,377^\circ)$\\
    $(\theta,\phi)_2$ & $(317^\circ,87^\circ)$\\
    \hline
    \end{tabular}
\end{table}

\subsection{Model results}
\label{sec:section43}
\subsubsection{Formation and evolution of tidal features}

To illustrate the evolution of the merging process between the galaxies, we show in the Appendix (Figs.~\ref{fig:morphology_evolution_stars_figure} and~\ref{fig:morphology_evolution_gas_figure}) snapshots of the stellar surface density and gas surface density, respectively, for the run with the interacting galaxies. Animations of these simulations are also presented in the Supplemental material. After the run, the whole model was rotated by $+20\degree$ in the $xy$ plane around the origin followed by a rotation of $-20\degree$ in the $yz$ plane still around the origin. The interacting galaxies rotate counterclockwise in the $xy$ plane adopted as the plane of the sky. Gal1int (\NGC2444) is easily identified in the $xy$ plane by it stellar disk seen edge-on.

\paragraph{The extended plume}

Stellar and gaseous disks of both galaxies start out almost featureless (Gal1int's gaseous disk is restricted to its central part). A first grazing encounter happens at $t=120\,\Myr$. Consequently, the stellar disk of Gal1int develops two strong spiral arms. Gal2int develops multiple transient and recurrent spiral arms at the approach of Gal1int. Right after the first encounter, its outer disk region appears lopsided (snapshots at $t=160\,\Myr$). The tidal-induced lopsidedness grows stronger as Gal2int continues interacting with Gal1int, ending up morphing in a long tidal tail and a much shorter counter tail, as shown in the snapshots at $t=235\,\Myr$. The inner disk of Gal2int is noticeably elliptical. This tidal tail grows longer as the interaction goes on but also becomes more diffuse, particularly at its tip, whereas the counter tail becomes more diffuse and shorter, ending up vanishing. The effect is much stronger for the stellar component. The tail has a helix-like or curled ribbon shape and is not straight (a bend is observable in the $xz$ plane). Indeed, as shown in the snapshot at $t=340\,\Myr$ for the gaseous component, the tail alternatively appears broad and narrow depending on the perspective (if broad in the $zy$ plane it is narrow in the $xy$ plane and vice-versa). The best match configuration of our numerical model to observations of \Arp143 occurs at $t=480$\,Myrs. At this time, the gaseous component of the tidal plume is seen as a $\rm\sim130\,kpc$ long tail in the plane of the sky. It is curved on the plane of the sky and broadens at about two-thirds of its length ($\rm\sim80\,kpc$), in agreement with \ho\ data and our Dragonfly images (Fig.~\ref{fig:DragonflyHIdeepimages_figure}). The broadening is not necessarily related to an expanding bubble-like shell caused by a burst of SF but might simply be, as suggested by the numerical model, a result of the intrinsic helix-like shape and the viewing angle of the extended plume (e.g, it is narrow at the corresponding position in the $zy$ plane). The plume is mainly gaseous with an estimated gas mass of $\rm9.4\times10^8M_{\odot}$,
%and $7\times 10^7 \rm M_{\odot}$ for its gaseous and stellar contents respectively. 
somehow higher than the value $\rm M_{HI}=3\times10^8M_{\odot}$ measured from observations \citep[][]{Appleton1987}. Besides, the appearance of the stellar counterpart of the plume as a thin filament at its base agrees well with observations too. Gal2int is located in front of Gal1int in the line of sight (l.o.s) and the plume goes behind Gal1int.

\paragraph{
Rings and spokes}

Gal1int and Gal2int undergo a second encounter at $t=440\,\Myr$, a slightly off-axis collision with their centres being located at the origin for Gal1int and at $(x,y,z)\equiv(-4.8,-2.6,1.1)\,\rm kpc$ for Gal2int. Matter in Gal1int and Gal2int experiences a brief radial inward pull towards their centres due to the additional gravitational attraction as they pass through one another. The shrinkage in size, shown in the snapshots at $t=460\,\Myr$, is most noticeable for Gal2int, the less massive galaxy, particularly for its gaseous component. As the intruder moves away, the extra gravitational acceleration diminishes, causing disk material to rebound. Numerous theoretical studies \citep[see, e.g,][for a review]{AppletonStruck1996} have addressed the issue of ring formation. It was shown that the response to the perturbation is faster in the inner disk and slower in the outer disk, so rebounding material meets material still moving inward, producing an expanding ring. The perturbation, in general, induces radial epicyclic oscillations to disk particles, with a frequency increasing with radial position and secondary inner rings can then form from a second rebound. Both Gal1int and Gal2int disks develop expanding rings visible in the snapshots at $t=440\,\Myr$. The ellipticities and PAs  of the rings are similar to those of the disks they are embedded in. The shapes of the rings are most likely inherited from the morphologies prior to the collision. \citet{Athanassoula1997} have shown that oblique impacts could generate eccentric rings in a circular disk. However, non-vertical passages targeting intrinsic elliptical disks would most likely produce rings with eccentricities and/or PAs that differ from those of their disks. In our model, the expansion of the rings is not azimuthally uniform; their shapes change with time and are not symmetrical, especially for Gal2int. For example, at $t=510\,\Myr$, Gal2int's ring is quite deformed. 

The kinematics of the expanding ring in Gal2int is discussed in detail here below. The collision between galaxies of comparable masses drive large amplitude perturbations, making the rings to expand together with their disks, giving the impression that the  galaxies are stretched. In fact, a stellar outer disk with a much lower surface density is always traced outside the rings in Gal1int and Gal2int. Outer disks can be observed too for the \Arp143 galaxies in the Dragonfly composite image in Fig.~\ref{fig:DragonflyHIdeepimages_figure}. According to this model, the rings intensities are maximum when they are young. They get broader and fade out as they expand, when material that falls back to the central parts of the galaxies is not compensated by the accumulation of swept up material. Therefore, ring galaxies would preferentially be identifiable when they are young or, conversely, surrounded by diffuse stellar disks beyond the rings. This is consistent with the results of the photometry analysis of \citet{Romano2008} who observe that rings in their sample of interacting galaxies are found to be located preferentially at around half-way through the stellar disk. 

As matter falls back and accumulates onto the nuclei, the stars form a spheroidal component supported by velocity dispersion, while the gas ends up forming a disk supported by rotation.
Then, the surface densities of the gaseous nuclear disks increase with time. It is worth mentioning that if the simulation is run with Gal2int being initially bulgeless, the resulting ring galaxy harbors a late-type central remnant. This is consistent with the results of \citet{Chen2018} who investigated, by means of numerical simulations, the impact of the bulge-to-disk mass ratio (B/D) parameter on the final properties of collisional ring galaxies and found that the density profile of the central remnant correlates with the progenitor's B/D.

\begin{figure}
    \includegraphics[width=\columnwidth]{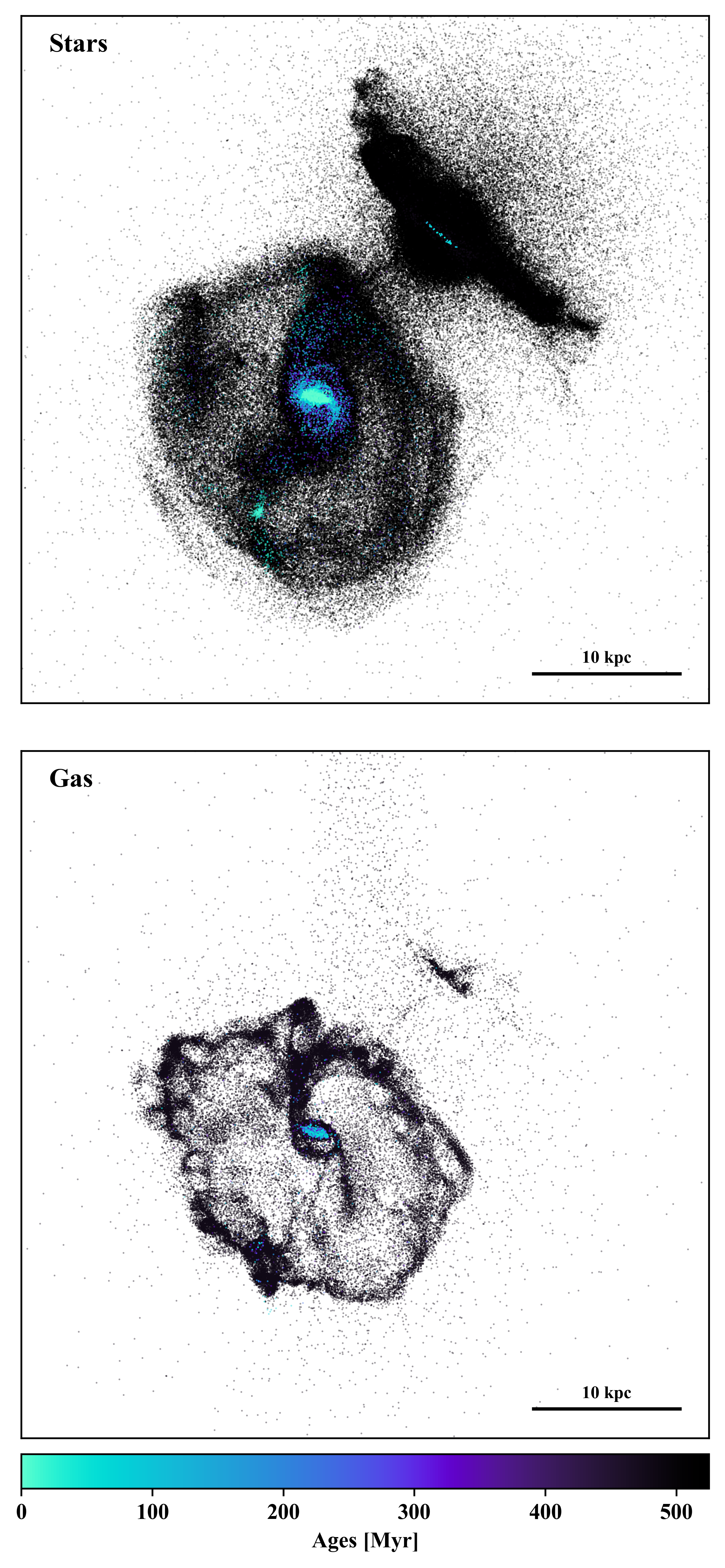}
\caption{Ages of the stars (top) and gas (bottom) at $t=480\,\Myr$.} For stars, the age is the time elapsed since the star formed. For gas, the age is the time elapsed since the gas was ejected by stellar outflows (winds and SNe). Large values correspond to stars and gas that were present at the beginning of the simulation.
    \label{fig:young_particles}
\end{figure}

\begin{figure}
    \includegraphics[scale=0.545]{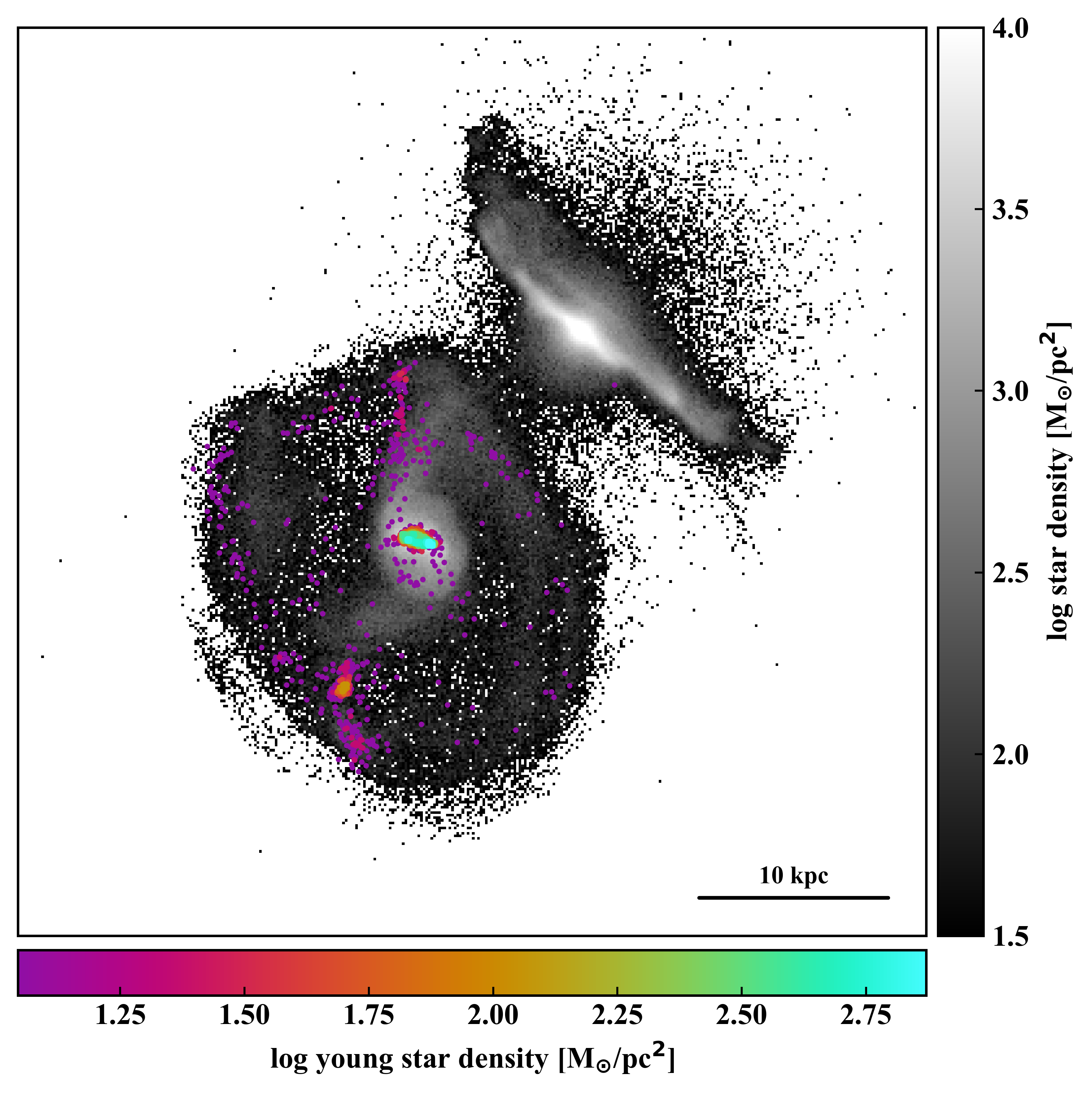}
\caption{Composite zoom-in surface density map showing, at $t=480\,\Myr$, the total stellar surface density (gray scale) and the stellar surface density of young stars, formed during the last $50\,\Myr$ (colour scale). The model predicts the presence of young stars in the nucleus of \NGC2445, as well as in the ring, as observed.}
    \label{fig:young_stars}
\end{figure}
 
To investigate the origin of the rings, we plot in Fig.~\ref{fig:young_particles} the age of stars and gas. For stars, the age is the time elapsed since the star formed; for gas, the age is the time elapsed since the gas was ejected by stellar winds of SNe explosions. An age of $480\,\Myr$ corresponds to stars and gas that were present in the initial conditions, at the beginning of the simulation. The youngest stars were formed when the system experienced a starburst during the second pericentre passage (see Fig.~\ref{fig:model_SF} below). In the top panel, we see a patchy outer ring and very compact inner ring in the centre of Gal2int, connected to the outer ring by spokes which are trailing with time. We also find some young stars in the centre of the gas-poor galaxy Gal1int. They are most likely second-generation stars that formed out of gas that was ejected by the stars present in the initial conditions. We find very little ``young gas,'' essentially concentrated in the centre of Gal2int and in a dense clump located in the lower part of the outer ring, where it intersects one of the spoke. Only $40\,\Myr$ separate the starburst and the present time, not giving enough time for the newly formed stars to produce significant outflows. Fig.~\ref{fig:young_stars} shows the surface density of all stars (grey scale) and stars recently formed (age $\leq50\,\Myr$, colour scale). The inner ring in Gal2int is seen as a thin and sharp circular structure bounded by the spokes. Many young stars are also found in the outer ring and along the spokes. In the corresponding gaseous surface density map (Fig.~\ref{fig:young_particles}, bottom panel), the inner ring can be recognized as a clumpy overdensity encompassing the nuclear disk.

\begin{figure*}
    \includegraphics[scale=1.20]{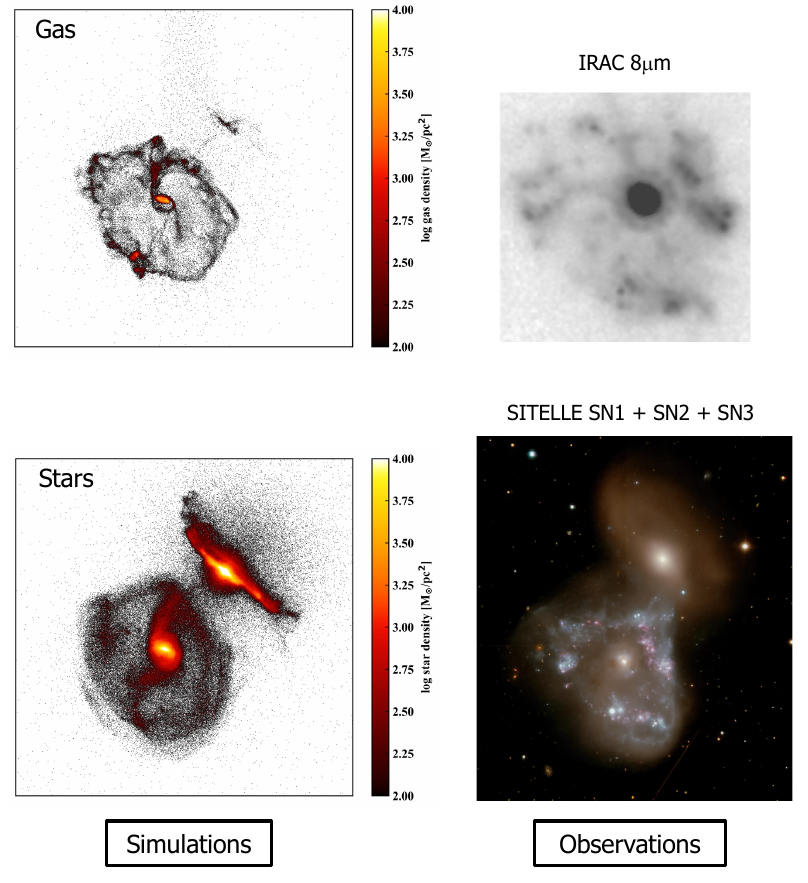}
    \caption{Zoom-in view of the gaseous (top left) and stellar (bottom left) surface density maps. For comparison purposes, archival \textit{Spitzer} and SITELLE composite coloured deep image are presented. In the optical image (bottom  right), the morphological structures such as the outer ring, spokes and inner ring in \NGC2445, easily recognizable by their blue colours tracing the young stellar clusters, are well reproduced by the model. The tidal features exhibit clumps in the gaseous component.}
    \label{fig:model_vs_observations}
\end{figure*}

Fig.~\ref{fig:model_vs_observations} show zoomed-in views (excluding the plume) of the stellar and gaseous surface density at $t=480\,\Myr$ for a comparison with observational data. A thin and partial ring around the nucleus of \NGC2445, with blue colours similar to those of its outer ring, can be identified in the SITELLE composite coloured deep image of \Arp143. We hypothesize that the invisible part of the probable inner ring is concealed and/or heavily extincted by the spokes and dust lanes in the eastern part of the disk of \NGC2445. A clumpy ring encloses the nucleus at the corresponding location in the archival \textit{Spitzer} mid-infrared \hbox{8 \textmu m} data. It is interesting to note that the model reproduces the clumpy nature of the outer ring and spokes as well. The model predicts that the inner rings expand much slower than the outer ones, and remain concentrated around the centre of the galaxies. The outer ring in Gal2int (and in \NGC2445) is approximately four times bigger and four times faster than the inner ring.

At best match, the numerical model reproduces well the morphology of \Arp143. The particular physiognomy of \NGC2444 is explained by the fact that Gal1int's ring resembles a thick, one-armed spiral that arches out of the plane of the stellar disk. Gal2int's ring has a projected radius of $\rm\sim10\,kpc$, well matching the size of the ring of \NGC2445 in SITELLE data (the observed maximum radial velocity was estimated at a radius of $9\,\rm kpc$). Its stellar and gaseous masses are roughly (ignoring material in the bridge and in the tidal plume) estimated to be $5.88\times10^{10}\,\rm M_{\odot}$ and $1.35\times10^{10}\,\rm M_{\odot}$, respectively, and the baryonic mass ratio between Gal2int and Gal1int is 0.99. This is close to the values derived from observations as well.  The corresponding values reported by \citet{Romano2008} are $6.05\times10^{10}\,\rm M_{\odot}$, $1.34\times10^{10}\,\rm M_{\odot}$, and 1.01. The projected distance ($\rm\sim18\,kpc$) between \NGC2444 and \NGC2445 is also well reproduced by the numerical model ($\rm\sim16\,kpc$).

\paragraph{Vertical structure.}

The perturbation is not limited to the plane of the disk.
A splash bridge connects Gal1int and Gal2int. Its stellar and gaseous contents have approximately masses of $2.48\times10^9\,\rm M_{\odot}$ and $0.072\times10^9\,\rm M_{\odot}$, respectively. Moreover, the stellar disks, when viewed edge-on, get thicker after the head-on collision. The effect is less important for the gaseous disks since gas is dissipative. This can be understood considering that the passages of the galaxies through one another also induce vertical oscillations. Indeed, the response in the $z$-direction of the extra gravitational pull is also stronger on the inner disk during the early phases of the collision whereas a similar effect occurs on the outer disk after the intruder passes through.

\paragraph{Merger remnant}

A third and fourth head-on collisions between Gal1int and Gal2int, closely spaced temporally, happens at $t=530\,\Myr$ and $t=570\,\Myr$, and leave the galaxies merged. The product of the merger evolves into an elliptical with shells and tidal tails. The extended tidal plume is still detectable in the gaseous component of the merger remnant. It is, thus, a long lived structure. 

\subsubsection{Kinematics}
\label{sec:modelkin}
\begin{figure}
    \includegraphics[width=\linewidth]{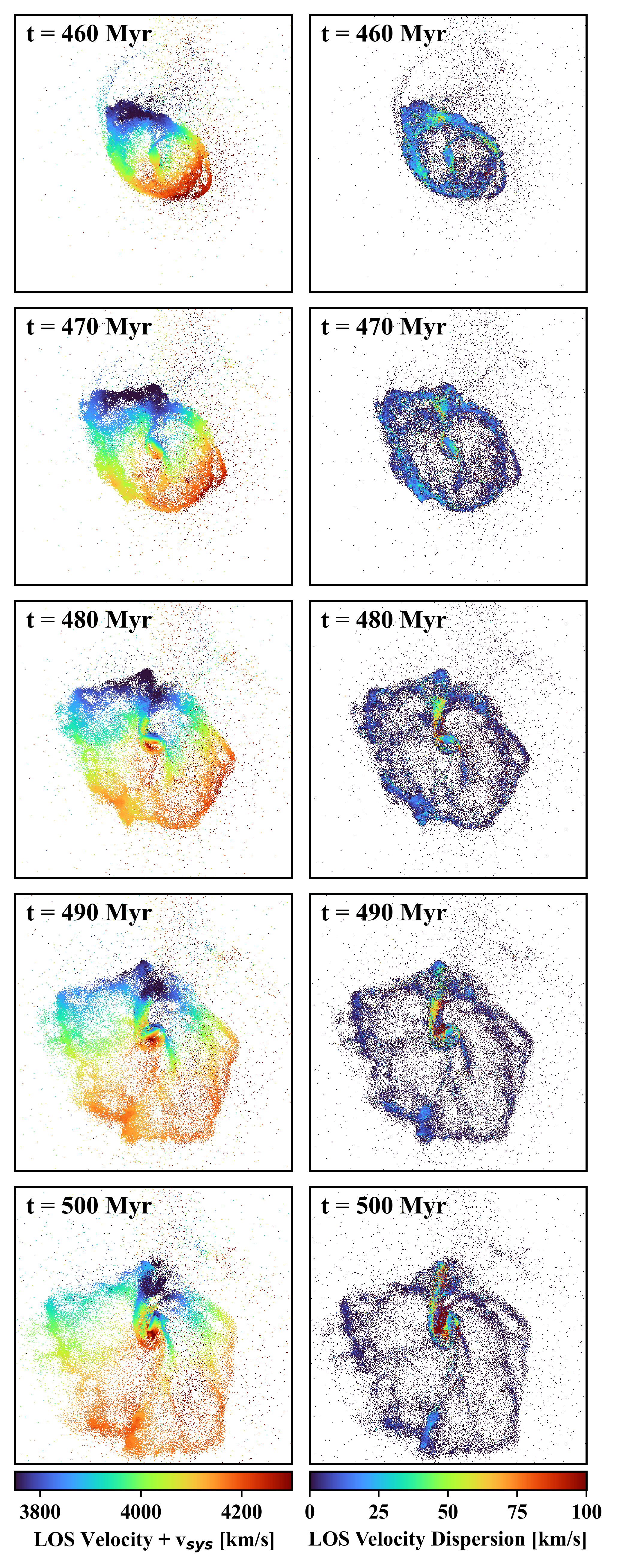}
    \caption{L.o.s. velocity and velocity dispersion of %ionized 
    gas in Gal2int, from $40\,\rm Myr$ before to $40\,\rm Myr$ after the present. Each panel is $\rm36\,kpc$ wide.}
    \label{fig:model_velocity_fields}
\end{figure}
 
Fig.~\ref{fig:model_velocity_fields} shows line-of-sight (l.o.s.) velocity and velocity dispersion, at five different times from $t=440\,\Myr$ to $t=520\,\Myr$, that is, around the time identified as the present. The average of the systemic velocities derived in Section~\ref{sec:section3} was added to the l.o.s velocity field to allow for a comparison with the data. The distribution of l.o.s velocities of Gal2int matches well ${\rm H\upalpha}$ velocities of \NGC2445. Nevertheless, the observed velocity dispersions are not well reproduced as the interstellar medium in the model is represented by a single-phase fluid whereas the different gaseous phases can be kinematically decoupled especially in galactic outflows produced by feedback events. High velocity dispersions ($\sim100\,\kms$) are predicted at the location of the inner ring where gas particles falling on the nucleus and gas particles drifted in the expanding inner ring are superimposed on the l.o.s. This is exactly what the observations show (Fig.~\ref{fig:specNGC2445nucvel}). The velocity dispersions in the outer ring of Gal2int are high at positions of star-forming complexes, reaching values of $\sim40\,\kms$. For comparison, velocity dispersion reaches maximum values of $\sim20\,\kms$ and $40\,\kms$ in the disk and nucleus of Gal2iso respectively. High dispersions in Gal2iso are also associated with sites of SF.

\begin{figure}
    \includegraphics[width=\linewidth]{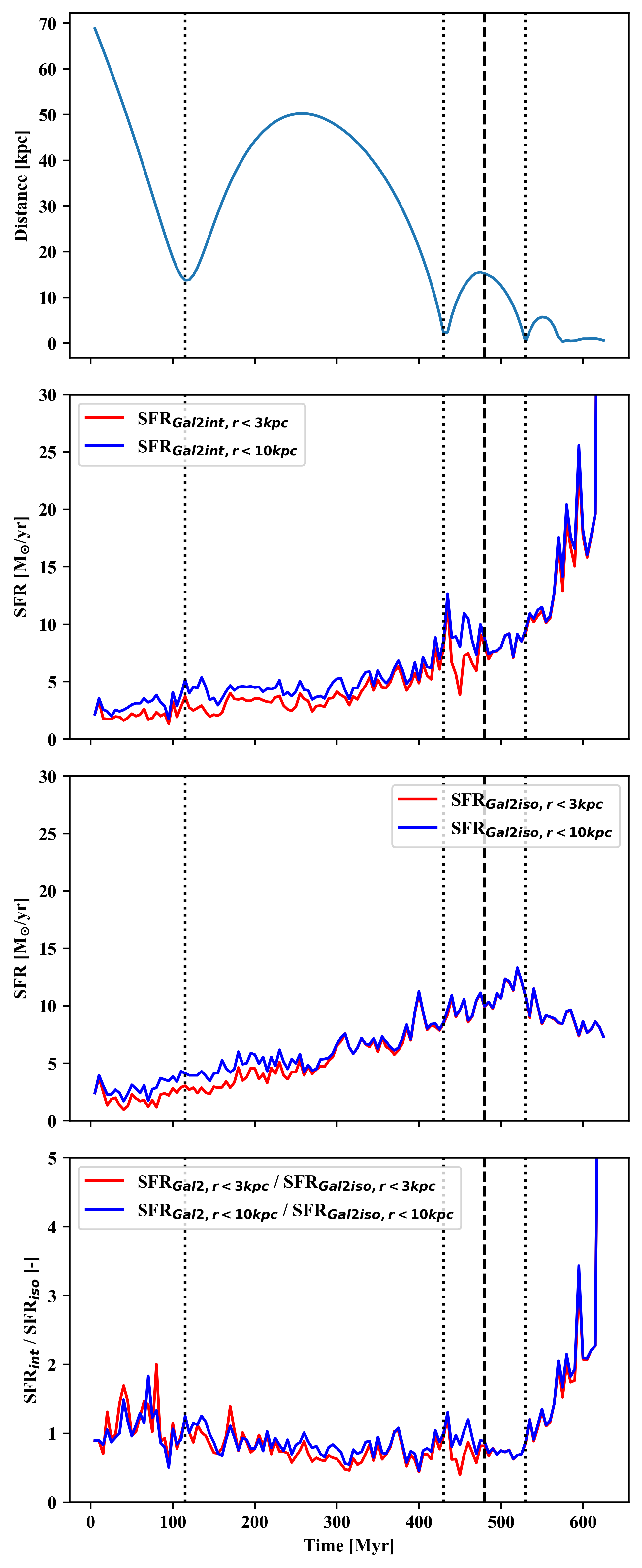}
    \caption{Top panel: temporal evolution of the separation between model interacting galaxies Gal1int and Gal2int. Second panel: evolution of SFR for interacting galaxy Gal2int, within $3\,\rm kpc$ and $10\,\rm kpc$ of the centre. Third panel: evolution of SFR for isolated galaxy Gal2iso. Bottom panel: Ratio of SFRs for galaxies Gal2int and Gal2iso. Vertical dotted lines indicate moments of pericentre passsages; the vertical dashed line marks present time (best match configuration).}
    \label{fig:model_SF}
\end{figure}

The expansion of the outer ring is not azimuthally uniform and the shape of the outer ring changes with time. A close look at the snapshots in Fig.~\ref{fig:morphology_evolution_stars_figure} and~\ref{fig:morphology_evolution_gas_figure} shows that the nucleus of Gal2int is knocked off after the head-on collision and is dragged afterwards towards Gal1int as the outer ring expands. At $t=440\,\Myr$, the nucleus of Gal2int is close to its eastern (left) outer ring edge. 

Also, the kinematic centre of the nuclear disk of Gal2int does not coincide with its centre of mass. The global velocity map at $t=480\,\rm\Myr$ is in general agreement with the observations.

\subsubsection{Star formation history}

To investigate the effects of the merging process on SF and to analyze how SF changes as a function of location, we show in Fig.~\ref{fig:model_SF} how the SFR  in Gal2 varies with time and radius,  when it interacts with Gal1 and when it is evolved in isolation. Broadly, the interacting galaxy experiences a series of bursts of SF that occur concurrently with closest passages. Latter bursts show more intense nuclear SF; SF is restricted to the inner $3\,\kpc$ of the remnant after the galaxies have merged, at the end of the run. In comparison, the SFR in Gal2iso increases smoothly, almost linearly. SF takes place both in the centre and in the outer regions until $t=290\,\Myr$, when the outer regions run out of dense gas. From that point, the SFR keeps increasing in the centre until $t=520\,\Myr$, when that region starts running out of gas. The SFR then slowly decreases, but is still around $7\,\rm M_\odot\,yr^{-1}$ at the end of the simulation, at $t=630\,\rm Myr$, possibly due to gas infall from the outer regions replenishing the centre.

At the time of the first encounter ($t=120\,\Myr$), the SFR in Gal2int increases by a factor of about 2, reaching $5\,\rm M_\odot\,yr^{-1}$. This takes place almost entirely in the outer regions. With a pericenter distance of $14\,\kpc$, the central $3\,\kpc$ is hardly affected by tides, while the outer regions are greatly affected, as Fig.~\ref{fig:morphology_evolution_gas_figure} shows.
A warp in the outer disk of Gal2int is formed which latter transforms into the tail and counter tail observable at $t=235\,\Myr$. From that point, the SFR remains roughly constant, at $4\,\rm M_\odot\,yr^{-1}$. 

At $t=300\,\Myr$, the SFR in Gal2int starts increasing again. At that point, the galaxies have pass their apocenter and are approaching again, increasing the strength of the tidal field. This happens almost entirely in the central $r<3\,\rm kpc$ corresponding the maximum radius of the inner ring). This is explained by the formation of spokes which funnel gas to the core, depleting the outer regions of gas. When the galaxies collide for a second time, at $t=440\,\Myr$, the SFR sharply increases and reaches a maximum of $13\,\rm M_\odot\,yr^{-1}$, exclusively in the centre. Interestingly, immediately after that, the SFR decreases in the centre, but SF activity reappears in the outer regions (gap between blue and red lines in second panel of Fig.~\ref{fig:model_SF}) as the formation of the ring moves gas from the centre to the outer regions, replenishing them. Eventually, the SFR starts increasing again in the centre, while the outer regions run out of gas again, this time due to gas consumption by the SF process. At $t=480\,\Myr$, which corresponds to the present, SF in the outer regions has stopped and from that point until the end of the simulation, all SF activity takes place in the centre.
The model does not predict SF in the extended tail but this is most likely due to the limited number of particles in the diffuse structure which does not allow to resolve adequately star-forming complexes.

As the interacting galaxies in our model approach one another, on their course towards a head-on collision, dense gas clumps form. Numerical simulations by \citet{Renaud2014} have shown that interactions enhance tidal compression and turbulence, generating an excess of dense gas in tidal features leading to intense SF. The tides are stronger at pericentre passages, causing the bursts of SF in our model.

In the bottom panel of Fig.~\ref{fig:model_SF}, we plot the ratio of SFR between Gal2int and Gal2iso, to highlight the effect of the interaction. The ratio remains close to one until the galaxies merge, but interestingly, the ratio is below one between the first two pericenter passages, indicating that the overall effect of tides is a {\it reduction\/} of the SFR during that period. In the presence of a tidal force field, different regions in the disk react differently: some contract while others are stretched, this latter effect being responsible, among other things, for the formation of the extended plume. The SFR will tend to increase in contracting regions, and decrease in stretched regions. In addition, a starburst is often followed by a period of low SF caused by the depletion of the gas reservoir, The final stellar mass in Gal2iso at present ($t=480\,\rm Myr$) is $5.93\times10^{10}M_\odot$, compared $5.88\times10^{10}M_\odot$ for Gal2int. Hence, the global effects of SF enhancement and suppression almost canceled out.

It is interesting to compare the star formation history of this system with the one of \NGC 2207/\IC 2163, another system of interacting galaxies that we have simulated in the past, using the same technique \citep{Poitras26}. In that previous study, the initial gas mass of \NGC 2207 was $8.75\times10^9\,M_\odot$, while in this study, the initial gas mass of Gal2, representing \NGC 2445, is $16.8\times10^9\,M_\odot$, a factor of 1.92 larger. In the case of \NGC 2207, the SFR slowly decreases until the galaxies reach pericenter. At that moment, the SFR increases, and reaches its peak value long after pericenter passage, when the galaxies have already reach their apocenter and started approaching each other for a second time. The same thing happened at the second pericenter passage: the SFR starts increasing again at that moment. The situation is different for \NGC 2445 (Fig.~\ref{fig:model_SF}, second panel). Not much happens during the first pericenter passage. Afterward, the SFR remains flat until we reached $t=300\,\Myr$, long before the second pericenter passage. At this point the SFR stars increasing, and reaches its peak value right at the second pericenter passage. This shows that when the gas content is low, as in \NGC 2207, the tides must be at their maximum values to have a significant effect, while in the case of a high gas content, the tides can have an effect long before they have reached their maximum values.

\subsubsection{Metallicity}

\begin{figure*}
    \includegraphics[scale=0.9]{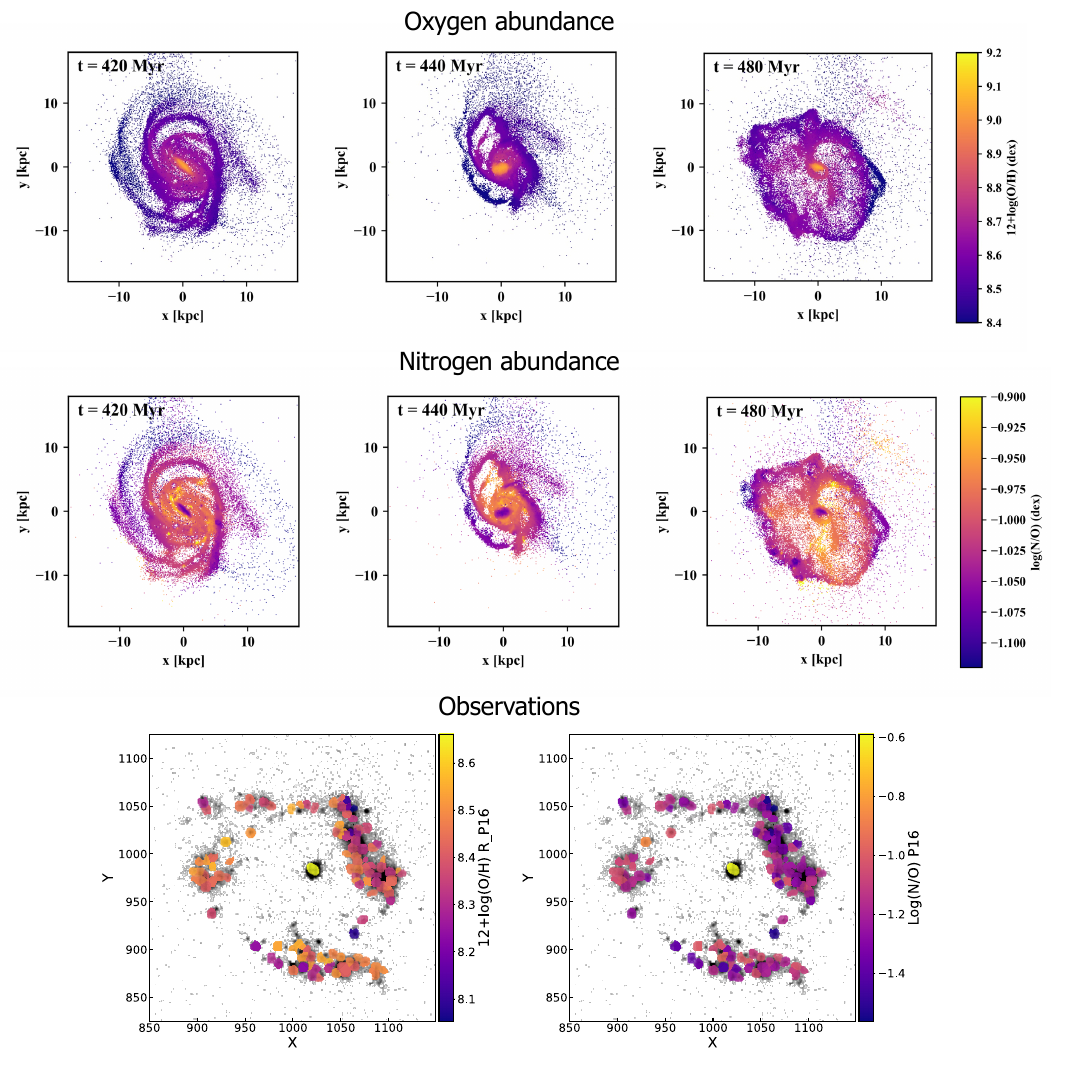}
\caption{Top row : Oxygen abundance maps of the numerical model before the head-on ($t=420\,\Myr$), right after the head-on collision ($t=440\,\Myr$), and at best match configuration ($t=480\,\Myr$). Middle row : Corresponding $\log\rm(N/O)$ maps. Bottom row : $12+\log\rm(O/H)$ and $\log\rm(N/O)$ derived from observations. Note the different scale used for log(N/O) maps from the simulations and observations.}
    \label{fig:model_metallicity_maps}
\end{figure*}

\begin{figure*}
    \includegraphics[scale=0.63]{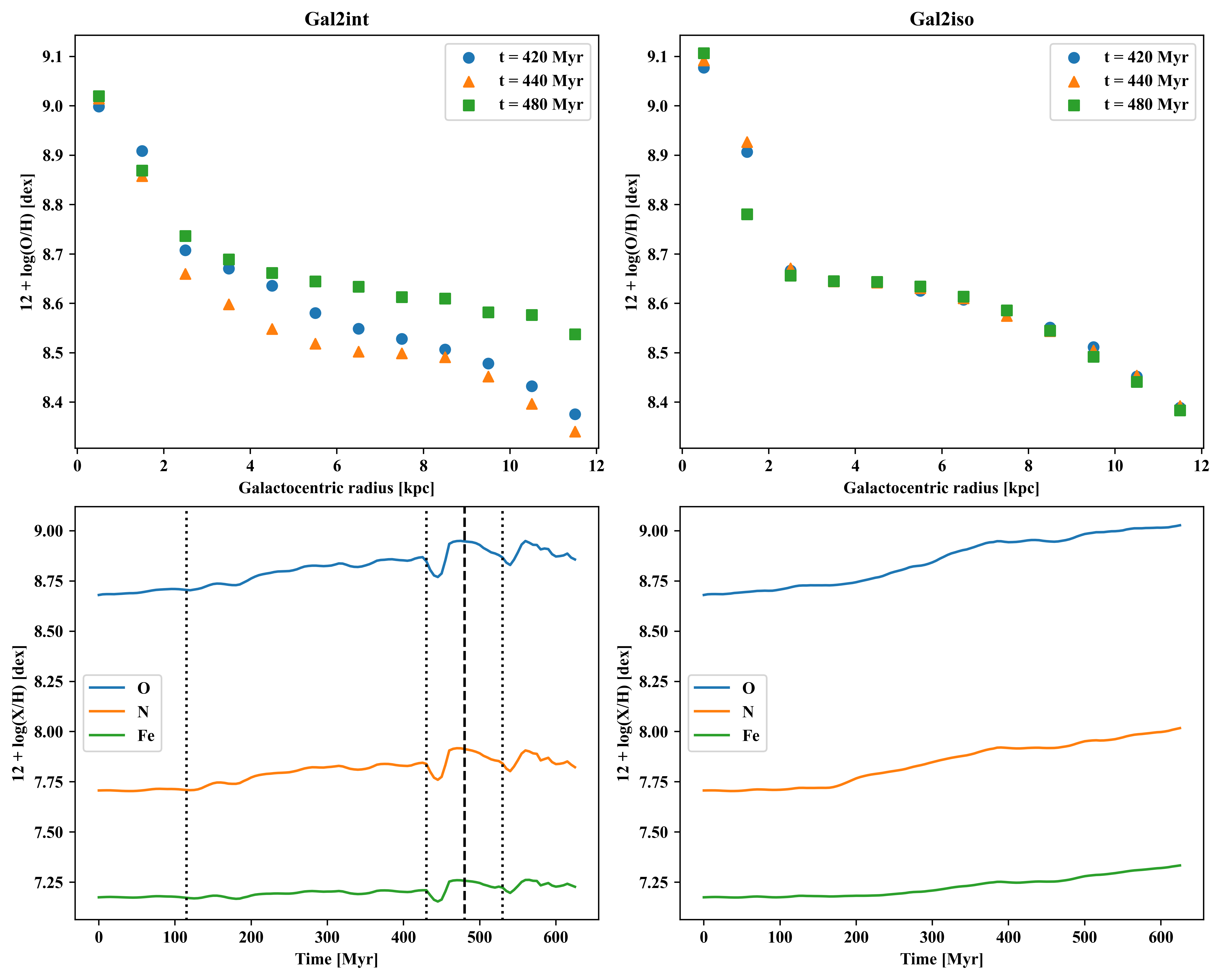}
    \caption{Top : Radial profiles of the oxygen abundance in Gal2int (left) and Gal2iso (right) before the outer expanding ring is formed ($t=420\,\Myr$), right after its formation ($t=440\,\Myr$), and at best match ($t=480\,\Myr$). Bottom : Temporal evolution of central ($r<4\,\rm kpc$) O, N, and Fe abundances in Gal2int (left) and Gal2iso (right). Vertical lines in bottom left panel have the same meaning as in Fig.~\ref{fig:model_SF}.}
    \label{fig:model_metallicity_evolution}
\end{figure*}

Taking advantage of the fact that the simulation code follows the evolution of several chemical elements, we investigate the effects of the interaction on the chemical properties of Gal2int during the ring phase.  The top row of Fig.~\ref{fig:model_metallicity_maps} shows the oxygen abundance maps at three different times: before the head-on collision ($t=420\,\Myr$), right after the head-on collision ($t=440\,\Myr$), and at the time the model best matches observations ($t=480\,\Myr$). In all the maps, the centre exhibits a metallicity $12+\log\rm(O/H)\approx8.8$. Before the head-on collision, gas with the lowest oxygen abundance values ($12+\log\rm(O/H)<8.55$) is found in the tidal arms of Gal2int. Some is still present in Gal2int's post-collision disk outskirts, beyond the nascent expanding ring while most of it is lost in the splash bridge as can be seen in projection across Gal2int's disk in the snapshot at $t=440\,\Myr$. By $t=480\,\Myr$, Gal2int's outer edge is swept up  and mixed in the expanding ring. The striking difference between the maps lies in the distribution of oxygen abundances in the outer parts ($r>4\rm\,kpc$). Both at $t=440\,\Myr$ and $t=480\,\Myr$, the expanding outer ring is characterized by an average $12+\log\rm(O/H)\simeq8.55$; a zone with an average $12+\log\rm(O/H)\simeq8.65$ separates the expanding outer ring and the centre. However, these specific areas are wider at $t=480\,\Myr$. Hence, it is the outer part of Gal2int's pre-collision disk that is essentially compressed/stretched when Gal2int shrinks/expands after the head-on collision. In Paper I, it was also found that the interaction in \Arp82 mainly affected the primary galaxy beyond a radius equivalent to its gaseous disk scale length, where extended tidal tails formed displaying constant metallicity over long patches. Gaseous aggregates with extremely low/high oxygen abundance values can be brought in the ring from the outskirts/central parts of Gal2int as a result of the complex kinematics at play, therefore creating strong local variations from the mean ring metallicity value. The best match model reproduces quite well the distribution of oxygen abundances in the ring derived from observations. In the nucleus, the oxygen abundance exceeds the observed value by about 0.25 dex. Feedback from a central supermassive black hole could lower the central SFR and reduce the central metal abundances. The model does not include a central supermassive black hole, and this could explain the discrepancy with the observations.

Nitrogen abundance relative to oxygen [log(N/O)] is shown in the middle row of Fig.~\ref{fig:model_metallicity_maps}. Sites of recent SF in all the maps have on average $\rm log(N/O)\simeq-1$. The values are slightly higher in their immediate surroundings because oxygen enrichment occurs on shorter time-scales than nitrogen enrichment in those sites of SF. Comparison to maps derived from observations (bottom row) shows that the model fails to reproduce values of $\rm log(N/O)>-0.9$ in the nucleus and in the ring. This is a consequence of the initial conditions used for the simulation (Section~\ref{sec:section42}). The $\rm[\alpha/Fe]$ ratios are identical for all alpha elements, including nitrogen and oxygen, implying that the $\rm N/O$ abundance ratio is initially equal to the solar value for all particles. This ratio evolves with time because chemical enrichment differently affects the abundances of N and O, but the spread of values only reaches $\rm0.1\,dex$ in the simulations. The much larger spread in $\rm N/O$, and the clear trend between the $\rm N/O$ ratio and the oxygen abundance seen in the observations strongly suggest the presence of an initial galactocentric $\rm N/O$ gradient in \NGC2445 before the interaction. 

Oxygen abundance radial profiles corresponding to the three maps are presented in the top row of Fig.~\ref{fig:model_metallicity_evolution}. The profiles are obtained by measuring average metallicities in nested elliptical annuli. The annuli are centered around the kinematic centre of the outer ring in the outer disk ($r>4\,\rm kpc$) whereas the centre of the nucleus is adopted for the central disk ($r<4\,\rm kpc$). The ellipticity and PA of the annuli at each time step are those that best fit the corresponding outer ring. The spokes region and the outer ring of Gal2int at $t=480\,\Myr$ is shallower in the corresponding radial distribution as expected from the stretching of the outer parts of Gal2int. The profiles at $t=420\,\Myr$ and $440\,\Myr$ are very similar 
except in the spoke region ($3\,\kpc<r<7\,\kpc$), where the profile at $t=440\,\Myr$ is steeper (same oxygen abundance values appear at smaller radii at $t=440\,\Myr$) since Gal2int shrinks right after the head-on collision. A close look at central oxygen abundances at the same times reveal that they are also slightly different. This can be explained by metal-poor gas infall. The evolution of the average central abundances (for oxygen, nitrogen, and iron) is depicted in the bottom row of Fig.~\ref{fig:model_metallicity_evolution}. As expected, decreases in elemental abundances mainly happen around pericentre passages. Infall of low-metallicity gas along the spokes contributes to the slight central metallicity depression at best match configuration. For comparison, central O, N, and Fe abundances in Gal2iso increase steadily and smoothly throughout the isolated galaxy simulation run. Outside the central region of Gal2iso, the metallicity distribution does not significantly change because SF and morphology are constant in the isolated galaxy. The evolution curve of central $12+\log\rm(Fe/H)$ is flatter than the evolution curve of central $12+\log\rm(O/H)$ since Fe is mainly produced by Type Ia SN whereas O is a byproduct of massive stars whose lifetimes are much shorter than Type Ia SN progenitors. The growth rate of N lies in between those of O and Fe, hinting that some of it is produced in a secondary way (if N was mainly of secondary origin, its growth rate would be similar to that of O). Since the production of secondary N is effective at $12+\log\rm(O/H)>8$ \citep[][]{Vincenzo2016}, it should therefore also happen in the outer ring, although significant enrichment and mixing should not be expected for the short lifetime of the ring phase in our model.

\citet{Bransford1998} analyzed oxygen abundances in a sample of northern ring galaxies. The galaxies in their sample have similar K-band luminosities, implying progenitors with uniform initial luminosities. Given that total luminosity is the main driver of mean oxygen abundance, we can assume that metal contents in the progenitors might have been similar too. The abundance values in the rings were found to exhibit rather uniform values despite having different ring sizes. This result is incompatible with the classical scenario \citep[][]{LyndsToomre1976} of a density wave propagating down abundance gradients, where larger rings would be expected to have lower abundances than smaller rings. It was suggested that there may exist an observational selection effect making the rings to be preferentially observed when their radii is close to one optical scale length, or that head-on collisions smooth out gradients seen in normal spiral galaxies (gas mixing between the primary and companion galaxies) but do not greatly affect their mean oxygen abundances. Rings being formed at a radius of about one disk scale length and being expanding material wave, as shown in our model, offer an alternative explanation to the observations of \citet{Bransford1998}. Recent 2D analysis of nebular abundances in the Cartwheel galaxy by \citet{Zaragoza-Cardiel2022} also favors a material wave-like nature to its ring. Besides, high resolution ultraviolet images of the Cartwheel galaxy obtained by \citet{Mayya2024} also suggest that a major fraction of formed stars got dragged in the ring wave. It is then likely that most ring galaxies are formed following head-on collisions with companions with comparable masses to form expanding material ring waves. The low metallicity at one disk scale length is conducive to the formation of progenitors of ULXs \citep[][]{Mapelli2010}.

\section{Discussion: global evolution of the system}
\label{sec:section5}

According to the numerical simulation, the ring structure in the star-forming galaxy changes its morphology during its evolution inasmuch as corresponding radial velocities are not azimuthally uniform. The companion too develops an expanding ring after the collision and is currently seen edge-on. The rings are brighter right after the collision. They broaden and fade out as they expand. Hence, the numerical model predicts that rings consistent with the formation scenario of \Arp143 (mass ratios of a few) would preferentially be recognizable when they are young and embedded in their hosts disks, especially if they are distant. The rings form at a radius of about one disk (gaseous) scale length. Considering the estimated young age of the ring structure in \NGC2445, its corresponding average metallicity did not significantly change during its evolution. 

The final stages of ring phase are characterized by the presence of spokes (formed only in the gas-rich galaxy) and secondary inner rings: we speculate that the semi-circular structure with blue colours which bounds the nucleus of \NGC2445 in SITELLE deep images is a secondary inner ring seen with its eastern part obscured. It is worth noting that the size ratio of the outer ring to the secondary inner ring in the numerical model is equal to the ratio of the corresponding expansion velocities as predicted by the analytical theory of \citet{Struck2010}. The secondary inner ring in \NGC2444 is not detected in the images as it is concealed by the edge-on galaxy's disk. The numerical model also suggests that the bridge that links the interacting galaxies in \Arp143 is a strong one, incorporating roughly a third of the baryonic content of \NGC2445.  

At closest approach, the head-on-collision that led to the formation of \Arp143 triggered a burst of SF essentially confined in the rings and associated with the formation of massive, dense and compact clumps with high velocity dispersions. Some of the biggest clumps might be progenitors of the ULXs detected in the system. At final stages of the ring phase of \NGC2445, as it becomes more and more diffuse and spokes funnel gas to the nucleus, its SF is dominated by the nuclear starburst. This gas funneling also induces a slight decrement of the nuclear metallicity. Future simulations incorporating supermassive black holes at the cores of the interacting galaxies will help to investigate the effects of the interaction on nuclear activity. One particular aspect of the numerical model is that it is different from the classical propagating density wave scenario, in that the bulk of formed stars/clusters are carried by the wave. Recent simulations of Cartwheel-like galaxies \citep[][]{Renaud2018} yielded similar results and also highlighted the transfer of SF from the ring to the nucleus during the evolution of the collisional ring galaxy, proposing that the location of SF be used as a tracer of dynamical stage of the interaction. Hence, even if \NGC2445 is morphologically similar to the Cartwheel galaxy (both are gas-rich collisional ring galaxies with spokes and double rings), the former should be relatively in a more advanced stage since the nucleus of the latter is almost devoid of H$\upalpha$ emission \citep[][]{Higdon1995b} despite its much older age estimated to be $\sim300\,\Myr$ based on \ho\ observations of \citet{Higdon1996}. In fact, the age of the outer ring is a strong function of the mass ratio of the colliders as more massive intruders induce faster expansion (higher ratios of radial to circular velocities) and therefore, faster evolution.

Of importance, the flyby also affects the morphology of the gas-rich colliding galaxy by changing its ellipticity and inducing some warping. Therefore, taking into account the possibility of multiple collisions in numerical models of collisional ring galaxies (at least in some cases) allows to have better insights on details (even minor ones) that characterize them. 

\section{Summary}
\label{sec:section6}

Optical spectral data cubes of \Arp143, obtained using the imaging spectrometer SITELLE, along with deep optical broad-band images obtained using the Dragonfly Telephoto array, are presented and analyzed. Numerical simulations that reproduce the properties of the pair of interacting galaxies were also run to investigate the effects induced by the interaction, using the chemodynamical evolution code GCD+. We highlight below the main outcomes of this work.

\begin{enumerate}

\item We detect nearly 230 \hh\ region complexes in \Arp143, mainly located in the ring of \NGC2445 but also in its nucleus. The total extinction-corrected \ha\ luminosity yields a SFR of about $3.8\,\rm M_\odot\,yr^{-1}$, nearly half of which 
(47\% of the \ha\ luminosity) is associated with the nucleus of \NGC2445. This current SFR is well reproduced by our simulation.

\item The optical counterpart of the extended \ho\ plume of \Arp143 is identified in the Dragonfly deep images. Three star-forming regions are seen by SITELLE in the plume, about 60\,kpc north of the galaxy pair. The numerical model reproduces very well the morphology and extent of this tidal feature and suggests that it formed roughly 300\,Myr ago as a result of a flyby between the colliding galaxies and is associated with debris from \NGC2445’s disk periphery. The simulation indicates that the plume might remain detectable in the merger remnant (for at least $\sim$240~Myr after its formation) although with very low surface brightness.

\item The dynamical study of the ionized gas reveals a rotating disk and an  expanding ring not centered on the galaxy’s optical center as identified by the nucleus. Assuming that a wave generated by the collision has been travelling through the disk at a constant speed, we infer an age of $\sim$75\,Myr for the ring. This agrees with the full numerical model of the interaction: the ring galaxy is indeed the result of a head-on collision that occurred some 50\,Myr ago with an early-type companion twice as massive, and the interaction displaced \NGC2445’s nucleus, offsetting it from the kinematic centre of the expanding ring.

\item The velocity dispersion of the ionized gas in the ring, particularly in the NW part that is close to the companion, is significantly higher than in unperturbed galaxies. 

\item A ridge of low surface brightness, low ionization and highly turbulent ionized gas is detected between the two galaxies, offset from the \hh\ region complexes from the NW part of \NGC2445's ring. Its origin is unclear.

\item Although we cannot exclude the presence of an AGN, the optical spectrum of the core of \NGC2445 is dominated by a starburst with a significant extinction: ${\rm A}_V\simeq3\,\rm mag$ (while we measure an average extinction of $\sim$1.3\,mag in the ring). We detect regions of high velocity dispersion to the SE and NW of the nucleus, which could be due to an outflow from the nucleus or, to the contrary, as suggested by the numerical simulations, to an inflow of material.

\item The metallicity spread found in the ring of \NGC2445, with an average value near 12+log(O/H)\,=\,8.5 and a possible negative gradient for the \hh\ region complexes in the SW part of the ring, is likely caused by the migration of the gaseous structures from different initial galactocentric radii into the expanding ring during the interaction. The N/O ratio observed is consistent with this interpretation and with the ongoing star-formation activity.

\item Finally, the simulations predict that the interacting galaxies are on the verge of merging after a third and fourth head-on collisions that will take place in $\sim50\,\Myr$and $\rm90\,\Myr$, respectively. 

\end{enumerate}

\section*{Acknowledgements}
We thank the referee, Divakara Mayya, for his constructive and detailed report, in particular for pointing out an issue with our initial simulations. This paper is based on data obtained with  SITELLE, a joint project of Universit\'e Laval, ABB, Universit\'e de Montr\'eal and the Canada-France-Hawaii Telescope (CFHT), with support from the Canada Foundation for Innovation, the National Sciences and Engineering Research Council of Canada (NSERC) and the Fonds de Recherche du Qu\'ebec -- Nature et Technologies (FRQNT). The authors wish to recognize and acknowledge the very significant cultural role that the summit of Mauna Kea has always had within the indigenous Hawaiian community. We are most grateful to have the opportunity to conduct observations from this mountain. We also thank Roberto Abraham for the fast-track access to the Dragonfly telescope array. LD, HM and CR are grateful to NSERC and FRQNT for financial support. SDP acknowledges financial support by MINECO under grants PID2023-149578NB-100 and PID2022-136598NB-C32. JMV and JIP acknowledge financial support from State Agency for Research of the Spanish MCIU through Center of Excellence Severo Ochoa’ award to the Instituto de Astrofísica de Andalucía.

LD and CR would like to thank the members of the astrophysics group of the Universidad de Granada for their hospitality and fruitful discussions during their sabbatical stay when this paper was completed.

\section*{Data Availability Statement}
The original SITELLE data cubes underlying this article can be downloaded from the Canadian Astronomy Data Centre web site at https://www.cadc-ccda.hia-iha.nrc-cnrc.gc.ca/. Flux and velocity maps extracted from these cubes will be shared on reasonable request to the authors.



\bibliographystyle{mnras}
\bibliography{references}

%%%%%%%%%%%%%%%%%%%%%%%%%%%%%%%%%%%%%%%%%%%%%%%%%%

%%%%%%%%%%%%%%%%% APPENDICES %%%%%%%%%%%%%%%%%%%%%

\appendix

\onecolumn

\section{Emission line fluxes and extinctions.}
\label{sec:appendixA}

Table~\ref{tab:tableA1} gives the flux of the main emission lines measured for the 227 \hh\ region complexes detected, along with their extinction calculated from their Balmer decrement (see Section~\ref{sec:extinc}). These fluxes are corrected for the global underlying stellar population (see Section~\ref{sec:stelpop}), but not for the extinction. Extinction values below 0 mag have been omitted in the table. Detected regions corresponding to HI and H2 are not in this table. Other missing emission regions have been considered as bad detections (i.e. noisy ZoI, bad $\chi^2$ of their profile, and/or small pixel number in their domains, see Section~\ref{sec:detection}).

%\centering
\begin{scriptsize}
\begin{longtable}{rrrccrrrrrrrr}
\caption{Positions and fluxes (uncorrected for reddening, in units of $10^{-16}$\,erg\,s$^{-1}$\,cm$^{-2}$) of the \hh\ region complexes detected in \Arp143.}
\label{tab:tableA1}\\
\hline
ID & X & Y & RA & DEC & \ool & \hb & \oool & \ha & \nnl & \si$\lambda$6716 & \si$\lambda$6731 & A$_{\rm V}$ \\
\hline
\endfirsthead
\caption*{Table \ref{tab:tableA1}: Cont.}\\
\hline
ID & X & Y & RA & DEC & \ool & \hb & \oool & \ha & \nnl & \si$\lambda$6716 & \si$\lambda$6731 & A$_{\rm V}$ \\
\hline
\endhead
\multicolumn{10}{c}{\ldots}
\endfoot
\endlastfoot
1 & 1093 & 861 & 7:46:53.1 & +39:00:14.5 & 3.39 ± 1.18 & 0.92 ± 0.41 & 1.42 ± 0.44 & 3.76 ± 0.38 & 1.35 ± 0.28 & 0.48 ± 0.27 & 0.87 ± 0.28 & 0.94 ± 0.31 \\ 2 & 1081 & 866 & 7:46:53.5 & +39:00:16.0 & 3.61 ± 0.71 & 1.00 ± 0.31 & 1.10 ± 0.34 & 3.82 ± 0.37 & 0.79 ± 0.25 & 0.61 ± 0.26 & 0.41 ± 0.26 & 0.77 ± 0.92 \\
3 & 1090 & 866 & 7:46:53.2 & +39:00:16.0 & 4.89 ± 0.82 & 1.19 ± 0.37 & 0.66 ± 0.40 & 4.14 ± 0.32 & 1.65 ± 0.24 & 0.51 ± 0.23 & 0.89 ± 0.23 & 0.52 ± 0.89 \\
4 & 1058 & 868 & 7:46:54.1 & +39:00:17.0 & 7.66 ± 1.35 & 2.45 ± 0.41 & 0.42 ± 0.44 & 6.44 ± 0.44 & 2.80 ± 0.34 & 1.91 ± 0.34 & 1.27 ± 0.33 & -- \\
5 & 1091 & 868 & 7:46:53.2 & +39:00:16.8 & -- & -- & 2.98 ± 1.56 & 3.68 ± 0.27 & 1.11 ± 0.20 & 0.96 ± 0.20 & 0.21 ± 0.20 & -- \\
6 & 1027 & 872 & 7:46:55.0 & +39:00:18.3 & 5.69 ± 1.00 & 1.52 ± 0.31 & 1.83 ± 0.33 & 6.82 ± 0.37 & 1.82 ± 0.26 & 1.92 ± 0.27 & 1.14 ± 0.27 & 1.19 ± 0.57 \\
7 & 1021 & 873 & 7:46:55.1 & +39:00:18.6 & 17.27 ± 1.30 & 4.51 ± 0.61 & 5.48 ± 0.67 & 16.61 ± 0.47 & 3.02 ± 0.32 & 3.85 ± 0.34 & 2.60 ± 0.34 & 0.67 ± 0.37 \\
8 & 1086 & 874 & 7:46:53.3 & +39:00:18.7 & 11.42 ± 1.30 & 3.54 ± 0.46 & 2.86 ± 0.50 & 17.58 ± 0.50 & 5.80 ± 0.36 & 3.72 ± 0.36 & 3.08 ± 0.36 & 1.46 ± 0.36 \\
9 & 1030 & 875 & 7:46:54.9 & +39:00:19.2 & 5.62 ± 0.93 & 2.04 ± 0.50 & 0.87 ± 0.40 & 6.87 ± 0.41 & 1.42 ± 0.28 & 2.16 ± 0.30 & 0.75 ± 0.29 & 0.43 ± 0.69 \\
10 & 1046 & 876 & 7:46:54.4 & +39:00:19.5 & 32.69 ± 1.56 & 9.66 ± 0.54 & 7.17 ± 0.58 & 33.56 ± 0.50 & 11.05 ± 0.36 & 7.00 ± 0.36 & 4.22 ± 0.36 & 0.51 ± 0.15 \\
11 & 1085 & 876 & 7:46:53.3 & +39:00:19.3 & 14.45 ± 1.51 & 4.34 ± 0.57 & 3.34 ± 0.53 & 18.87 ± 0.39 & 5.97 ± 0.28 & 4.38 ± 0.29 & 3.05 ± 0.29 & 1.11 ± 0.36 \\
12 & 1055 & 877 & 7:46:54.2 & +39:00:20.0 & 26.17 ± 1.25 & 8.12 ± 0.49 & 6.79 ± 0.53 & 44.11 ± 0.61 & 12.44 ± 0.43 & 7.92 ± 0.43 & 6.16 ± 0.43 & 1.71 ± 0.17 \\
13 & 1001 & 879 & 7:46:55.7 & +39:00:20.6 & 32.60 ± 1.59 & 9.66 ± 0.79 & 8.51 ± 0.77 & 48.09 ± 0.60 & 14.94 ± 0.43 & 6.89 ± 0.43 & 4.72 ± 0.43 & 1.47 ± 0.22 \\
14 & 1094 & 879 & 7:46:53.1 & +39:00:20.3 & 3.72 ± 1.01 & 1.21 ± 0.36 & 1.76 ± 0.42 & 5.33 ± 0.30 & 1.59 ± 0.22 & 1.40 ± 0.22 & 0.96 ± 0.22 & 1.14 ± 0.84 \\
15 & 1024 & 880 & 7:46:55.1 & +39:00:20.8 & 14.74 ± 1.14 & 2.92 ± 0.51 & 4.34 ± 0.61 & 14.84 ± 0.50 & 3.42 ± 0.35 & 3.34 ± 0.37 & 2.79 ± 0.37 & 1.53 ± 0.48 \\
16 & 1038 & 880 & 7:46:54.7 & +39:00:20.9 & 11.28 ± 0.90 & 2.95 ± 0.48 & 2.51 ± 0.46 & 9.79 ± 0.58 & 1.84 ± 0.37 & 2.04 ± 0.39 & 2.11 ± 0.39 & 0.39 ± 0.47 \\
17 & 1052 & 881 & 7:46:54.3 & +39:00:21.2 & 80.57 ± 3.41 & 23.67 ± 0.99 & 22.39 ± 1.06 & 132.10 ± 1.07 & 37.73 ± 0.76 & 21.62 ± 0.77 & 14.07 ± 0.77 & 1.78 ± 0.11 \\
18 & 1087 & 881 & 7:46:53.3 & +39:00:21.0 & 10.64 ± 1.01 & 3.35 ± 0.39 & 1.95 ± 0.42 & 11.52 ± 0.36 & 3.65 ± 0.26 & 2.88 ± 0.27 & 2.03 ± 0.26 & 0.48 ± 0.32 \\
19 & 1011 & 882 & 7:46:55.4 & +39:00:21.5 & 43.57 ± 2.12 & 9.14 ± 0.89 & 21.09 ± 1.20 & 47.81 ± 0.65 & 8.27 ± 0.45 & 7.25 ± 0.46 & 5.16 ± 0.46 & 1.61 ± 0.26 \\
20 & 1018 & 883 & 7:46:55.2 & +39:00:22.1 & 8.30 ± 0.97 & 2.14 ± 0.43 & 1.71 ± 0.40 & 7.70 ± 0.35 & 2.74 ± 0.26 & 2.27 ± 0.26 & 1.25 ± 0.26 & 0.61 ± 0.56 \\
21 & 1065 & 883 & 7:46:53.9 & +39:00:21.8 & 23.53 ± 1.66 & 6.09 ± 0.48 & 6.31 ± 0.52 & 31.59 ± 0.55 & 9.04 ± 0.39 & 6.69 ± 0.40 & 4.70 ± 0.40 & 1.58 ± 0.22 \\
22 & 1039 & 884 & 7:46:54.6 & +39:00:22.2 & 5.62 ± 0.66 & 1.49 ± 0.38 & 0.99 ± 0.34 & 6.36 ± 0.40 & 1.26 ± 0.26 & 1.85 ± 0.28 & 1.23 ± 0.27 & 1.06 ± 0.72 \\
23 & 986 & 885 & 7:46:56.1 & +39:00:22.7 & 15.29 ± 1.10 & 4.14 ± 0.64 & 9.96 ± 0.87 & 19.99 ± 0.36 & 3.31 ± 0.24 & 2.03 ± 0.25 & 1.97 ± 0.25 & 1.39 ± 0.42 \\
24 & 999 & 886 & 7:46:55.7 & +39:00:22.8 & 20.59 ± 2.33 & 4.20 ± 0.78 & 4.84 ± 0.84 & 5.16 ± 0.22 & 1.32 ± 0.15 & 0.90 ± 0.16 & 0.76 ± 0.16 & -- \\
25 & 1038 & 887 & 7:46:54.7 & +39:00:23.3 & 4.95 ± 0.71 & 1.58 ± 0.44 & 1.16 ± 0.40 & 5.38 ± 0.37 & 2.03 ± 0.27 & 1.13 ± 0.27 & 0.66 ± 0.27 & 0.46 ± 0.78 \\
26 & 1060 & 887 & 7:46:54.0 & +39:00:23.0 & 20.40 ± 1.42 & 5.47 ± 0.34 & 5.19 ± 0.37 & 24.69 ± 0.46 & 8.12 ± 0.33 & 5.22 ± 0.34 & 3.56 ± 0.33 & 1.21 ± 0.18 \\
27 & 1075 & 887 & 7:46:53.6 & +39:00:23.1 & 8.93 ± 1.15 & 2.42 ± 0.62 & 2.61 ± 0.65 & 12.39 ± 0.46 & 4.11 ± 0.34 & 2.34 ± 0.34 & 2.13 ± 0.34 & 1.55 ± 0.71 \\
28 & 1068 & 890 & 7:46:53.8 & +39:00:23.9 & 5.69 ± 0.86 & 2.03 ± 0.34 & 1.49 ± 0.36 & 8.32 ± 0.40 & 2.60 ± 0.29 & 2.41 ± 0.30 & 1.64 ± 0.30 & 0.96 ± 0.47 \\
29 & 1086 & 890 & 7:46:53.3 & +39:00:23.8 & 3.34 ± 0.46 & 0.73 ± 0.26 & 0.82 ± 0.28 & 3.20 ± 0.20 & 1.13 ± 0.20 & 0.85 ± 0.21 & 0.78 ± 0.21 & 1.14 ± 1.01 \\
30 & 969 & 891 & 7:46:56.6 & +39:00:24.8 & 2.30 ± 1.12 & 1.14 ± 0.65 & 1.22 ± 0.69 & 3.70 ± 0.40 & 1.23 ± 0.30 & 0.20 ± 0.29 & 0.69 ± 0.30 & 0.34 ± 1.79 \\
31 & 978 & 891 & 7:46:56.3 & +39:00:24.7 & 7.18 ± 0.95 & 1.15 ± 0.34 & 1.54 ± 0.36 & 4.92 ± 0.41 & 1.18 ± 0.29 & 1.16 ± 0.30 & 0.87 ± 0.29 & 1.07 ± 0.84 \\
32 & 1010 & 892 & 7:46:55.4 & +39:00:24.9 & 15.72 ± 1.46 & 5.46 ± 0.94 & 4.83 ± 0.92 & 15.75 ± 0.53 & 4.72 ± 0.38 & 3.41 ± 0.38 & 2.28 ± 0.38 & 0.02 ± 0.47 \\
33 & 1030 & 892 & 7:46:54.9 & +39:00:24.6 & 10.92 ± 1.02 & 3.31 ± 0.79 & 2.45 ± 0.71 & 11.33 ± 0.50 & 4.05 ± 0.36 & 2.55 ± 0.36 & 1.94 ± 0.36 & 0.47 ± 0.66 \\
35 & 985 & 893 & 7:46:56.1 & +39:00:25.1 & 3.78 ± 1.53 & 1.48 ± 0.77 & 0.35 ± 0.59 & 3.09 ± 0.34 & 0.84 ± 0.24 & 1.10 ± 0.26 & 0.50 ± 0.25 & -- \\
36 & 1018 & 893 & 7:46:55.2 & +39:00:25.3 & 23.39 ± 2.15 & 5.76 ± 0.52 & 4.28 ± 0.55 & 31.84 ± 0.67 & 9.97 ± 0.49 & 7.75 ± 0.49 & 4.67 ± 0.49 & 1.75 ± 0.25 \\
37 & 1051 & 893 & 7:46:54.3 & +39:00:24.9 & 12.28 ± 1.15 & 4.40 ± 0.39 & 2.73 ± 0.41 & 19.00 ± 0.42 & 6.38 ± 0.31 & 3.88 ± 0.31 & 2.77 ± 0.31 & 1.09 ± 0.24 \\
39 & 1013 & 895 & 7:46:55.3 & +39:00:25.7 & 9.87 ± 1.25 & 3.69 ± 0.69 & 2.08 ± 0.59 & 12.88 ± 0.52 & 3.95 ± 0.37 & 2.81 ± 0.38 & 2.02 ± 0.37 & 0.52 ± 0.52 \\
40 & 1034 & 895 & 7:46:54.8 & +39:00:25.8 & 4.28 ± 0.98 & 1.44 ± 0.52 & 0.78 ± 0.43 & 7.44 ± 0.42 & 2.93 ± 0.31 & 2.03 ± 0.31 & 1.27 ± 0.31 & 1.57 ± 1.02 \\
41 & 1043 & 896 & 7:46:54.5 & +39:00:26.0 & 7.90 ± 1.17 & 2.54 ± 0.34 & 1.80 ± 0.37 & 12.45 ± 0.47 & 4.10 ± 0.34 & 2.36 ± 0.34 & 1.01 ± 0.34 & 1.43 ± 0.38 \\
42 & 1008 & 898 & 7:46:55.5 & +39:00:26.8 & 8.25 ± 1.26 & 2.28 ± 0.45 & 0.67 ± 0.49 & 7.82 ± 0.29 & 2.65 ± 0.21 & 1.93 ± 0.22 & 1.33 ± 0.21 & 0.47 ± 0.55 \\
43 & 1001 & 899 & 7:46:55.7 & +39:00:27.2 & 7.94 ± 0.84 & 2.36 ± 0.54 & 1.98 ± 0.52 & 6.54 ± 0.34 & 2.48 ± 0.25 & 2.02 ± 0.25 & 0.98 ± 0.25 & -- \\
44 & 1013 & 899 & 7:46:55.4 & +39:00:27.0 & 7.91 ± 1.08 & 1.73 ± 0.36 & 1.40 ± 0.38 & 12.06 ± 0.43 & 4.24 ± 0.31 & 3.42 ± 0.32 & 2.01 ± 0.31 & 2.38 ± 0.57 \\
45 & 1092 & 899 & 7:46:53.2 & +39:00:26.8 & 2.52 ± 0.58 & 0.41 ± 0.39 & 0.21 ± 0.41 & 1.82 ± 0.20 & 0.69 ± 0.20 & 0.85 ± 0.21 & 0.35 ± 0.21 & 1.14 ± 4.65 \\
47 & 1010 & 902 & 7:46:55.4 & +39:00:28.2 & 4.65 ± 0.90 & 1.74 ± 0.43 & 1.27 ± 0.40 & 8.14 ± 0.31 & 2.85 ± 0.23 & 2.01 ± 0.23 & 1.56 ± 0.23 & 1.31 ± 0.69 \\
48 & 985 & 903 & 7:46:56.1 & +39:00:28.7 & 3.89 ± 1.02 & 1.32 ± 0.44 & 1.83 ± 0.51 & 4.40 ± 0.45 & 1.35 ± 0.32 & 0.96 ± 0.33 & 0.51 ± 0.33 & 0.41 ± 0.98 \\
49 & 993 & 903 & 7:46:55.9 & +39:00:28.5 & 4.92 ± 1.08 & 1.62 ± 0.41 & 0.60 ± 0.44 & 5.15 ± 0.38 & 2.23 ± 0.29 & 0.75 ± 0.28 & 1.14 ± 0.28 & 0.27 ± 0.71 \\
50 & 961 & 904 & 7:46:56.8 & +39:00:29.0 & 9.60 ± 1.46 & 1.77 ± 0.36 & 1.30 ± 0.39 & 8.65 ± 0.38 & 1.97 ± 0.27 & 2.04 ± 0.28 & 1.44 ± 0.28 & 1.42 ± 0.57 \\
51 & 1077 & 905 & 7:46:53.6 & +39:00:28.8 & 2.12 ± 0.54 & -- & 0.43 ± 0.29 & 1.61 ± 0.20 & 0.76 ± 0.21 & 0.34 ± 0.22 & 0.78 ± 0.22 & -- \\
52 & 1006 & 906 & 7:46:55.5 & +39:00:29.3 & 8.37 ± 0.92 & 3.18 ± 0.60 & 3.33 ± 0.62 & 14.09 ± 0.43 & 4.96 ± 0.32 & 2.67 ± 0.31 & 2.11 ± 0.31 & 1.16 ± 0.52 \\
53 & 957 & 908 & 7:46:56.9 & +39:00:30.1 & 1.79 ± 0.73 & 0.69 ± 0.36 & 0.67 ± 0.37 & 2.30 ± 0.25 & 0.60 ± 0.18 & 0.12 ± 0.18 & 0.48 ± 0.19 & 0.39 ± 1.59 \\
54 & 1014 & 908 & 7:46:55.3 & +39:00:30.0 & 5.69 ± 0.97 & 1.47 ± 0.40 & 0.22 ± 0.43 & 7.47 ± 0.45 & 3.04 ± 0.34 & 2.26 ± 0.34 & 1.58 ± 0.34 & 1.52 ± 0.77 \\
55 & 969 & 909 & 7:46:56.6 & +39:00:30.7 & 3.73 ± 0.63 & 1.09 ± 0.35 & 1.08 ± 0.37 & 4.83 ± 0.63 & 2.15 ± 0.47 & 1.30 ± 0.46 & 0.79 ± 0.45 & 1.15 ± 0.95 \\
56 & 1002 & 909 & 7:46:55.6 & +39:00:30.4 & 5.02 ± 0.83 & 1.48 ± 0.36 & 1.15 ± 0.38 & 4.39 ± 0.32 & 1.65 ± 0.24 & 0.95 ± 0.23 & 0.43 ± 0.23 & 0.09 ± 0.69 \\
57 & 1027 & 909 & 7:46:55.0 & +39:00:30.3 & 6.69 ± 0.69 & 1.77 ± 0.36 & 1.17 ± 0.39 & 12.30 ± 0.47 & 4.04 ± 0.34 & 2.90 ± 0.34 & 2.37 ± 0.34 & 2.36 ± 0.56 \\
58 & 953 & 912 & 7:46:57.0 & +39:00:31.6 & 3.65 ± 0.67 & 0.88 ± 0.31 & 1.89 ± 0.41 & 4.12 ± 0.35 & 0.45 ± 0.24 & 1.11 ± 0.26 & 0.60 ± 0.25 & 1.32 ± 1.03 \\
59 & 1006 & 914 & 7:46:55.5 & +39:00:32.0 & 2.38 ± 0.50 & -- & 0.83 ± 0.25 & 3.56 ± 0.18 & 1.36 ± 0.18 & 0.81 ± 0.19 & 0.63 ± 0.19 & -- \\
60 & 1039 & 915 & 7:46:54.6 & +39:00:32.3 & 0.86 ± 0.59 & 0.05 ± 0.29 & 0.11 ± 0.31 & 1.22 ± 0.21 & 0.10 ± 0.21 & 0.73 ± 0.22 & 0.42 ± 0.22 & -- \\
62 & 1003 & 916 & 7:46:55.6 & +39:00:32.7 & 2.37 ± 0.93 & 1.42 ± 0.28 & 0.55 ± 0.30 & 5.12 ± 0.33 & 1.92 ± 0.25 & 1.44 ± 0.25 & 0.89 ± 0.24 & 0.60 ± 0.57 \\
63 & 1064 & 917 & 7:46:53.9 & +39:00:32.9 & 10.88 ± 1.01 & 2.32 ± 0.35 & 3.84 ± 0.37 & 12.49 ± 0.45 & 1.63 ± 0.31 & 2.36 ± 0.32 & 1.28 ± 0.32 & 1.68 ± 0.41 \\
64 & 945 & 918 & 7:46:57.2 & +39:00:33.6 & 1.88 ± 0.60 & 0.56 ± 0.35 & 0.91 ± 0.37 & 2.04 ± 0.39 & 0.53 ± 0.27 & -- & -- & 0.64 ± 2.08 \\
67 & 1001 & 920 & 7:46:55.7 & +39:00:33.9 & 1.42 ± 0.31 & 0.90 ± 0.22 & -- & 1.97 ± 0.15 & 0.82 ± 0.16 & 0.44 ± 0.16 & 0.11 ± 0.16 & -- \\
68 & 1063 & 923 & 7:46:53.9 & +39:00:34.7 & 4.08 ± 0.60 & 1.32 ± 0.28 & 0.76 ± 0.30 & 4.72 ± 0.38 & 0.77 ± 0.27 & 1.90 ± 0.30 & 0.67 ± 0.28 & 0.60 ± 0.61 \\
69 & 1000 & 926 & 7:46:55.7 & +39:00:36.0 & 4.82 ± 0.92 & 1.14 ± 0.80 & 1.56 ± 0.94 & 6.21 ± 0.40 & 2.47 ± 0.30 & 1.27 ± 0.30 & 0.90 ± 0.30 & 1.71 ± 2.34 \\
71 & 1073 & 931 & 7:46:53.7 & +39:00:37.3 & 5.81 ± 0.95 & 1.74 ± 0.52 & 1.53 ± 0.51 & 5.82 ± 0.44 & 1.11 ± 0.31 & 1.92 ± 0.33 & 1.21 ± 0.33 & 0.40 ± 0.85 \\
72 & 968 & 934 & 7:46:56.6 & +39:00:38.7 & 1.07 ± 0.34 & -- & 0.48 ± 0.26 & 1.34 ± 0.16 & 0.48 ± 0.16 & -- & -- & -- \\
73 & 975 & 936 & 7:46:56.4 & +39:00:39.4 & 1.78 ± 1.47 & 1.10 ± 0.91 & -- & 3.20 ± 0.50 & 2.31 ± 0.44 & 2.14 ± 0.44 & 0.12 ± 0.39 & 0.02 ± 3.27 \\
74 & 1053 & 936 & 7:46:54.2 & +39:00:39.0 & 4.90 ± 1.66 & 1.08 ± 0.39 & 0.68 ± 0.42 & 5.69 ± 0.59 & 2.74 ± 0.45 & 1.35 ± 0.42 & 1.18 ± 0.42 & 1.62 ± 1.06 \\
75 & 925 & 937 & 7:46:57.8 & +39:00:39.7 & 1.30 ± 0.29 & 0.37 ± 0.17 & 0.39 ± 0.19 & 1.00 ± 0.13 & 0.32 ± 0.13 & 0.39 ± 0.14 & 0.09 ± 0.14 & -- \\
76 & 915 & 938 & 7:46:58.0 & +39:00:40.1 & 18.10 ± 1.10 & 5.02 ± 0.49 & 6.19 ± 0.54 & 25.92 ± 0.44 & 6.30 ± 0.31 & 5.23 ± 0.32 & 3.22 ± 0.32 & 1.57 ± 0.27 \\
77 & 1092 & 940 & 7:46:53.1 & +39:00:40.0 & 6.37 ± 1.23 & 1.97 ± 0.91 & -- & 4.11 ± 0.36 & 1.01 ± 0.25 & 1.47 ± 0.27 & 0.94 ± 0.26 & -- \\
78 & 1095 & 940 & 7:46:53.1 & +39:00:39.9 & 5.33 ± 0.96 & 1.10 ± 0.27 & 0.68 ± 0.29 & 4.63 ± 0.36 & 1.11 ± 0.25 & 1.33 ± 0.26 & 0.57 ± 0.26 & 1.02 ± 0.71 \\
79 & 921 & 943 & 7:46:57.9 & +39:00:41.7 & 4.37 ± 0.79 & 1.06 ± 0.24 & 0.81 ± 0.26 & 5.42 ± 0.26 & 2.09 ± 0.19 & 0.89 ± 0.19 & 0.43 ± 0.19 & 1.54 ± 0.64 \\
80 & 1133 & 943 & 7:46:52.0 & +39:00:40.9 & 1.87 ± 0.80 & 0.28 ± 0.29 & 1.05 ± 0.31 & 2.13 ± 0.27 & 0.32 ± 0.18 & 0.11 ± 0.19 & -- & -- \\
81 & 926 & 944 & 7:46:57.8 & +39:00:42.0 & 3.48 ± 0.96 & 1.02 ± 0.53 & -- & 1.73 ± 0.28 & 0.57 ± 0.21 & 0.85 ± 0.23 & 0.49 ± 0.21 & -- \\
82 & 1098 & 946 & 7:46:53.0 & +39:00:41.9 & 6.04 ± 0.81 & 1.03 ± 0.37 & 1.46 ± 0.40 & 4.63 ± 0.36 & 1.48 ± 0.26 & 0.88 ± 0.26 & 1.41 ± 0.27 & 1.20 ± 1.04 \\
83 & 915 & 951 & 7:46:58.0 & +39:00:44.5 & 4.73 ± 1.03 & 1.65 ± 0.34 & 1.24 ± 0.37 & 3.37 ± 0.56 & 0.50 ± 0.37 & 0.82 ± 0.40 & 0.67 ± 0.40 & -- \\
84 & 1102 & 952 & 7:46:52.8 & +39:00:44.1 & 21.17 ± 1.12 & 5.30 ± 0.52 & 8.20 ± 0.62 & 25.79 ± 0.42 & 5.19 ± 0.29 & 4.18 ± 0.30 & 2.39 ± 0.30 & 1.41 ± 0.27 \\
85 & 1084 & 954 & 7:46:53.4 & +39:00:44.6 & 2.09 ± 0.71 & 0.40 ± 0.38 & 0.26 ± 0.37 & 2.89 ± 0.31 & 1.30 ± 0.23 & 0.21 ± 0.22 & 0.45 ± 0.22 & 2.47 ± 4.78 \\
87 & 1109 & 957 & 7:46:52.7 & +39:00:45.6 & 3.47 ± 0.66 & 1.25 ± 0.38 & 0.83 ± 0.33 & 3.31 ± 0.28 & 0.70 ± 0.20 & 1.40 ± 0.22 & 0.61 ± 0.21 & -- \\
88 & 1083 & 959 & 7:46:53.4 & +39:00:46.2 & 6.38 ± 0.90 & 2.26 ± 0.39 & 1.41 ± 0.42 & 6.02 ± 0.37 & 1.69 ± 0.27 & 1.59 ± 0.28 & 1.42 ± 0.28 & -- \\
89 & 1101 & 959 & 7:46:52.9 & +39:00:46.2 & 19.40 ± 1.26 & 7.79 ± 0.60 & 9.98 ± 0.67 & 46.57 ± 0.57 & 11.28 ± 0.40 & 7.19 ± 0.41 & 6.03 ± 0.41 & 1.96 ± 0.21 \\
90 & 1078 & 964 & 7:46:53.5 & +39:00:47.9 & 0.97 ± 0.13 & 0.35 ± 0.10 & 0.56 ± 0.10 & 1.09 ± 0.11 & 0.25 ± 0.08 & 0.58 ± 0.10 & 0.40 ± 0.09 & 0.23 ± 0.82 \\
91 & 1100 & 964 & 7:46:52.9 & +39:00:47.9 & 8.89 ± 0.58 & 2.83 ± 0.22 & 2.57 ± 0.24 & 12.61 ± 0.28 & 3.73 ± 0.20 & 2.66 ± 0.21 & 2.04 ± 0.21 & 1.17 ± 0.22 \\
92 & 1097 & 965 & 7:46:53.0 & +39:00:48.2 & 19.04 ± 1.06 & 6.34 ± 0.36 & 5.82 ± 0.38 & 21.19 ± 0.32 & 5.05 ± 0.23 & 5.22 ± 0.24 & 3.15 ± 0.23 & 0.41 ± 0.16 \\
93 & 909 & 966 & 7:46:58.2 & +39:00:49.3 & 52.89 ± 2.49 & 16.65 ± 0.97 & 17.89 ± 1.02 & 79.75 ± 0.78 & 20.19 ± 0.54 & 12.37 ± 0.55 & 8.11 ± 0.55 & 1.37 ± 0.16 \\
94 & 1093 & 966 & 7:46:53.1 & +39:00:48.3 & 33.41 ± 1.48 & 9.39 ± 0.72 & 10.47 ± 0.77 & 32.17 ± 0.45 & 7.12 ± 0.31 & 7.56 ± 0.33 & 5.91 ± 0.33 & 0.47 ± 0.21 \\
96 & 925 & 968 & 7:46:57.8 & +39:00:50.0 & 2.48 ± 0.39 & 0.39 ± 0.24 & -- & 3.60 ± 0.16 & 1.37 ± 0.16 & 0.87 ± 0.17 & 0.64 ± 0.17 & 3.13 ± 1.91 \\
97 & 896 & 969 & 7:46:58.6 & +39:00:50.4 & 3.41 ± 0.36 & 1.68 ± 0.24 & 1.08 ± 0.26 & 2.23 ± 0.13 & 0.42 ± 0.14 & 0.74 ± 0.14 & -- & -- \\
99 & 1070 & 969 & 7:46:53.7 & +39:00:49.7 & 8.20 ± 1.12 & 1.82 ± 0.30 & 3.48 ± 0.32 & 9.87 ± 0.48 & 2.28 ± 0.33 & 2.38 ± 0.35 & 1.52 ± 0.34 & 1.70 ± 0.47 \\
100 & 916 & 971 & 7:46:58.0 & +39:00:50.9 & 13.84 ± 1.38 & 4.27 ± 0.56 & 4.24 ± 0.57 & 19.76 ± 0.40 & 5.45 ± 0.28 & 3.55 ± 0.28 & 2.68 ± 0.28 & 1.27 ± 0.35 \\
101 & 1087 & 971 & 7:46:53.2 & +39:00:50.1 & 102.60 ± 3.44 & 28.60 ± 1.82 & 59.86 ± 2.39 & 126.90 ± 0.85 & 23.65 ± 0.58 & 16.32 ± 0.60 & 12.08 ± 0.60 & 1.17 ± 0.17 \\
102 & 905 & 972 & 7:46:58.3 & +39:00:51.0 & 30.99 ± 1.99 & 11.81 ± 0.92 & 10.97 ± 0.91 & 41.69 ± 0.61 & 10.41 ± 0.42 & 7.17 ± 0.43 & 4.96 ± 0.43 & 0.55 ± 0.21 \\
103 & 1079 & 972 & 7:46:53.5 & +39:00:50.3 & 14.33 ± 1.23 & 4.70 ± 0.48 & 6.70 ± 0.51 & 14.48 ± 0.47 & 3.86 ± 0.33 & 4.43 ± 0.35 & 2.45 ± 0.34 & 0.19 ± 0.29 \\
104 & 1103 & 972 & 7:46:52.8 & +39:00:50.4 & 133.10 ± 4.34 & 42.36 ± 1.77 & 33.24 ± 1.66 & 196.10 ± 1.51 & 54.80 ± 1.06 & 31.09 ± 1.07 & 22.33 ± 1.07 & 1.28 ± 0.11 \\
105 & 899 & 973 & 7:46:58.5 & +39:00:51.4 & 5.13 ± 0.46 & 2.25 ± 0.35 & 2.87 ± 0.39 & 5.35 ± 0.31 & 1.32 ± 0.21 & 1.32 ± 0.22 & 1.08 ± 0.22 & -- \\
106 & 940 & 973 & 7:46:57.4 & +39:00:51.5 & 1.25 ± 0.39 & 1.06 ± 0.36 & 0.84 ± 0.34 & 2.60 ± 0.34 & 0.70 ± 0.23 & 0.36 ± 0.23 & 0.80 ± 0.24 & -- \\
107 & 884 & 974 & 7:46:58.9 & +39:00:51.9 & 1.87 ± 0.30 & -- & 0.33 ± 0.19 & 0.88 ± 0.12 & 0.38 ± 0.13 & 0.05 ± 0.13 & -- & -- \\
108 & 1067 & 974 & 7:46:53.8 & +39:00:51.3 & 3.92 ± 0.59 & 1.45 ± 0.19 & 1.30 ± 0.20 & 4.75 ± 0.28 & 1.29 ± 0.20 & 1.80 ± 0.22 & 0.76 ± 0.21 & 0.35 ± 0.38 \\
109 & 932 & 976 & 7:46:57.6 & +39:00:52.3 & 5.86 ± 1.06 & 1.20 ± 0.27 & 1.05 ± 0.29 & 7.60 ± 0.35 & 2.76 ± 0.26 & 1.66 ± 0.26 & 1.59 ± 0.26 & 2.12 ± 0.63 \\
110 & 1087 & 976 & 7:46:53.2 & +39:00:51.7 & 62.98 ± 2.81 & 14.11 ± 0.91 & 16.32 ± 0.98 & 53.50 ± 0.88 & 11.18 ± 0.60 & 14.78 ± 0.64 & 9.71 ± 0.63 & 0.74 ± 0.18 \\
111 & 923 & 977 & 7:46:57.8 & +39:00:52.6 & 15.98 ± 1.88 & 4.76 ± 0.75 & 4.14 ± 0.73 & 25.94 ± 0.60 & 8.07 ± 0.43 & 5.46 ± 0.44 & 3.80 ± 0.43 & 1.71 ± 0.43 \\
112 & 909 & 978 & 7:46:58.2 & +39:00:53.2 & 8.38 ± 1.14 & 3.02 ± 0.57 & 2.85 ± 0.57 & 17.37 ± 0.49 & 4.55 ± 0.33 & 2.67 ± 0.34 & 1.47 ± 0.34 & 1.86 ± 0.52 \\
113 & 1067 & 978 & 7:46:53.8 & +39:00:52.6 & 10.58 ± 0.90 & 2.07 ± 0.39 & 2.28 ± 0.41 & 11.52 ± 0.36 & 3.42 ± 0.26 & 3.02 ± 0.26 & 2.42 ± 0.26 & 1.77 ± 0.52 \\
114 & 903 & 980 & 7:46:58.4 & +39:00:53.9 & 11.72 ± 1.39 & 4.32 ± 0.58 & 4.89 ± 0.62 & 17.52 ± 0.53 & 5.01 ± 0.37 & 3.52 ± 0.38 & 2.10 ± 0.37 & 0.93 ± 0.37 \\
116 & 1113 & 980 & 7:46:52.5 & +39:00:53.1 & 4.79 ± 0.70 & 1.52 ± 0.26 & 0.88 ± 0.28 & 6.51 ± 0.35 & 1.51 ± 0.24 & 1.87 ± 0.26 & 1.34 ± 0.25 & 1.07 ± 0.48 \\
117 & 1081 & 981 & 7:46:53.4 & +39:00:53.3 & 29.08 ± 1.51 & 7.12 ± 0.69 & 8.63 ± 0.76 & 27.01 ± 0.57 & 6.02 ± 0.40 & 7.54 ± 0.42 & 5.50 ± 0.41 & 0.75 ± 0.27 \\
118 & 1102 & 981 & 7:46:52.9 & +39:00:53.2 & 82.33 ± 3.03 & 19.61 ± 0.98 & 16.59 ± 0.94 & 94.77 ± 0.94 & 23.16 ± 0.66 & 20.51 ± 0.68 & 14.67 ± 0.68 & 1.39 ± 0.14 \\
119 & 897 & 982 & 7:46:58.5 & +39:00:54.4 & 6.45 ± 0.75 & 2.16 ± 0.25 & 1.40 ± 0.27 & 5.07 ± 0.30 & 1.86 ± 0.22 & 1.00 ± 0.22 & 0.52 ± 0.22 & -- \\
120 & 932 & 982 & 7:46:57.6 & +39:00:54.2 & 0.66 ± 0.26 & 0.51 ± 0.11 & 0.20 ± 0.12 & 1.91 ± 0.17 & 0.55 ± 0.12 & 0.36 ± 0.12 & 0.55 ± 0.13 & 0.71 ± 0.65 \\
121 & 887 & 983 & 7:46:58.8 & +39:00:54.9 & 1.00 ± 0.35 & 0.04 ± 0.10 & -- & 0.90 ± 0.16 & 0.18 ± 0.10 & 0.31 ± 0.11 & 0.13 ± 0.11 & -- \\
122 & 936 & 983 & 7:46:57.4 & +39:00:54.6 & 2.27 ± 0.54 & 0.93 ± 0.20 & -- & 4.84 ± 0.30 & 1.99 ± 0.23 & 1.11 ± 0.22 & 0.44 ± 0.22 & 1.60 ± 0.62 \\
123 & 1068 & 983 & 7:46:53.8 & +39:00:54.1 & 24.75 ± 1.50 & 6.41 ± 0.56 & 7.98 ± 0.62 & 33.70 ± 0.58 & 11.13 ± 0.42 & 8.55 ± 0.42 & 6.34 ± 0.42 & 1.62 ± 0.24 \\
124 & 1109 & 983 & 7:46:52.7 & +39:00:53.9 & 3.37 ± 0.37 & 1.00 ± 0.18 & 0.66 ± 0.19 & 4.20 ± 0.20 & 1.62 ± 0.15 & 1.51 ± 0.16 & 0.98 ± 0.15 & 1.02 ± 0.50 \\
125 & 930 & 984 & 7:46:57.6 & +39:00:55.0 & 3.36 ± 0.53 & 0.86 ± 0.32 & 0.59 ± 0.35 & 3.83 ± 0.20 & 1.64 ± 0.21 & 0.65 ± 0.22 & 0.30 ± 0.22 & 1.18 ± 1.07 \\
126 & 1061 & 984 & 7:46:54.0 & +39:00:54.5 & 9.46 ± 0.99 & 2.71 ± 0.33 & 2.36 ± 0.36 & 8.87 ± 0.46 & 2.43 ± 0.32 & 3.20 ± 0.35 & 2.39 ± 0.34 & 0.35 ± 0.36 \\
127 & 1021 & 985 & 7:46:55.1 & +39:00:54.8 & 43.14 ± 5.51 & 93.62 ± 3.54 & 13.57 ± 2.56 & 851.10 ± 9.03 & 396.40 ± 6.70 & 87.79 ± 6.22 & 89.58 ± 6.26 & 3.08 ± 0.10 \\
128 & 1078 & 985 & 7:46:53.5 & +39:00:54.5 & 16.17 ± 0.96 & 3.72 ± 0.47 & 5.34 ± 0.54 & 14.42 ± 0.40 & 3.65 ± 0.28 & 4.94 ± 0.30 & 3.62 ± 0.29 & 0.80 ± 0.34 \\
129 & 1093 & 986 & 7:46:53.1 & +39:00:55.0 & 55.38 ± 2.69 & 11.86 ± 0.92 & 10.06 ± 0.89 & 48.33 ± 0.77 & 13.09 ± 0.55 & 13.78 ± 0.57 & 9.47 ± 0.56 & 0.94 ± 0.21 \\
130 & 941 & 987 & 7:46:57.3 & +39:00:56.0 & 2.33 ± 0.80 & 0.53 ± 0.30 & 0.42 ± 0.32 & 2.59 ± 0.34 & 1.40 ± 0.28 & 1.24 ± 0.28 & 0.78 ± 0.27 & 1.44 ± 1.79 \\
131 & 917 & 988 & 7:46:58.0 & +39:00:56.2 & 3.95 ± 0.68 & 1.40 ± 0.40 & 1.63 ± 0.43 & 6.97 ± 0.51 & 2.58 ± 0.36 & 1.29 ± 0.36 & 0.70 ± 0.36 & 1.47 ± 0.82 \\
132 & 1083 & 988 & 7:46:53.4 & +39:00:55.6 & 14.75 ± 1.06 & 3.18 ± 0.42 & 2.89 ± 0.42 & 11.29 ± 0.36 & 2.75 ± 0.25 & 3.96 ± 0.27 & 2.58 ± 0.26 & 0.57 ± 0.37 \\
133 & 935 & 992 & 7:46:57.5 & +39:00:57.6 & 3.72 ± 0.91 & 1.52 ± 0.56 & 1.56 ± 0.58 & 4.34 ± 0.41 & 1.20 ± 0.29 & 1.41 ± 0.30 & 1.18 ± 0.30 & -- \\
134 & 945 & 992 & 7:46:57.2 & +39:00:57.5 & 7.24 ± 4.32 & 2.24 ± 1.05 & 1.70 ± 0.96 & 4.49 ± 0.39 & 1.62 ± 0.29 & 1.06 ± 0.29 & 0.71 ± 0.28 & -- \\
135 & 1068 & 992 & 7:46:53.8 & +39:00:57.0 & 14.14 ± 1.27 & 3.59 ± 0.44 & 5.39 ± 0.52 & 14.31 ± 0.42 & 3.73 ± 0.29 & 4.87 ± 0.31 & 2.68 ± 0.30 & 0.88 ± 0.34 \\
136 & 903 & 993 & 7:46:58.4 & +39:00:58.2 & 6.87 ± 0.77 & 2.17 ± 0.55 & 1.68 ± 0.51 & 8.76 ± 0.36 & 2.77 ± 0.26 & 2.19 ± 0.26 & 1.69 ± 0.26 & 0.91 ± 0.70 \\
137 & 1095 & 993 & 7:46:53.0 & +39:00:57.1 & 5.52 ± 0.57 & 1.05 ± 0.34 & 1.77 ± 0.42 & 4.36 ± 0.29 & 1.52 ± 0.22 & 2.10 ± 0.23 & 0.95 ± 0.22 & 0.98 ± 0.91 \\
138 & 910 & 994 & 7:46:58.2 & +39:00:58.4 & 7.12 ± 0.82 & 2.48 ± 0.41 & 2.56 ± 0.43 & 10.30 ± 0.37 & 3.37 ± 0.27 & 2.51 ± 0.27 & 1.63 ± 0.27 & 0.99 ± 0.46 \\
139 & 952 & 995 & 7:46:57.0 & +39:00:58.4 & 1.94 ± 0.74 & 0.83 ± 0.29 & 0.27 ± 0.31 & 3.64 ± 0.38 & 1.65 ± 0.29 & 0.84 ± 0.29 & 0.90 ± 0.29 & 1.15 ± 1.03 \\
140 & 962 & 995 & 7:46:56.7 & +39:00:58.4 & 2.12 ± 1.12 & 0.13 ± 0.34 & 0.39 ± 0.36 & 2.83 ± 0.47 & 1.15 ± 0.36 & 1.15 ± 0.37 & 0.52 ± 0.35 & -- \\
141 & 1076 & 995 & 7:46:53.5 & +39:00:57.9 & 5.59 ± 0.53 & 1.33 ± 0.28 & 1.44 ± 0.29 & 5.16 ± 0.28 & 1.13 ± 0.19 & 1.49 ± 0.20 & 1.50 ± 0.21 & 0.82 ± 0.58 \\
142 & 905 & 996 & 7:46:58.3 & +39:00:58.9 & 3.76 ± 0.83 & 1.21 ± 0.25 & 0.77 ± 0.27 & 6.56 ± 0.30 & 2.12 ± 0.22 & 1.67 ± 0.22 & 1.38 ± 0.22 & 1.70 ± 0.58 \\
143 & 898 & 997 & 7:46:58.5 & +39:00:59.4 & 3.92 ± 1.06 & 0.69 ± 0.22 & 0.43 ± 0.24 & 3.67 ± 0.30 & 1.69 ± 0.23 & 0.81 ± 0.22 & 0.62 ± 0.22 & 1.66 ± 0.93 \\
144 & 916 & 997 & 7:46:58.0 & +39:00:59.2 & 8.03 ± 1.06 & 2.97 ± 0.59 & 3.57 ± 0.64 & 17.36 ± 0.43 & 4.58 ± 0.30 & 3.68 ± 0.31 & 1.94 ± 0.31 & 1.91 ± 0.54 \\
145 & 935 & 999 & 7:46:57.5 & +39:00:59.8 & 1.82 ± 0.96 & -- & 0.67 ± 0.23 & 1.97 ± 0.29 & 1.11 ± 0.24 & 0.96 ± 0.24 & 0.57 ± 0.23 & -- \\
146 & 1066 & 999 & 7:46:53.8 & +39:00:59.2 & 18.58 ± 1.18 & 5.17 ± 0.53 & 6.88 ± 0.60 & 17.24 ± 0.49 & 4.31 ± 0.34 & 4.16 ± 0.36 & 3.38 ± 0.35 & 0.40 ± 0.28 \\
147 & 1057 & 1000 & 7:46:54.1 & +39:00:59.8 & 7.93 ± 0.67 & 3.00 ± 0.37 & 2.63 ± 0.36 & 7.11 ± 0.38 & 2.15 ± 0.27 & 1.60 ± 0.27 & 1.18 ± 0.27 & -- \\
148 & 917 & 1001 & 7:46:58.0 & +39:01:00.4 & 2.49 ± 0.70 & 0.59 ± 0.31 & 0.29 ± 0.26 & 3.83 ± 0.22 & 1.19 ± 0.16 & 0.26 ± 0.16 & 0.64 ± 0.16 & 2.16 ± 1.58 \\
149 & 1073 & 1005 & 7:46:53.6 & +39:01:01.2 & 112.10 ± 3.86 & 28.43 ± 1.57 & 47.33 ± 1.93 & 154.50 ± 0.85 & 33.41 ± 0.59 & 22.09 ± 0.61 & 14.74 ± 0.61 & 1.71 ± 0.15 \\
150 & 1103 & 1006 & 7:46:52.8 & +39:01:01.4 & 3.89 ± 1.15 & 0.94 ± 0.37 & 0.76 ± 0.40 & 4.12 ± 0.38 & 1.18 ± 0.27 & 1.64 ± 0.29 & 1.14 ± 0.28 & 1.14 ± 1.16 \\
151 & 881 & 1007 & 7:46:59.0 & +39:01:02.6 & 2.18 ± 1.62 & -- & 0.99 ± 0.67 & 1.45 ± 0.38 & 0.71 ± 0.30 & 0.51 ± 0.30 & 0.54 ± 0.30 & -- \\
152 & 899 & 1007 & 7:46:58.5 & +39:01:02.5 & 1.17 ± 0.48 & -- & 0.59 ± 0.35 & 1.42 ± 0.19 & 0.39 ± 0.20 & 0.06 ± 0.20 & 0.26 ± 0.21 & -- \\
153 & 1063 & 1007 & 7:46:53.9 & +39:01:02.0 & 57.07 ± 2.23 & 14.35 ± 0.98 & 21.37 ± 1.16 & 67.34 ± 0.56 & 15.93 ± 0.39 & 12.34 ± 0.40 & 8.44 ± 0.40 & 1.32 ± 0.18 \\
154 & 1084 & 1007 & 7:46:53.3 & +39:01:01.8 & 9.42 ± 0.56 & 2.90 ± 0.56 & 2.75 ± 0.56 & 10.41 ± 0.54 & 2.63 ± 0.37 & 3.62 ± 0.40 & 3.46 ± 0.40 & 0.60 ± 0.54 \\
155 & 919 & 1010 & 7:46:57.9 & +39:01:03.4 & 2.81 ± 0.83 & 1.12 ± 0.45 & 0.71 ± 0.40 & 4.35 ± 0.51 & 0.89 ± 0.34 & 1.32 ± 0.36 & 0.05 ± 0.34 & 0.82 ± 1.21 \\
156 & 1088 & 1011 & 7:46:53.2 & +39:01:03.0 & 6.45 ± 0.74 & 1.20 ± 0.32 & 1.31 ± 0.34 & 7.11 ± 0.38 & 1.95 ± 0.27 & 2.74 ± 0.29 & 1.95 ± 0.28 & 1.94 ± 0.75 \\
157 & 930 & 1013 & 7:46:57.6 & +39:01:04.3 & 5.52 ± 1.22 & 3.55 ± 0.80 & 3.33 ± 0.79 & 15.36 ± 0.42 & 4.98 ± 0.31 & 3.09 ± 0.31 & 2.89 ± 0.31 & 1.10 ± 0.62 \\
158 & 1066 & 1013 & 7:46:53.8 & +39:01:03.8 & 60.91 ± 2.05 & 15.30 ± 0.95 & 22.55 ± 1.12 & 79.36 ± 0.73 & 17.44 ± 0.50 & 15.51 ± 0.52 & 11.29 ± 0.52 & 1.58 ± 0.17 \\
159 & 922 & 1015 & 7:46:57.8 & +39:01:04.9 & 2.13 ± 0.75 & 1.74 ± 0.43 & 0.63 ± 0.34 & 4.15 ± 0.40 & 0.90 ± 0.27 & 1.43 ± 0.29 & 0.31 ± 0.27 & -- \\
160 & 937 & 1015 & 7:46:57.4 & +39:01:05.1 & 2.52 ± 0.54 & 1.12 ± 0.32 & 0.65 ± 0.34 & 5.20 ± 0.37 & 2.70 ± 0.29 & 1.50 ± 0.28 & 1.24 ± 0.28 & 1.28 ± 0.80 \\
161 & 1089 & 1016 & 7:46:53.2 & +39:01:04.5 & 2.82 ± 0.47 & 0.27 ± 0.22 & 0.58 ± 0.29 & 1.85 ± 0.19 & 0.48 ± 0.13 & 0.32 ± 0.14 & 0.47 ± 0.14 & 2.37 ± 3.21 \\
162 & 1062 & 1017 & 7:46:53.9 & +39:01:05.2 & 27.64 ± 1.07 & 6.39 ± 0.43 & 8.74 ± 0.50 & 24.40 ± 0.32 & 5.39 ± 0.22 & 5.52 ± 0.22 & 3.60 ± 0.22 & 0.76 ± 0.18 \\
163 & 1057 & 1019 & 7:46:54.1 & +39:01:05.8 & 116.60 ± 3.69 & 29.79 ± 1.87 & 56.05 ± 2.38 & 125.40 ± 0.97 & 23.98 ± 0.66 & 19.38 ± 0.69 & 12.27 ± 0.68 & 1.02 ± 0.17 \\
164 & 1018 & 1020 & 7:46:55.2 & +39:01:06.4 & 1.90 ± 0.56 & -- & -- & 1.35 ± 0.24 & 0.98 ± 0.24 & 1.24 ± 0.26 & 0.39 ± 0.26 & -- \\
165 & 998 & 1021 & 7:46:55.7 & +39:01:06.7 & 1.09 ± 0.34 & 0.35 ± 0.21 & 0.85 ± 0.22 & 1.00 ± 0.15 & 0.51 ± 0.16 & -- & -- & -- \\
166 & 1003 & 1021 & 7:46:55.6 & +39:01:06.7 & 2.25 ± 0.44 & 0.63 ± 0.55 & 2.06 ± 0.84 & 3.61 ± 0.42 & 1.27 ± 0.31 & 0.97 ± 0.31 & 0.14 ± 0.30 & 1.84 ± 3.58 \\
167 & 1052 & 1021 & 7:46:54.2 & +39:01:06.4 & 28.07 ± 1.47 & 7.62 ± 0.66 & 12.81 ± 0.81 & 19.91 ± 0.51 & 6.61 ± 0.37 & 5.96 ± 0.38 & 4.38 ± 0.37 & -- \\
168 & 1061 & 1021 & 7:46:54.0 & +39:01:06.5 & 32.93 ± 1.16 & 6.45 ± 0.53 & 10.57 ± 0.65 & 37.80 ± 0.69 & 6.28 ± 0.44 & 7.90 ± 0.46 & 5.14 ± 0.46 & 1.91 ± 0.23 \\
169 & 925 & 1022 & 7:46:57.7 & +39:01:07.4 & 1.22 ± 0.37 & -- & 0.29 ± 0.20 & 1.29 ± 0.14 & 0.17 ± 0.14 & 0.48 ± 0.15 & 0.10 ± 0.15 & -- \\
170 & 956 & 1022 & 7:46:56.9 & +39:01:07.3 & 6.62 ± 1.13 & 2.81 ± 0.56 & 2.54 ± 0.55 & 12.52 ± 0.43 & 3.80 ± 0.31 & 2.77 ± 0.31 & 1.06 ± 0.31 & 1.18 ± 0.55 \\
171 & 1066 & 1022 & 7:46:53.8 & +39:01:06.6 & 37.06 ± 1.39 & 8.08 ± 0.55 & 10.21 ± 0.62 & 35.66 ± 0.58 & 7.28 ± 0.40 & 8.76 ± 0.42 & 5.95 ± 0.41 & 1.15 ± 0.19 \\
172 & 1086 & 1022 & 7:46:53.3 & +39:01:06.7 & 6.02 ± 1.01 & 0.85 ± 0.32 & 1.14 ± 0.34 & 5.25 ± 0.52 & 1.19 ± 0.35 & 1.66 ± 0.37 & 1.36 ± 0.37 & 2.04 ± 1.09 \\
174 & 1047 & 1023 & 7:46:54.3 & +39:01:07.1 & 18.03 ± 1.07 & 2.96 ± 0.33 & 4.45 ± 0.35 & 12.05 ± 0.44 & 3.62 ± 0.32 & 4.25 ± 0.33 & 2.90 ± 0.33 & 0.94 ± 0.31 \\
175 & 935 & 1024 & 7:46:57.5 & +39:01:08.1 & 0.73 ± 0.35 & -- & 0.05 ± 0.22 & 0.98 ± 0.12 & 0.08 ± 0.12 & 0.33 ± 0.12 & -- & -- \\
176 & 1069 & 1026 & 7:46:53.8 & +39:01:07.8 & 13.28 ± 1.28 & 3.21 ± 0.48 & 3.36 ± 0.50 & 11.42 ± 0.64 & 2.74 ± 0.42 & 4.15 ± 0.46 & 3.19 ± 0.45 & 0.57 ± 0.43 \\
177 & 1097 & 1026 & 7:46:52.9 & +39:01:07.9 & 2.75 ± 0.52 & 0.72 ± 0.24 & 0.17 ± 0.26 & 2.77 ± 0.63 & 0.15 ± 0.37 & 1.02 ± 0.43 & 0.34 ± 0.40 & 0.77 ± 1.15 \\
178 & 970 & 1027 & 7:46:56.5 & +39:01:08.8 & 0.71 ± 0.34 & -- & 0.49 ± 0.19 & 0.69 ± 0.13 & 0.32 ± 0.13 & 0.15 ± 0.13 & 0.09 ± 0.13 & -- \\
179 & 1077 & 1027 & 7:46:53.5 & +39:01:08.4 & 14.05 ± 1.39 & 2.39 ± 0.56 & 3.09 ± 0.63 & 14.28 ± 0.96 & 4.41 ± 0.63 & 5.32 ± 0.68 & 3.78 ± 0.66 & 1.96 ± 0.66 \\
180 & 957 & 1029 & 7:46:56.8 & +39:01:09.4 & 1.71 ± 0.39 & 0.89 ± 0.48 & 0.12 ± 0.34 & 2.18 ± 0.23 & 0.42 ± 0.17 & 0.96 ± 0.18 & 0.85 ± 0.18 & -- \\
182 & 1049 & 1029 & 7:46:54.3 & +39:01:08.9 & 10.52 ± 0.93 & 2.91 ± 0.53 & 3.28 ± 0.56 & 8.01 ± 0.48 & 2.91 ± 0.35 & 3.11 ± 0.37 & 2.33 ± 0.36 & -- \\
183 & 1034 & 1031 & 7:46:54.7 & +39:01:09.7 & 1.91 ± 0.42 & 0.53 ± 0.27 & 0.18 ± 0.29 & 1.44 ± 0.17 & 0.29 ± 0.17 & 0.89 ± 0.17 & 0.40 ± 0.18 & -- \\
185 & 888 & 1035 & 7:46:58.8 & +39:01:11.6 & 1.29 ± 0.61 & 0.47 ± 0.16 & 0.54 ± 0.17 & 1.64 ± 0.29 & 0.42 ± 0.20 & 0.26 ± 0.20 & 0.60 ± 0.21 & 0.52 ± 1.09 \\
187 & 977 & 1036 & 7:46:56.3 & +39:01:11.7 & 4.96 ± 1.05 & 0.93 ± 0.38 & 1.25 ± 0.40 & 4.94 ± 0.43 & 2.06 ± 0.33 & 1.80 ± 0.34 & 1.52 ± 0.33 & 1.66 ± 1.18 \\
188 & 1087 & 1038 & 7:46:53.2 & +39:01:11.9 & 2.72 ± 0.58 & 0.68 ± 0.35 & 0.32 ± 0.37 & 2.56 ± 0.19 & 1.08 ± 0.19 & 0.77 ± 0.20 & 0.55 ± 0.20 & 0.71 ± 1.52 \\
189 & 912 & 1039 & 7:46:58.1 & +39:01:12.8 & 2.71 ± 0.84 & 0.72 ± 0.25 & 0.50 ± 0.23 & 1.62 ± 0.23 & 0.50 ± 0.17 & 0.65 ± 0.18 & 0.10 ± 0.17 & -- \\
190 & 919 & 1039 & 7:46:57.9 & +39:01:12.7 & 4.41 ± 1.11 & 0.12 ± 0.30 & -- & 2.90 ± 0.38 & 1.25 ± 0.30 & 1.59 ± 0.32 & 1.14 ± 0.30 & -- \\
191 & 982 & 1039 & 7:46:56.2 & +39:01:12.7 & -- & -- & 0.32 ± 0.18 & 1.68 ± 0.21 & 0.62 ± 0.15 & 0.29 ± 0.15 & 0.49 ± 0.16 & -- \\
192 & 1027 & 1039 & 7:46:54.9 & +39:01:12.5 & 2.37 ± 0.58 & 0.47 ± 0.27 & 1.03 ± 0.29 & 1.72 ± 0.22 & 0.50 ± 0.23 & 0.46 ± 0.24 & 0.81 ± 0.24 & 0.65 ± 1.81 \\
193 & 1063 & 1039 & 7:46:53.9 & +39:01:12.1 & 12.69 ± 1.14 & 3.09 ± 0.44 & 4.24 ± 0.51 & 12.19 ± 0.54 & 2.49 ± 0.34 & 3.75 ± 0.37 & 2.78 ± 0.36 & 0.85 ± 0.41 \\
194 & 961 & 1040 & 7:46:56.7 & +39:01:13.0 & 2.89 ± 0.78 & -- & 0.74 ± 0.29 & 3.31 ± 0.44 & 1.17 ± 0.32 & 1.21 ± 0.33 & 1.12 ± 0.33 & -- \\
195 & 948 & 1041 & 7:46:57.1 & +39:01:13.2 & 4.53 ± 1.12 & 0.73 ± 0.43 & 0.48 ± 0.38 & 1.52 ± 0.32 & 0.33 ± 0.23 & 0.54 ± 0.25 & 0.40 ± 0.25 & -- \\
196 & 904 & 1042 & 7:46:58.3 & +39:01:13.8 & 3.65 ± 0.45 & 1.24 ± 0.28 & 0.29 ± 0.30 & 1.97 ± 0.18 & 0.43 ± 0.18 & 0.66 ± 0.19 & -- & -- \\
198 & 989 & 1042 & 7:46:56.0 & +39:01:13.5 & 0.90 ± 0.47 & 0.98 ± 0.33 & 0.65 ± 0.35 & 4.38 ± 0.21 & 1.48 ± 0.21 & 0.85 ± 0.22 & 0.73 ± 0.23 & 1.19 ± 0.94 \\
199 & 1064 & 1042 & 7:46:53.9 & +39:01:13.2 & 3.57 ± 0.38 & 0.28 ± 0.17 & 1.16 ± 0.25 & 5.14 ± 0.23 & 1.01 ± 0.15 & 1.33 ± 0.16 & 0.89 ± 0.16 & 4.96 ± 1.85 \\
200 & 922 & 1043 & 7:46:57.8 & +39:01:14.1 & 2.70 ± 0.52 & 0.74 ± 0.42 & 0.36 ± 0.34 & 2.70 ± 0.23 & 0.72 ± 0.17 & 1.02 ± 0.18 & 0.47 ± 0.17 & 0.63 ± 1.74 \\
201 & 1057 & 1043 & 7:46:54.1 & +39:01:13.5 & 23.14 ± 1.25 & 5.59 ± 0.50 & 7.40 ± 0.57 & 24.85 ± 0.51 & 6.69 ± 0.36 & 7.44 ± 0.37 & 5.44 ± 0.37 & 1.17 ± 0.25 \\
203 & 1006 & 1045 & 7:46:55.5 & +39:01:14.4 & 7.40 ± 1.19 & 2.63 ± 0.35 & 1.08 ± 0.38 & 9.52 ± 0.40 & 3.35 ± 0.29 & 2.15 ± 0.29 & 0.91 ± 0.28 & 0.62 ± 0.38 \\
204 & 1013 & 1045 & 7:46:55.3 & +39:01:14.5 & 2.03 ± 0.27 & 0.79 ± 0.25 & 0.09 ± 0.27 & 2.91 ± 0.13 & 1.21 ± 0.14 & 0.78 ± 0.14 & 0.64 ± 0.14 & 0.66 ± 0.88 \\
205 & 1077 & 1045 & 7:46:53.5 & +39:01:14.1 & 11.27 ± 1.16 & 3.97 ± 0.92 & 1.98 ± 0.72 & 15.71 ± 0.46 & 5.60 ± 0.34 & 2.67 ± 0.33 & 1.99 ± 0.33 & 0.86 ± 0.63 \\
206 & 910 & 1046 & 7:46:58.2 & +39:01:15.2 & 7.78 ± 1.10 & 2.19 ± 0.52 & 2.11 ± 0.52 & 7.25 ± 0.35 & 1.62 ± 0.25 & 2.36 ± 0.27 & 1.46 ± 0.26 & 0.38 ± 0.66 \\
207 & 1031 & 1046 & 7:46:54.8 & +39:01:14.7 & 8.50 ± 0.90 & 1.42 ± 0.43 & 1.14 ± 0.46 & 8.45 ± 0.42 & 2.93 ± 0.31 & 1.94 ± 0.31 & 2.00 ± 0.31 & 1.96 ± 0.86 \\
209 & 1000 & 1047 & 7:46:55.7 & +39:01:15.2 & 4.17 ± 0.73 & 1.10 ± 0.39 & 1.36 ± 0.43 & 3.30 ± 0.40 & 1.14 ± 0.30 & 1.24 ± 0.31 & 1.24 ± 0.31 & 0.11 ± 1.04 \\
210 & 1062 & 1047 & 7:46:53.9 & +39:01:14.8 & 8.67 ± 0.69 & 1.82 ± 0.31 & 1.97 ± 0.33 & 13.35 ± 0.46 & 2.05 ± 0.30 & 4.09 ± 0.33 & 2.32 ± 0.32 & 2.50 ± 0.47 \\
211 & 961 & 1048 & 7:46:56.7 & +39:01:15.7 & 5.58 ± 0.88 & 1.65 ± 0.31 & 0.79 ± 0.33 & 4.73 ± 0.26 & 1.48 ± 0.18 & 0.90 ± 0.18 & 0.46 ± 0.18 & -0.00 ± 0.53 \\
212 & 1023 & 1048 & 7:46:55.0 & +39:01:15.3 & 6.99 ± 0.97 & 1.43 ± 0.29 & 1.08 ± 0.31 & 5.76 ± 0.38 & 1.68 ± 0.27 & 1.26 ± 0.28 & 1.31 ± 0.28 & 0.91 ± 0.57 \\
213 & 955 & 1049 & 7:46:56.9 & +39:01:15.9 & 5.53 ± 0.43 & 1.11 ± 0.24 & 1.55 ± 0.25 & 4.79 ± 0.25 & 1.21 ± 0.18 & 0.88 ± 0.18 & 0.89 ± 0.18 & 1.09 ± 0.59 \\
214 & 1057 & 1049 & 7:46:54.1 & +39:01:15.6 & 16.09 ± 0.70 & 3.67 ± 0.43 & 4.38 ± 0.47 & 22.11 ± 0.46 & 4.02 ± 0.31 & 5.57 ± 0.32 & 3.91 ± 0.32 & 1.98 ± 0.32 \\
215 & 923 & 1050 & 7:46:57.8 & +39:01:16.3 & 2.94 ± 0.56 & 0.78 ± 0.23 & -- & 3.40 ± 0.23 & 1.27 ± 0.18 & 1.20 ± 0.18 & 0.72 ± 0.18 & 1.13 ± 0.85 \\
216 & 945 & 1050 & 7:46:57.2 & +39:01:16.3 & 12.56 ± 1.21 & 4.74 ± 0.55 & 4.36 ± 0.55 & 11.02 ± 0.46 & 2.93 ± 0.32 & 2.62 ± 0.33 & 1.73 ± 0.33 & -- \\
217 & 1009 & 1050 & 7:46:55.4 & +39:01:15.9 & 4.62 ± 0.48 & 1.28 ± 0.28 & 0.99 ± 0.30 & 4.39 ± 0.37 & 1.46 ± 0.27 & 1.43 ± 0.28 & 1.25 ± 0.28 & 0.47 ± 0.63 \\
218 & 1052 & 1050 & 7:46:54.2 & +39:01:15.8 & 73.56 ± 2.39 & 17.99 ± 1.03 & 28.27 ± 1.23 & 77.31 ± 0.75 & 16.54 ± 0.52 & 12.11 ± 0.53 & 7.75 ± 0.53 & 1.08 ± 0.15 \\
219 & 977 & 1051 & 7:46:56.3 & +39:01:16.4 & 4.99 ± 0.71 & 1.66 ± 0.30 & 1.37 ± 0.32 & 9.76 ± 0.31 & 2.78 ± 0.22 & 2.60 ± 0.23 & 0.87 ± 0.23 & 1.92 ± 0.49 \\
220 & 919 & 1052 & 7:46:57.9 & +39:01:16.9 & 3.81 ± 0.86 & 2.51 ± 0.60 & -- & 4.01 ± 0.30 & 1.24 ± 0.22 & 1.35 ± 0.23 & 0.60 ± 0.22 & -- \\
221 & 951 & 1052 & 7:46:57.0 & +39:01:16.9 & 3.31 ± 0.32 & 1.16 ± 0.29 & 1.88 ± 0.35 & 4.71 ± 0.24 & 1.08 ± 0.16 & 0.70 ± 0.17 & 0.67 ± 0.17 & 0.92 ± 0.70 \\
222 & 1015 & 1052 & 7:46:55.2 & +39:01:16.5 & 6.75 ± 1.14 & 1.36 ± 0.23 & 1.11 ± 0.24 & 7.11 ± 0.33 & 2.10 ± 0.23 & 2.07 ± 0.24 & 1.12 ± 0.24 & 1.61 ± 0.47 \\
223 & 1067 & 1053 & 7:46:53.8 & +39:01:16.8 & 6.70 ± 1.11 & 2.38 ± 0.75 & 1.65 ± 0.66 & 7.20 ± 0.55 & 2.19 ± 0.38 & 2.06 ± 0.39 & 2.43 ± 0.40 & 0.14 ± 0.89 \\
224 & 906 & 1054 & 7:46:58.2 & +39:01:17.8 & 26.81 ± 1.67 & 6.00 ± 0.63 & 9.96 ± 0.77 & 24.40 ± 0.47 & 4.61 ± 0.32 & 3.44 ± 0.33 & 3.34 ± 0.33 & 0.93 ± 0.29 \\
225 & 1043 & 1054 & 7:46:54.4 & +39:01:17.2 & 18.91 ± 1.38 & 4.57 ± 0.52 & 7.43 ± 0.56 & 12.23 ± 0.51 & 3.87 ± 0.37 & 3.06 ± 0.38 & 2.61 ± 0.38 & -- \\
226 & 915 & 1055 & 7:46:58.0 & +39:01:18.0 & 5.24 ± 0.51 & 0.69 ± 0.28 & 1.44 ± 0.30 & 3.78 ± 0.25 & 1.08 ± 0.18 & 1.37 ± 0.19 & 1.01 ± 0.19 & 1.72 ± 1.15 \\
227 & 970 & 1055 & 7:46:56.5 & +39:01:17.8 & 30.10 ± 1.26 & 10.27 ± 0.71 & 14.72 ± 0.83 & 65.35 ± 0.60 & 15.04 ± 0.42 & 8.85 ± 0.43 & 6.43 ± 0.43 & 2.13 ± 0.19 \\
228 & 1055 & 1055 & 7:46:54.1 & +39:01:17.3 & 14.08 ± 0.94 & 2.68 ± 0.48 & 3.35 ± 0.53 & 18.35 ± 0.42 & 3.23 ± 0.29 & 4.35 ± 0.30 & 2.70 ± 0.30 & 2.32 ± 0.48 \\
229 & 956 & 1058 & 7:46:56.9 & +39:01:18.7 & 11.14 ± 1.17 & 4.04 ± 0.54 & 7.76 ± 0.69 & 18.99 ± 0.37 & 4.36 ± 0.26 & 3.55 ± 0.27 & 2.87 ± 0.27 & 1.32 ± 0.36 \\
230 & 1052 & 1058 & 7:46:54.2 & +39:01:18.3 & 9.46 ± 1.10 & 2.43 ± 0.46 & 2.67 ± 0.48 & 7.62 ± 0.40 & 1.86 ± 0.27 & 2.37 ± 0.29 & 1.54 ± 0.28 & 0.24 ± 0.53 \\
231 & 905 & 1059 & 7:46:58.3 & +39:01:19.3 & 5.09 ± 0.91 & 0.84 ± 0.33 & 2.17 ± 0.46 & 3.93 ± 0.30 & 0.91 ± 0.21 & 0.66 ± 0.22 & 0.90 ± 0.22 & 1.32 ± 1.13 \\
232 & 1038 & 1059 & 7:46:54.6 & +39:01:18.7 & 3.38 ± 0.79 & 1.17 ± 0.49 & 0.51 ± 0.40 & 2.94 ± 0.38 & 1.58 ± 0.31 & 1.13 ± 0.30 & 1.13 ± 0.30 & -- \\
234 & 1048 & 1061 & 7:46:54.3 & +39:01:19.4 & 3.99 ± 0.50 & 1.21 ± 0.24 & 0.31 ± 0.26 & 3.77 ± 0.40 & 1.75 ± 0.30 & 1.42 ± 0.30 & 1.06 ± 0.30 & 0.22 ± 0.62 \\
236 & 945 & 1062 & 7:46:57.2 & +39:01:20.2 & 4.78 ± 0.89 & 1.19 ± 0.39 & 1.09 ± 0.41 & 4.74 ± 0.38 & 1.45 ± 0.27 & 0.94 ± 0.27 & 1.01 ± 0.28 & 0.87 ± 0.93 \\
237 & 1023 & 1062 & 7:46:55.0 & +39:01:19.8 & 2.94 ± 0.74 & -- & 0.61 ± 0.34 & 2.49 ± 0.34 & 1.15 ± 0.26 & 1.37 ± 0.28 & 0.36 ± 0.25 & -- \\
239 & 958 & 1064 & 7:46:56.8 & +39:01:20.7 & 0.62 ± 0.31 & -- & 0.28 ± 0.19 & 1.04 ± 0.10 & 0.32 ± 0.10 & 0.24 ± 0.11 & 0.20 ± 0.11 & -- \\
240 & 969 & 1064 & 7:46:56.5 & +39:01:20.8 & 2.90 ± 1.26 & 1.08 ± 0.34 & 0.69 ± 0.37 & 4.08 ± 0.39 & 1.24 ± 0.28 & 1.31 ± 0.29 & 0.47 ± 0.29 & 0.73 ± 0.92 \\
241 & 978 & 1064 & 7:46:56.3 & +39:01:20.8 & 1.75 ± 0.39 & 0.55 ± 0.24 & 0.17 ± 0.25 & 1.48 ± 0.15 & 0.62 ± 0.15 & 0.32 ± 0.16 & 0.11 ± 0.16 & -- \\
242 & 936 & 1067 & 7:46:57.4 & +39:01:21.8 & 2.01 ± 0.48 & -- & 0.30 ± 0.31 & 1.72 ± 0.24 & 0.37 ± 0.24 & 0.92 ± 0.25 & 0.32 ± 0.25 & -- \\
243 & 1009 & 1069 & 7:46:55.4 & +39:01:22.3 & 2.47 ± 0.36 & 0.84 ± 0.23 & 0.12 ± 0.25 & 1.10 ± 0.15 & 0.60 ± 0.15 & 0.58 ± 0.16 & 0.24 ± 0.16 & -- \\
245 & 1035 & 1071 & 7:46:54.7 & +39:01:22.8 & 3.41 ± 0.47 & 1.29 ± 0.30 & 0.58 ± 0.33 & 1.42 ± 0.19 & 0.46 ± 0.20 & 0.07 ± 0.21 & 0.22 ± 0.21 & -- \\
246 & 1065 & 1071 & 7:46:53.8 & +39:01:22.5 & 1.81 ± 0.90 & 1.83 ± 0.92 & 0.77 ± 0.67 & 2.50 ± 0.44 & 0.72 ± 0.30 & -- & 1.14 ± 0.34 & -- \\
247 & 1003 & 1072 & 7:46:55.5 & +39:01:23.1 & 3.19 ± 0.50 & -- & -- & 1.50 ± 0.24 & 0.77 ± 0.24 & 0.87 ± 0.25 & -- & -- \\
\hline
\end{longtable}
\end{scriptsize}

\newpage 

\section{Snapshots of the simulation.}
\label{sec:appendixB}

Figures~\ref{fig:morphology_evolution_stars_figure} and~\ref{fig:morphology_evolution_gas_figure} show snapshots of 
the stellar surface density and gas surface density, respectively, for the run with the interacting galaxies. Animations of the same simulation are presented in the on-line version of the paper.

\begin{figure*}
    \includegraphics[width=\linewidth]{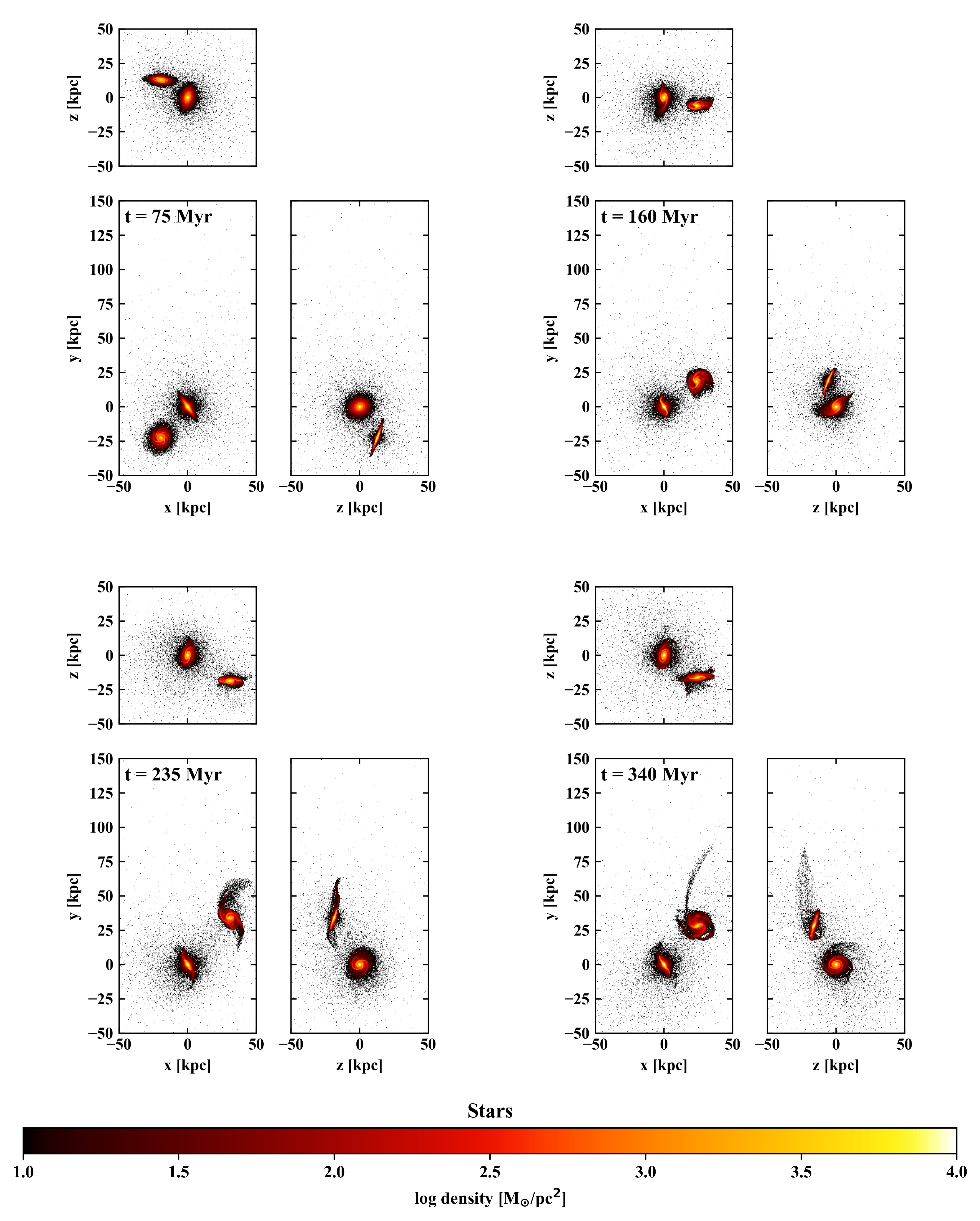}
    \caption{Stellar surface density maps from the numerical model at selected time steps indicated in the $xy$ plane, chosen to be the plane of the sky. The best match configuration occurs at $t=480\,\Myr$.}
    \label{fig:morphology_evolution_stars_figure}
\end{figure*}

\begin{figure*}
 \ContinuedFloat\includegraphics[width=\linewidth]{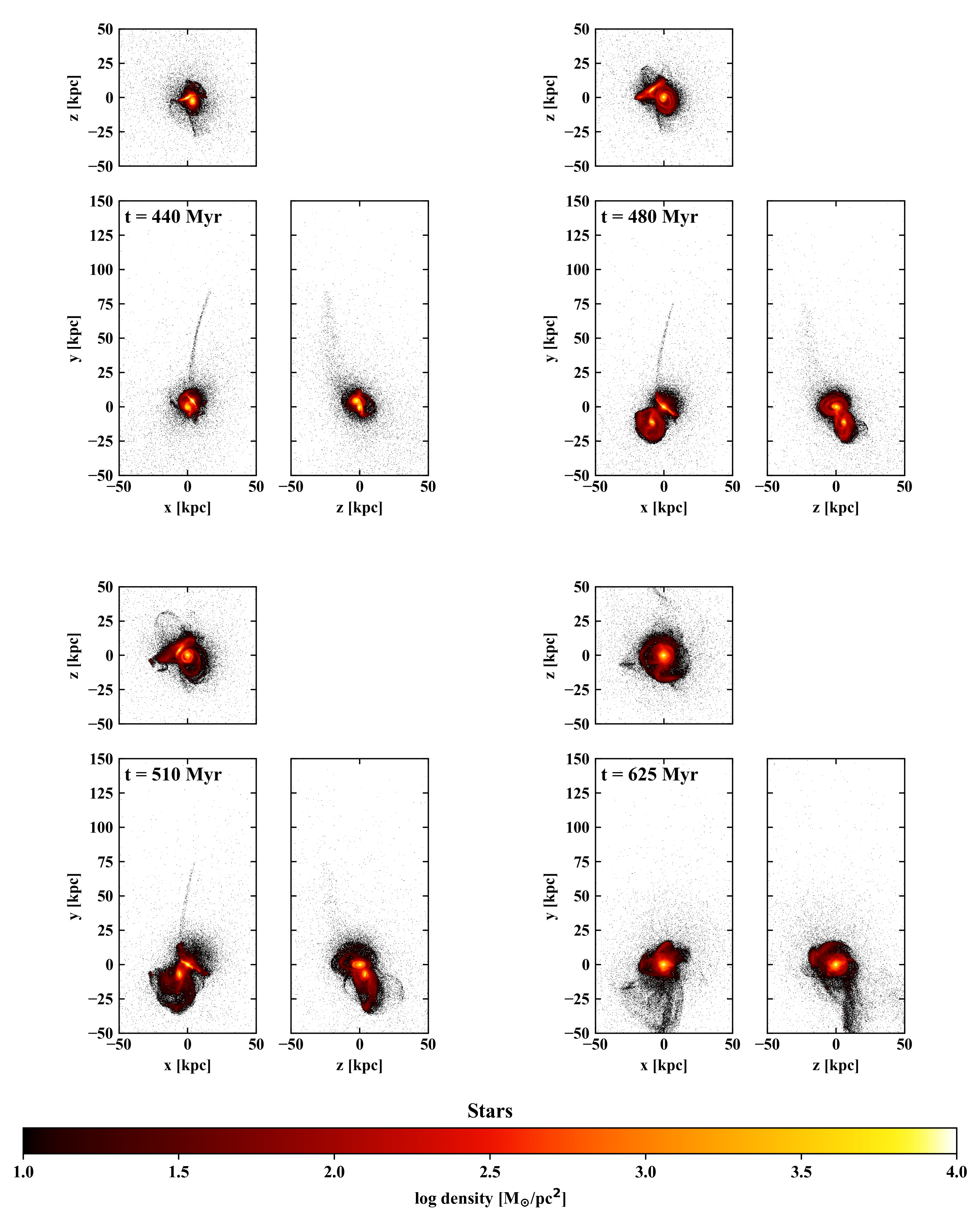}
    \caption{Cont.}
\end{figure*}

\begin{figure*}
    \includegraphics[width=\linewidth]{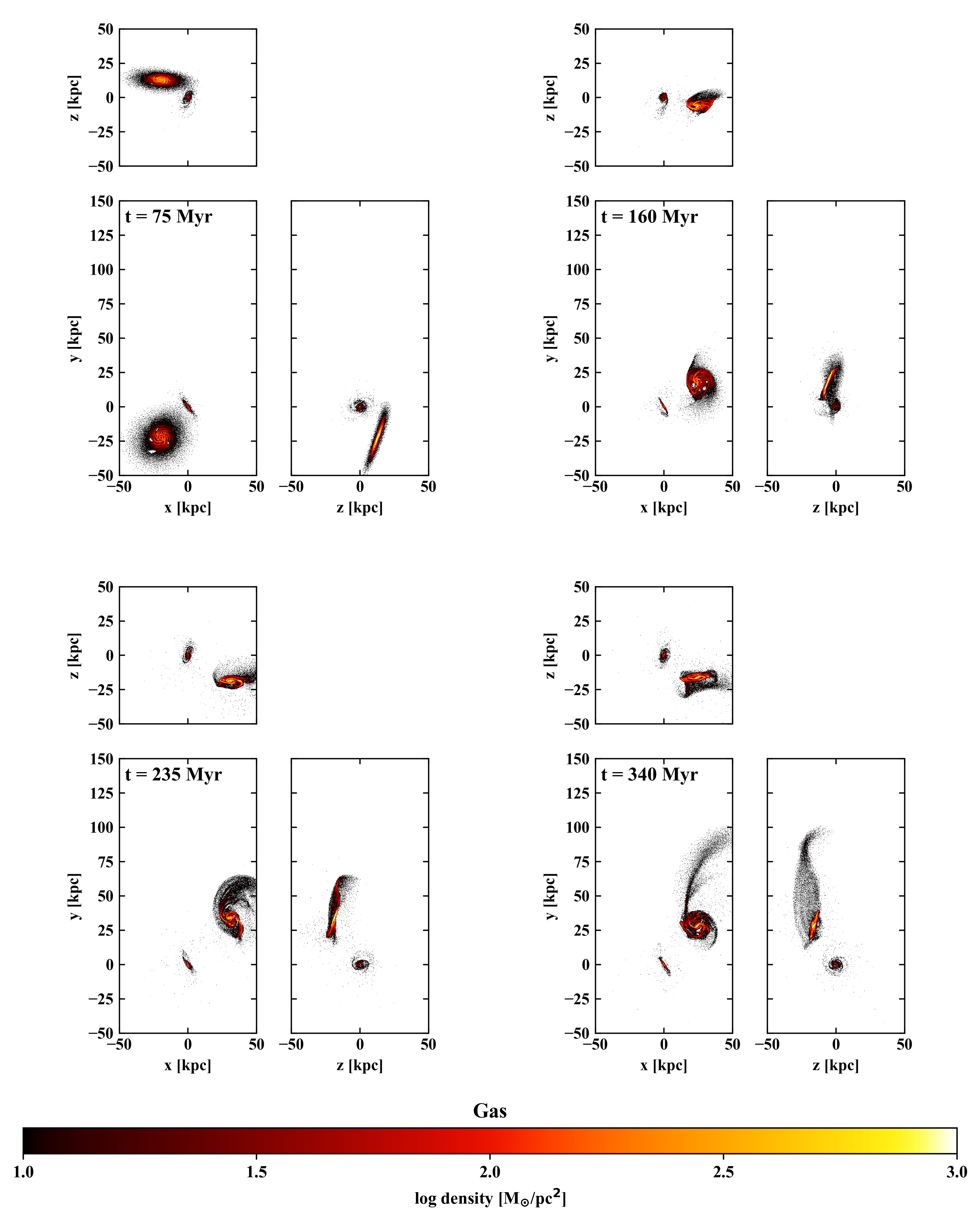}
    \caption{Same as Fig.~\ref{fig:morphology_evolution_stars_figure} but for the gaseous component.}
    \label{fig:morphology_evolution_gas_figure}
\end{figure*}

\begin{figure*}
 \ContinuedFloat\includegraphics[width=\linewidth]{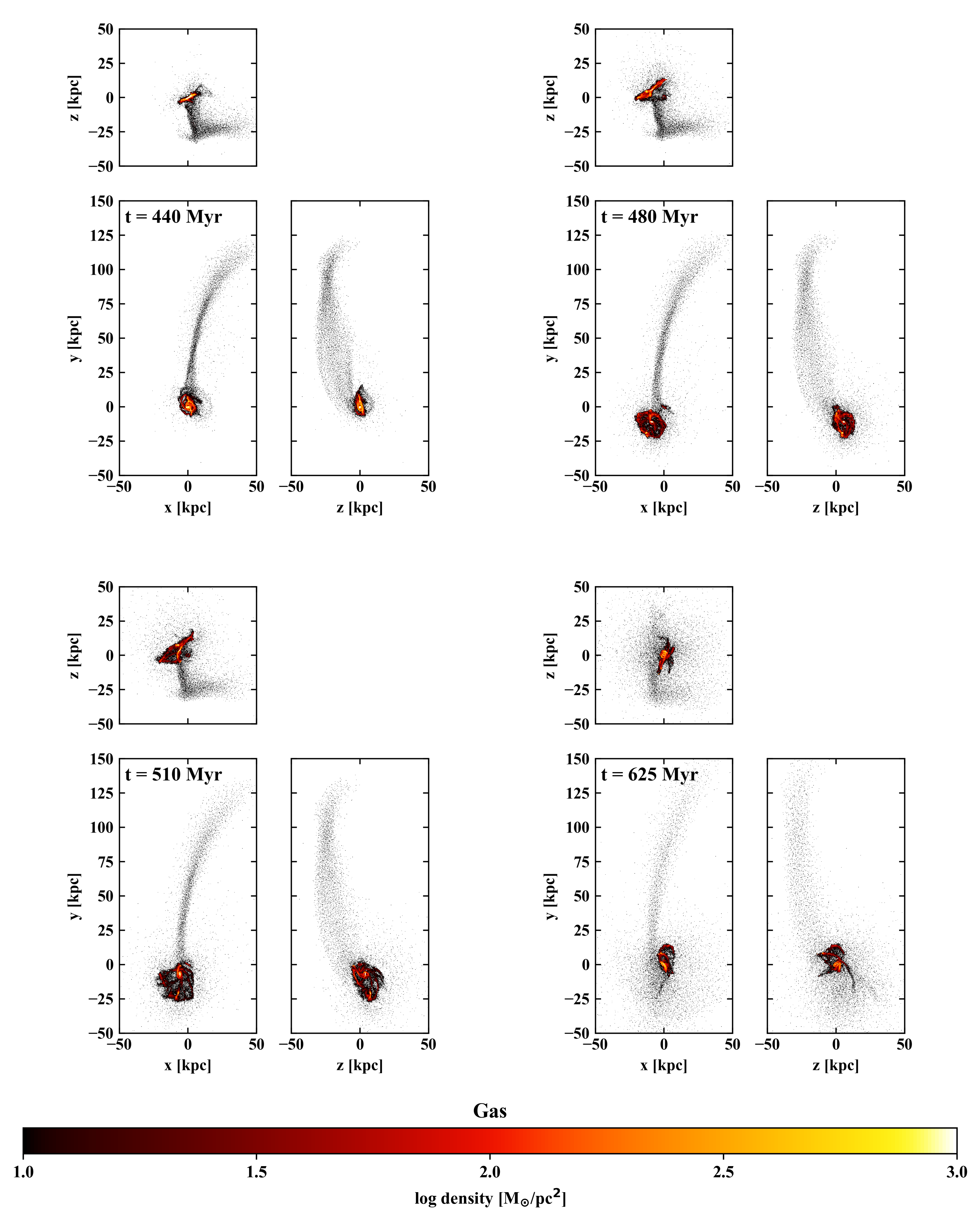}
    \caption{Cont.}
\end{figure*}

%\onecolumn
%\begin{longtable}{cccccccccc}
%\caption{Positions and fluxes (uncorrected for reddening) of the HII region complexes detected in Arp 143. Fluxes are in units of $10^{-16}\,\mathrm{erg\,s^{-1}\,cm^{-2}}$.}\label{tab:tableA1}
%%\centering
%\\ \hline
%\small \makecell{Region \\ ID} & 
%\small \makecell{R.A} &
%\small \makecell{Dec} &
%\small \makecell{[OII]$\uplambda$3727} &
%\small \makecell{H$\upbeta$} &
%\small \makecell{[OIII]$\uplambda$5007} &
%\small \makecell{H$\upalpha$} &
%\small \makecell{[NII]$\uplambda$6583} &
%\small \makecell{[SII]$\uplambda$6716} &
%\small \makecell{[SII]$\uplambda$6731} \\
%\hline
%\endfirsthead
%\caption*{Table \ref{tab:tableA1}: Cont.} \\
%\hline
%\small \makecell{Region \\ ID} & 
%\small \makecell{R.A} &
%\small \makecell{Dec} &
%\small \makecell{[OII]$\uplambda$3727} &
%\small \makecell{H$\upbeta$} &
%\small \makecell{[OIII]$\uplambda$5007} &
%\small \makecell{H$\upalpha$} &
%\small \makecell{[NII]$\uplambda$6583} &
%\small \makecell{[SII]$\uplambda$6716} &
%\small \makecell{[SII]$\uplambda$6731} \\
%\hline
%\endhead
%\multicolumn{10}{c}{\ldots}
% \endfoot
%\endlastfoot
%\small \makecell {1} &
%\small \makecell { 116.72192} &
%\small \makecell { 39.00437} &	\small \makecell {2.28$\pm$0.68} &
%\small \makecell {0.15$\pm$0.25} &
%\small \makecell {0.03$\pm$0.26} &
%\small \makecell {2.78$\pm$0.35} &
%\small \makecell {1.08$\pm$0.26} &
%\small \makecell {0.36$\pm$0.25} &
%\small \makecell {0.50$\pm$0.26} \\
%    \hline
%\end{longtable}
%\twocolumn
%%%%%%%%%%%%%%%%%%%%%%%%%%%%%%%%%%%%%%%%%%%%%%%%%%

%%%%%%%%%%%%%%%%%%%%%%%%%%%%%%%%%%%%%%%%%%%%%%%%%%

% Don't change these lines
\bsp	% typesetting comment
\label{lastpage}
\end{document}